%Paper: hep-th/9204064
%From: foda@mundoe.maths.mu.oz.au (Omar Foda RBA5)
%Date: Tue, 21 Apr 92 19:43:20 EST

%\input plainorder
\message{Cross-reference macros, B. Davies, version 3 March 1992.}
\par
% these macros replace symbolic labels by numbers for equations, references
% and other nominated things such as theorems, lemmas, etc
% compatible with plain TeX
\par
% the following commands are defined herein
\par
% \refeq
% \numberby
% \prefixby
% \order
% \printlabels
\par
% \refto
% \refis
% \listreferences
% \refbylabel
\par
% the following commands are re-defined herein
\par
% \eqno & \leqno            - so that they automatically pick up labels used
% \eqalignno & \leqalignno  - as equation numbers and replace them by numbers

% \beginsection  - so that it is not an outer definition and
%                  can be used as an argument of \numberby
\par
%                         Equation numbering
\par
% \eqno(label) & \leqno(label) is used for displayed equations
% &(label) is used for \eqalignn & \leqalignno
\par
% the string "label" is replaced by an equation number
% labels may have a subsidiary label separated by a colon
% eg. \eqno(red:a) \eqno(red:b) will give (5a) (5b) if red is 5
\par
% \eqno() and &() produce an equation number but no label
\par
% \refeq{label} - use this for referring to equations

% labels may be separated by commas, and pairs of labels by a hyphen
% eg. \refeq{label}, \refeq{l1,l2-l3,l4 a,l4 b} etc are valid
\par
%                  Section and/or subsection numbers
\par
% if section numbers are wanted, eq (2.1), then use \numberby{\command}
% where \command defines the beginning of a new section and nothing else.
\par
% for section and subsection numbers use \numberby{\command1,\command2}
\par
% to switch off again use \numberby{}
\par
%              Prefix - for example prefix A in appendices
\par
% \prefixby{string} will give a prefix "string" for all numbers
\par
% \numberby{\section}\prefixby{A} at the beginning of the first appendix.
% - section count is reset and the prefix A is added: e.g. (A1.5)
\par
% \numberby{}\prefixby{A} would give (A1), (A2) etc.
\par
%              Viewing the labels instead of the numbers
\par
% \printlabels does just that
\par
% Important:  but it must be used before use of \order since these definitions
% cannot not be re-made

% note that commands used as labels are turned into strings of characters
% with the \ replaced by @ - for example \alpha will print as a Greek
% alpha using \printlabels; as a label it is redefined to be @label.
\par
%                     Numbering of arbitrary types
\par
% the command \order{string} defines two new commands, \string, \refstring.
\par
% for example, \order{thm} can be used to number theorems automatically.
% \thm{label} is then replaced by the next theorem number and
% \refthm{label} is replaced by the corresponding number.
\par
% labels are stored with their type, so \eqno(abc) is distinct from \thm{abc}
\par
% prefixes and subsidiary labels work as for equation numbers
\par
%                        Forward referencing
\par
% a file jobname.xrf is produced for referencing labels not yet defined
% after changes, run twice to get such forward references correct
\par
%                    Citing literature references
\par
% \refto can have any number of labels which must be separated by commas
% each new label generates a new number in the order they are first read
\par
% labels are replaced by numbers in ascending order regardless of label order
% three or more consecutive numbers are replaced by firstnumber--lastnumber
\par
%                     Providing reference details
\par
% \refis{ref1} <reference details> \par
\par
% only cited references are used - input may be a master file
% note that a blank line is equivalent to \par
\par
% \listreferences - used for listing references after last \refis
% \refbylabel - the labels are used, no numbers are produced
\par
\catcode`@=11
\par
% look for xrf file from previous run
\par
\newif\if@xrf\@xrffalse   % this becomes true once the xrf file is set up
\def\l@bel #1 #2 #3>>{\expandafter\gdef\csname @@#1#2\endcsname{#3}}
\immediate\newread\xrffile
\def\xrf@n#1#2{\expandafter\expandafter\expandafter
\csname immediate\endcsname\csname #1\endcsname\xrffile#2}
\def\xrf@@n{\if@xrf\relax\else%
  \expandafter\xrf@n{openin}{ = \jobname.xrf}\relax%
  \ifeof\xrffile%
    \message{ no file \jobname.xrf - run again for correct forward references
}%
  \else%
    \expandafter\xrf@n{closein}{}\relax%
    \catcode`@=11 \input\jobname.xrf \catcode`@=12%
  \fi\global\@xrftrue%
  \expandafter\expandafter\csname immediate\endcsname%
  \csname  newwrite\endcsname\xrffile%
  \expandafter\xrf@n{openout}{ = \jobname.xrf}\relax\fi}
\par
% general macros which define \<string> and \ref<string> from \order{string}
% eg. \order{thm} defines \thm{label} and \refthm{label}
\par
\newcount\t@g
\par
\def\order#1{%
  \expandafter\expandafter\csname newcount\endcsname
  \csname t@g#1\endcsname\csname t@g#1\endcsname=0
  \expandafter\expandafter\csname newcount\endcsname
  \csname t@ghd#1\endcsname\csname t@ghd#1\endcsname=0
\par
  \expandafter\def\csname #1\endcsname##1{\xrf@@n\csname n@#1\endcsname##1:>}
\par
  \expandafter\def\csname n@#1\endcsname##1:##2>%
    {\def\n@xt{##1}\ifx\n@xt\empty%
     \expandafter\csname n@@#1\endcsname##1:##2:>
     \else\def\n@xt{##2}\ifx\n@xt\empty%
     \expandafter\csname n@@#1\endcsname\unp@ck##1 >:##2:>\else%
     \expandafter\csname n@@#1\endcsname\unp@ck##1 >:##2>\fi\fi}
\par
  \expandafter\def\csname n@@#1\endcsname##1:##2:>%
    {\edef\t@g{\csname t@g#1\endcsname}\edef\t@@ghd{\csname t@ghd#1\endcsname}%
     \ifnum\t@@ghd=\t@ghd\else\global\t@@ghd=\number\t@ghd\global\t@g=0\fi%
     \ifunc@lled{@#1}{##1}\global\advance\t@g by 1%
       {\def\n@xt{##1}\ifx\n@xt\empty%
       \else\writ@new{#1}{##1}{\pret@g\t@ghead\number\t@g}\expandafter%
       \xdef\csname @#1##1\endcsname{\pret@g\t@ghead\number\t@g}\fi}%
       {\pret@g\t@ghead\number\t@g}%
     \else\def\n@xt{##1}%
       \w@rnmess#1,\n@xt>\csname @#1##1\endcsname%
     \fi##2}%
\par
  \expandafter\def\csname ref#1\endcsname##1{\xrf@@n%
     \expandafter\each@rg\csname #1cite\endcsname{##1}}
\par
  \expandafter\def\csname #1cite\endcsname##1:##2,%
    {\def\n@xt{##2}\ifx\n@xt\empty%
     \csname #1cit@\endcsname##1:##2:,\else%
       \csname #1cit@\endcsname##1:##2,\fi}
\par
  \expandafter\def\csname #1cit@\endcsname##1:##2:,%
    {\def\n@xt{\unp@ck##1 >}\ifunc@lled{@#1}{\n@xt}%
      {\expandafter\ifx\csname @@#1\n@xt\endcsname\relax%
       \und@fmess#1,\n@xt>>>\n@xt<<%
       \else\csname @@#1\n@xt\endcsname##2\fi}%
     \else\csname @#1\n@xt\endcsname##2%
     \fi}}
\par
% for dealing with list of arguments separated by commas and
% dashes, and applying a specified control sequence to each
% so we can have \refeq{a,b}, etc
\par
\def\each@rg#1#2{{\let\thecsname=#1\expandafter\first@rg#2,\end,}}
\def\first@rg#1,{\callr@nge{#1}\apply@rg}
\def\apply@rg#1,{\ifx\end#1\let\n@xt=\relax%
\else,\callr@nge{#1}\let\n@xt=\apply@rg\fi\n@xt}
\par
\def\callr@nge#1{\calldor@nge#1-\end-}
\def\callr@ngeat#1\end-{#1}
\def\calldor@nge#1-#2-{\ifx\end#2\thecsname#1:,%
  \else\thecsname#1:,\hbox{\rm--}\thecsname#2:,\callr@ngeat\fi}
\par
% for turning any labels of the form \alpha into @alpha
\par
\def\unp@ck#1 #2>{\unp@@k#1@> @>>}
\def\unp@@k#1 #2>>{\ifx#2@\@np@@k#1\else\@np@@k#1@> \unp@@k#2>>\fi}
\def\@np@@k#1#2#3>{\ifx#2@\@@np@@k#1>\else\@@np@@k#1>\@np@@k#2#3>\fi}
\def\@@np@@k#1>{\ifcat#1\alpha\expandafter\@@np@@@k\string#1>\else#1\fi}
\def\@@np@@@k#1#2>{@#2}
\par
% for writing labels to jobname.xrf
\par
\def\writ@new#1#2#3{\xrf@@n\immediate\write\xrffile
  {\noexpand\l@bel #1 #2 {#3}>>}}
\par
% for checking if labels are undefined
\par
\def\ifunc@lled#1#2{\expandafter\ifx\csname #1#2\endcsname\relax}
\def\und@fmess#1#2,#3>{\ifx#1@%
  \message{ eqn label >>#3<< is undefined }\else
  \message{ #1#2 label >>#3<< is undefined }\fi}
\def\w@rnmess#1#2,#3>{\ifx#1@%
  \message{ Warning - duplicate eqn label >>#3<< }\else
  \message{ Warning - duplicate #1#2 label >>#3<< }\fi}
\par
% to precede equation numbers by a taghead
\par
\def\t@ghead{}
\newcount\t@ghd\t@ghd=0
\def\taghead#1{\gdef\t@ghead{#1}\global\advance\t@ghd by 1}
\par
% define equation labels with an @qn
\par
\order{@qn}
\par
% now define \eqno(label), \leqno(label), \refeq(label)
% also redefine \eqalinno & \leqalinno so that &() is picked up
\par
\let\eqno@@=\eqno
\def\eqno(#1){\xrf@@n\eqno@@\hbox{{\rm(}$\@qn{#1}${\rm)}}}
\par
\let\leqno@@=\leqno
\def\leqno(#1){\xrf@@n\leqno@@\hbox{{\rm(}$\@qn{#1}${\rm)}}}
\par
\def\refeq#1{\xrf@@n{{\rm(}$\ref@qn{#1}${\rm)}}}
\par
% the following only differs from plain TeX by picking up equation labels
\par
\def\eqalignno#1{\xrf@@n\displ@y \tabskip=\centering
  \halign to\displaywidth{\hfil$\displaystyle{##}$\tabskip=0pt
   &$\displaystyle{{}##}$\hfil\tabskip=\centering
   &\llap{$\eqaln@##$}\tabskip=0pt\crcr
   #1\crcr}}
\par
\def\leqalignno#1{\xrf@@n\displ@y \tabskip=\centering
  \halign to\displaywidth{\hfil$\displaystyle{##}$\tabskip=0pt
   &$\displaystyle{{}##}$\hfil\tabskip=\centering
    &\kern-\displaywidth\rlap{$\eqaln@##$}\tabskip\displaywidth\crcr
   #1\crcr}}
\par
\def\eqaln@#1#2{\relax\ifcat#1(\expandafter\eqno@\else\fi#1#2}
\def\eqno@(#1){\xrf@@n\hbox{{\rm(}$\@qn{#1}${\rm)}}}
\par
% macro to make order use section and/or subsection numbers as a prefix
\par
\def\n@@me#1#2>{#2}
\def\numberby#1{\xrf@@n
  \ifx\s@ction\undefined\else
  \expandafter\let\csname\s@@ve\endcsname=\s@ction\fi
  \ifx\subs@ction\undefined\else
  \expandafter\let\csname\subs@@ve\endcsname=\subs@ction\fi
  \numb@rby#1,>#1>}
\def\numb@rby#1,#2>#3>{\def\n@xt{#1}\ifx\n@xt\empty\taghead{}\else
  \def\n@xt{#2}\ifx\n@xt\empty\n@by#3>\else\n@@by#3>\fi\fi}
\def\n@by#1>{\ifx\s@cno\undefined\expandafter\expandafter
  \csname newcount\endcsname\csname s@cno\endcsname
  \csname s@cno\endcsname=0\else\s@cno=0\fi
  \xdef\s@@ve{\expandafter\n@@me\string#1>}
  \let\s@ction=#1\def#1{\global\advance\s@cno by 1
  \taghead{\number\s@cno.}\s@ction}}
\def\n@@by#1,#2>{\ifx\s@cno\undefined\expandafter\expandafter
  \csname newcount\endcsname\csname s@cno\endcsname
  \csname s@cno\endcsname=0\else\s@cno=0\fi
  \ifx\subs@cno\undefined\expandafter\expandafter
  \csname newcount\endcsname\csname subs@cno\endcsname
  \csname subs@cno\endcsname=0\else\subs@cno=0\fi
  \xdef\s@@ve{\expandafter\n@@me\string#1>}
  \let\s@ction=#1\def#1{\global\advance\s@cno by 1\global\subs@cno=0
  \taghead{\number\s@cno.}\s@ction}
  \xdef\subs@@ve{\expandafter\n@@me\string#2>}
  \let\subs@ction=#2\def#2{\global\advance\subs@cno by 1
  \taghead{\number\s@cno.\number\subs@cno.}\subs@ction}}
\par
% macro to add additional prefix to numbers
\par
\def\pret@g{}
\def\prefixby#1{\gdef\pret@g{#1}}
\par
% now for the literature referencing macros - these differ because
% the reference generates the numbers (rather than the object being referenced)
\par
% replace reference label by number
\par
\newcount\r@fcount\r@fcount=0
\newcount\r@fcurr
\newcount\r@fone
\newcount\r@ftwo
\newif\ifc@te\c@tefalse
\newif\ifr@feat
\par
\def\refto#1{{\rm[}\def\s@p{}\refn@te#1>>\refc@te#1>>{\rm]}}
\par
\def\refn@te#1>>{\refn@@te#1,>>}
\par
\def\refn@@te#1,#2>>{\r@fnote{\expandafter\unp@ck\str@pbl#1 >> >}%
   \def\n@xt{#2}\ifx\n@xt\empty\else\refn@@te#2>>\fi}
\par
\def\refc@te#1>>{\r@fcurr=0\r@featfalse\def\s@ve{}%
  {\loop\ifnum\r@fcurr<\r@fcount\advance\r@fcurr by 1\c@tefalse%
   \expandafter\refc@@te\number\r@fcurr>>#1,>>%
   \ifc@te\expandafter\refe@t\number\r@fcurr>>\fi\repeat\s@ve}}
\par
\def\refc@@te#1>>#2,#3>>{\def\n@xt{\expandafter\unp@ck\str@pbl#2 >> >}%
   \expandafter\refc@@@te\csname r@f\n@xt\endcsname>>#1>>%
   \def\n@xt{#3}\ifx\n@xt\empty\else\refc@@te#1>>#3>>\fi}
\par
\def\refc@@@te#1>>#2>>{\ifnum#2=#1\relax\c@tetrue\fi}
\par
\def\refe@t#1>>{\ifr@feat\ifnum\r@fone=\r@ftwo\res@cond#1>>%
   \else\reth@rd#1>>\fi\else\r@feattrue\ref@rst#1>>\fi}
\par
\def\ref@rst#1>>{\r@feattrue\r@fone=#1\r@ftwo=#1%
   \s@p\expandafter\relax\number\r@fone}%
\par
\def\res@cond#1>>{\advance\r@ftwo by 1\def\n@xt{#1}%
   \expandafter\ifnum\n@xt=\number\r@ftwo%
   \edef\s@ve{,\expandafter\relax\number\r@ftwo}\else,\ref@rst#1>>\fi}%
\par
\def\reth@rd#1>>{\advance\r@ftwo by 1\def\n@xt{#1}%
   \expandafter\ifnum\n@xt=\number\r@ftwo%
   \edef\s@ve{--\expandafter\relax\number\r@ftwo}\def\s@p{,}\else%
   \s@ve\def\s@ve{}\ref@rst#1>>\fi}%
\par
\def\r@fnote#1%
  {\ifunc@lled{r@f}{#1}\global\advance\r@fcount by 1%
   \expandafter\xdef\csname r@f#1\endcsname{\number\r@fcount}%
   \expandafter\gdef\csname r@ftext\number\r@fcount\endcsname%
   {\message{ Reference #1 to be supplied }%
   Reference $#1$ to be supplied\par}\fi}
\par
\def\str@pbl#1 #2>>{#1#2}
\par
% read list of references and match with those cited
\par
\def\refis#1 #2\par{\def\n@xt{\unp@ck#1 >}\r@fis\n@xt>>#2>>}
\def\r@fis#1>>#2>>{\ifunc@lled{r@f}{#1}\else
   \expandafter\gdef\csname r@ftext\csname
r@f#1\endcsname\endcsname{#2\par}\fi}
\par
% produce the references
\par
\def\listreferences{\global\r@fcurr=0%
  {\loop\ifnum\r@fcurr<\r@fcount\global\advance\r@fcurr by 1%
   \numr@f\number\r@fcurr>>\csname r@ftext\number\r@fcurr\endcsname>>%
   \repeat}}
\par
\def\numr@f#1>>#2>>{\vbox{\noindent\hang\hangindent=30truept%
   {\hbox to 30truept{\rm[#1]\hfill}}#2}\smallskip\par}
\par
% macro to print labels instead of numbers
\par
\def\printlabels{\global\@xrftrue\def\s@me##1{$##1$}
  \def\@qn##1{##1}\def\refeq##1{{\rm(}$##1${\rm)}}\refbylabel
  \def\listreferences{\relax}\def\referencefile{\relax}
  \def\order##1{\expandafter\let\csname ##1\endcsname=\s@me
                \expandafter\let\csname ref##1\endcsname=\s@me}}
\par
\def\refbylabel{\def\refto##1{[$##1$]}%
  \def\refis##1 ##2\par{\numr@f$##1$>>##2>>}\def\numr@f##1>>##2>>%
  {\noindent\hang\hangindent=30truept{{\rm[##1]}\ }##2\par}}
\par
% this only differs from plain TeX by not being outer
\par
\def\beginsection#1\par{\vskip0pt plus.3\vsize\penalty-250
  \vskip0pt plus -.3\vsize\bigskip\vskip\parskip
  \message{#1}\leftline{\bf#1}\nobreak\smallskip\noindent}
\par
\catcode`@=12
\par
%\endinput
\par
\par
\def\XXZ{{X\hskip-2pt X\hskip-2pt Z}}
\par
\def\title{Diagonalization of the XXZ Hamiltonian \cr
by Vertex Operators}
\def\author{Brian Davies${}^a$, Omar Foda${}^b$, Michio Jimbo${}^c$,\cr
Tetsuji Miwa${}^d$ and Atsushi Nakayashiki${}^e$\cr
{}\cr
${}^a$
Mathematics Department, the Faculties \cr
The Australian National University \cr
GPO Box 4, Canberra ACT 2601, Australia \cr
{}\cr
${}^b$
Institute for Theoretical Physics, University of Nijmegen, \cr
6525 ED Nijmegen, The Netherlands \cr
and \cr
Department of Mathematics, University of Melbourne, \cr
Parkville, Victoria 3052, Australia \cr
{}\cr
${}^c$
Department of Mathematics,Faculty of Science,\cr
Kyoto University, Kyoto 606, Japan \cr
{}\cr
${}^d$
Research Institute for Mathematical Sciences,\cr
Kyoto University, Kyoto 606, Japan\cr
{}\cr
${}^e$
The Graduate School of Science and Technology, \cr
Kobe University, Rokkodai, Kobe 657, Japan \cr
}
\par
\def\rhead{Diagonalization of $\XXZ$ Hamiltonian}
\def\lhead{B. Davies et al.}

%This is the style file for RIMS preprint.
\hsize=5.3truein
\vsize=7.8truein
\baselineskip=10pt
\font\twelvebf=cmbx12
\nopagenumbers
\headline={\ifnum\pageno=1 \hss\sl RIMS\hss \else \hdline\fi}
\def\hdline{\ifodd\pageno\rightheadline \else\leftheadline\fi}
\def\rightheadline{\tenrm\hfil{\it \rhead}\hfil\folio}
\def\leftheadline{\tenrm\folio\hfil{\it \lhead}\hfil}
\voffset=2\baselineskip
\vglue 2cm
%
%title
%
\centerline{\twelvebf
\vbox{
\halign{\hfil # \hfil\cr
\title\crcr}}}
%
%endtitle
%
\bigskip
%
%author
%
\centerline{\tenrm
\vbox{
\halign{\hfil#\hfil\cr
\author\crcr}}}
%
%endauthor
%
\vglue 1cm
%
%Figure
%
\def\Figure(#1|#2|#3)
{\midinsert
\vskip #2
\hsize 9cm
\raggedright
\noindent
{\bf Figure #1\quad} #3
\endinsert}
%
%Table
%
\def\Table #1. \size #2 \caption #3
{\midinsert
\vskip #2
\hsize 7cm
\raggedright
\noindent
{\bf Table #1.} #3
\endinsert}
%
%\address
%
\def\address #1
{\vglue 0.5cm
\halign{\quad\it##\hfil\cr
#1\crcr}}
%
%\eq
%
\def\eq#1\endeq
{$$\eqalignno{#1}$$}
%\qbox
%
\def\qbox#1{\quad\hbox{#1}\quad}
%
%\Proof
%
\def\Proof{\smallskip\noindent {\sl Proof.\quad}}
%
%\goth
%
\font\germ=eufm10
\def\goth#1{\hbox{\germ #1}}
%
%\Remark
%
\def\Remark{\smallskip\noindent {\sl Remark.\quad}}
%
%\Example
%
\def\Example{\smallskip\noindent {\sl Example.\quad}}
%
%\Definition
%
\def\Definition#1.#2{\smallskip\noindent {\sl Definition #1.#2\quad}}
%
%\subsec
%
\def\subsec(#1|#2){\medskip\noindent{#1}\hskip8pt{\sl {#2}}\quad}
%
%\qed
%
\def\qed{\qquad$\Fsquare(.2cm,{})$}
%
%\abstract
%
\def\abstract#1\endabstract{
\bigskip
\itemitem{{}}
{\bf Abstract.}
\quad
#1
\bigskip
}
\par
\def\m@th{\mathsurround=0pt}
\par
\def\fsquare(#1,#2){
\hbox{\vrule$\hskip-0.4pt\vcenter to #1{\normalbaselines\m@th
\hrule\vfil\hbox to #1{\hfill$\scriptstyle #2$\hfill}\vfil\hrule}$\hskip-0.4pt
\vrule}}
\par
\def\addsquare(#1,#2){\hbox{$
	\dimen1=#1 \advance\dimen1 by -0.8pt
	\vcenter to #1{\hrule height0.4pt depth0.0pt%
	\hbox to #1{%
	\vbox to \dimen1{\vss%
	\hbox to \dimen1{\hss$\scriptstyle~#2~$\hss}%
	\vss}%
	\vrule width0.4pt}%
	\hrule height0.4pt depth0.0pt}$}}
\par
\def\Fsquare(#1,#2){
\hbox{\vrule$\hskip-0.4pt\vcenter to #1{\normalbaselines\m@th
\hrule\vfil\hbox to #1{\hfill$#2$\hfill}\vfil\hrule}$\hskip-0.4pt
\vrule}}
\par
\def\Addsquare(#1,#2){\hbox{$
	\dimen1=#1 \advance\dimen1 by -0.8pt
	\vcenter to #1{\hrule height0.4pt depth0.0pt%
	\hbox to #1{%
	\vbox to \dimen1{\vss%
	\hbox to \dimen1{\hss$~#2~$\hss}%
	\vss}%
	\vrule width0.4pt}%
	\hrule height0.4pt depth0.0pt}$}}
\par
\def\hfourbox(#1,#2,#3,#4){%

\fsquare(0.3cm,#1)\addsquare(0.3cm,#2)\addsquare(0.3cm,#3)\addsquare(0.3cm,#4)}
\par
\def\Hfourbox(#1,#2,#3,#4){%

\Fsquare(0.4cm,#1)\Addsquare(0.4cm,#2)\Addsquare(0.4cm,#3)\Addsquare(0.4cm,#4)}
\par
\def\hthreebox(#1,#2,#3){%
	\fsquare(0.3cm,#1)\addsquare(0.3cm,#2)\addsquare(0.3cm,#3)}
\par
\def\htwobox(#1,#2){%
	\fsquare(0.3cm,#1)\addsquare(0.3cm,#2)}
\par
\def\vfourbox(#1,#2,#3,#4){%
%	\hbox{
	\normalbaselines\m@th\offinterlineskip
	\vtop{\hbox{\fsquare(0.3cm,#1)}
	      \vskip-0.4pt
	      \hbox{\fsquare(0.3cm,#2)}
	      \vskip-0.4pt
	      \hbox{\fsquare(0.3cm,#3)}
	      \vskip-0.4pt
	      \hbox{\fsquare(0.3cm,#4)}}}%}
\par
\def\Vfourbox(#1,#2,#3,#4){%
%	\hbox{
	\normalbaselines\m@th\offinterlineskip
	\vtop{\hbox{\Fsquare(0.4cm,#1)}
	      \vskip-0.4pt
	      \hbox{\Fsquare(0.4cm,#2)}
	      \vskip-0.4pt
	      \hbox{\Fsquare(0.4cm,#3)}
	      \vskip-0.4pt
	      \hbox{\Fsquare(0.4cm,#4)}}}%}
\par
\def\vthreebox(#1,#2,#3){%
%	\hbox{
	\normalbaselines\m@th\offinterlineskip
	\vtop{\hbox{\fsquare(0.3cm,#1)}
	      \vskip-0.4pt
	      \hbox{\fsquare(0.3cm,#2)}
	      \vskip-0.4pt
	      \hbox{\fsquare(0.3cm,#3)}}}%}
\par
\def\vtwobox(#1,#2){%
%	\hbox{
	\normalbaselines\m@th\offinterlineskip
	\vtop{\hbox{\fsquare(0.3cm,#1)}
	      \vskip-0.4pt
	      \hbox{\fsquare(0.3cm,#2)}}}%}
\par
\def\Hthreebox(#1,#2,#3){%
	\Fsquare(0.4cm,#1)\Addsquare(0.4cm,#2)\Addsquare(0.4cm,#3)}
\par
\def\Htwobox(#1,#2){%
	\Fsquare(0.4cm,#1)\Addsquare(0.4cm,#2)}
\par
\def\Vthreebox(#1,#2,#3){%
	\normalbaselines\m@th\offinterlineskip
	\vtop{\hbox{\Fsquare(0.4cm,#1)}
	      \vskip-0.4pt
	      \hbox{\Fsquare(0.4cm,#2)}
	      \vskip-0.4pt
	      \hbox{\Fsquare(0.4cm,#3)}}}
\par
\def\Vtwobox(#1,#2){%
	\normalbaselines\m@th\offinterlineskip
	\vtop{\hbox{\Fsquare(0.4cm,#1)}
	      \vskip-0.4pt
	      \hbox{\Fsquare(0.4cm,#2)}}}
\par
\def\twoone(#1,#2,#3){%
	\normalbaselines\m@th\offinterlineskip
	\vtop{\hbox{\htwobox({#1},{#2})}
	      \vskip-0.4pt
	      \hbox{\fsquare(0.3cm,#3)}}}
\par
\par
\def\Twoone(#1,#2,#3){%
	\hbox{
	\normalbaselines\m@th\offinterlineskip
	\vtop{\hbox{\Htwobox({#1},{#2})}
	      \vskip-0.4pt
	      \hbox{\Fsquare(0.4cm,#3)}}}}
\par
\def\threeone(#1,#2,#3,#4){%
	\normalbaselines\m@th\offinterlineskip
	\vtop{\hbox{\hthreebox({#1},{#2},{#3})}
	      \vskip-0.4pt
	      \hbox{\fsquare(0.3cm,#4)}}}
\par
\par
\def\Threeone(#1,#2,#3,#4){%
	\normalbaselines\m@th\offinterlineskip
	\vtop{\hbox{\Hthreebox({#1},{#2},{#3})}
	      \vskip-0.4pt
	      \hbox{\Fsquare(0.4cm,#4)}}}
\par
\def\Threetwo(#1,#2,#3,#4,#5){%
	\normalbaselines\m@th\offinterlineskip
	\vtop{\hbox{\Hthreebox({#1},{#2},{#3})}
	      \vskip-0.4pt
	      \hbox{\Htwobox({#4},{#5})}}}
\par
\def\threetwo(#1,#2,#3,#4,#5){%
	\normalbaselines\m@th\offinterlineskip
	\vtop{\hbox{\hthreebox({#1},{#2},{#3})}
	      \vskip-0.4pt
	      \hbox{\htwobox({#4},{#5})}}}
\par
\def\twotwo(#1,#2,#3,#4){%
	\normalbaselines\m@th\offinterlineskip
	\vtop{\hbox{\htwobox({#1},{#2})}
	      \vskip-0.4pt
	      \hbox{\htwobox({#3},{#4})}}}
\par
\def\Twotwo(#1,#2,#3,#4){%
	\normalbaselines\m@th\offinterlineskip
	\vtop{\hbox{\Htwobox({#1},{#2})}
	      \vskip-0.4pt
	      \hbox{\Htwobox({#3},{#4})}}}
\par
\def\twooneone(#1,#2,#3,#4){%
	\normalbaselines\m@th\offinterlineskip
	\vtop{\hbox{\htwobox({#1},{#2})}
	      \vskip-0.4pt
	      \hbox{\fsquare(0.4cm,#3)}
	      \vskip-0.4pt
	      \hbox{\fsquare(0.4cm,#4)}}}
\par
\def\Twooneone(#1,#2,#3,#4){%
	\normalbaselines\m@th\offinterlineskip
	\vtop{\hbox{\Htwobox({#1},{#2})}
	      \vskip-0.4pt
	      \hbox{\Fsquare(0.4cm,#3)}
	      \vskip-0.4pt
	      \hbox{\Fsquare(0.4cm,#4)}}}
\par
\def\a{\fsquare(0.3cm){1}\addsquare(0.3cm)(2)\addsquare(0.3cm)(3)}
\par
\def\b{\hbox{%
	\normalbaselines\m@th\offinterlineskip
	\vtop{\hbox{\fsquare(0.3cm){2}}\vskip-0.4pt\hbox{\fsquare(0.3cm){2}}}}}
\par
\def\c{\hbox{\normalbaselines\m@th\offinterlineskip%
	\vtop{\hbox{\a}\vskip-0.4pt\hbox{\b}}}}
\par
%This is the test of box.$X_{\a}$
%\b
%$X_{\c}-C$
\par
\dimen1=0.4cm\advance\dimen1 by -0.8pt
\def\ffsquare#1{%
	\fsquare(0.4cm,\hbox{#1})}
\par
\def\naga{%
	\hbox{$\vcenter to 0.4cm{\normalbaselines\m@th
	\hrule\vfil\hbox to 1.2cm{\hfill$\cdots$\hfill}\vfil\hrule}$}}
\par
\def\vnaga{\normalbaselines\m@th\baselineskip0pt\offinterlineskip%
	\vrule\vbox to 1.2cm{\vskip7pt\hbox to
\dimen1{$\hfil\vdots\hfil$}\vfil}\vrule}
\par

\par
\def\dvbox{\hbox{\normalbaselines\m@th\baselineskip0pt\offinterlineskip\vbox{%
	  \hbox{$\ffsquare 1$}\vskip-0.4pt\hbox{$\vnaga$}
\vskip-0.4pt\hbox{$\ffsquare N$}}}}
\par
%\section
%
\def\sec(#1){Sect.\hskip2pt#1}
\par
\def\sqr#1#2#3{{\vcenter{\vskip-#3pt\vbox{\hrule height.#2pt
   \vskip-.3pt\hbox{\vrule width.#2pt height#1pt \kern#1pt
   \vrule width.#2pt}
   \vskip-.3pt\hrule height.#2pt}}}}

\def\rightup#1{\smash{\mathop{\hbox to 7mm{\rightarrowfill}}
\limits^{\lower 3pt #1}}}
\def\mathpalette#1#2{\mathchoice{#1\displaystyle{#2}}
   {#1\textstyle{#2}}{#1\scriptstyle{#2}}{#1\scriptscriptstyle{#2}}}
\def\c@ncel#1#2{\ooalign{$\hfil#1\mkern1mu/\hfil$\crcr$#1#2$}}
\def\notni{\mathrel{\mathpalette{\c@ncel}{\ni}}}
\def\uelement{\mathop{\hbox{
\setbox1=\hbox{$\cup$}
\dimen1=\wd1
\dimen3=0.2pt
\divide \dimen1 by 2
\advance \dimen1 by \dimen3
\dimen5=\dimen1
\advance \dimen5 by -0.4pt
$\copy1\hskip-\dimen1 \vrule height\ht1 depth\dp1 width0.4pt\hskip \dimen5$
}}}
\par
\newdimen\ex
\ex.2326ex
\newdimen\z
\z0pt
\par
\def\varinjlim{\mathop{\vtop{\ialign{$##$\cr
\hfil{\fam\z lim}\hfil\cr\noalign{\nointerlineskip}%
{-}\mkern-6mu\cleaders\hbox{$\mkern-2mu{-}\mkern-2mu$}\hfill
\mkern-6mu{\to}\cr\noalign{\nointerlineskip\kern-\ex}\cr}}}}
\par
\def\varprojlim{\mathop{\vtop{\ialign{$##$\cr
\hfil{\fam\z lim}\hfil\cr\noalign{\nointerlineskip}%
{\leftarrow}\mkern-6mu\cleaders\hbox{$\mkern-2mu{-}\mkern-2mu$}\hfill
\mkern-6mu{-}\cr\noalign{\nointerlineskip\kern-\ex}\cr}}}}
\par

\par
\def\downtiefill{$\braceld\leaders\vrule\hfill\bracerd$}

\def\overtie#1{\mathop{\vbox{\ialign{##\crcr
    \noalign{\kern3pt}\downtiefill\crcr
    \noalign{\kern3pt\nointerlineskip}
    $\hfil\displaystyle{#1}\hfil$\crcr}}}\limits}
\par
\bigskip
\par
\centerline{\bf\it Dedicated to Professors Huzihiro Araki and
Noboru Nakanishi}
\centerline{\bf\it on the occasion of their sixtieth birthdays}
\par
\bigskip
\par
\order{thm}
\order{prop}
\def\C{{\bf C}}
\def\R{{\bf R}}
\def\bR{\overline{R}}
\def\Q{{\bf Q}}
\def\Z{{\bf Z}}

\def\H{{\cal H}}

\def\F{{\cal F}}

\def\P{{\cal P}}
\def\eL{{\cal  L}}
\def\la{\lambda}
\def\La{\Lambda}
\def\Ga{\Gamma}

\def\Th{\Theta}

\def\Vh{\widehat{V}}
\def\Lh{\widehat{L}}
\def\ep{\varepsilon}

\def\Hom{\hbox{Hom}}
\def\End{\hbox{End}}
\def\id{\hbox{id}}
\def\tr{\hbox{tr}}

\def\slt{\goth{sl}_2}
\def\slth{\widehat{\goth{sl}}(2)\hskip 1pt}
\def\uq{U_q(\goth{g})}
\def\goto#1{{\buildrel #1 \over \longrightarrow}}
\def\br#1{\langle #1 \rangle}
\def\bra#1{\langle #1 |}
\def\ket#1{|#1\rangle}
\def\brak#1#2{\langle #1|#2\rangle}

\def\vac{|\hbox{vac}\rangle}
\def\dvac{\langle \hbox{vac}|}

\def\wt{\hbox{wt}\,}

\def\Phih{\widetilde{\Phi}}

\def\ft{\tilde{f}}
\def\et{\tilde{e}}
\def\fti{\tilde{f}_i}
\def\eti{\tilde{e}_i}
%
%\goth
\font\germ=eufm10
\def\goth#1{\hbox{\germ #1}}
%
%Figure
%
\def\Figure(#1|#2|#3)
{\midinsert
\vskip #2
\hsize 9cm
\raggedright
\noindent
{\bf Figure #1\quad} #3
\endinsert}
%
%Table
%
\def\Table #1. \size #2 \caption #3
{\midinsert
\vskip #2
\hsize 7cm
\raggedright
\noindent
{\bf Table #1.} #3
\endinsert}
%\eq
%
\def\sectiontitle#1\par{\vskip0pt plus.1\vsize\penalty-250
 \vskip0pt plus-.1\vsize\bigskip\vskip\parskip
 \message{#1}\leftline{\bf#1}\nobreak\vglue 5pt}
\def\qed{\hbox{${\vcenter{\vbox{
    \hrule height 0.4pt\hbox{\vrule width 0.4pt height 6pt
    \kern5pt\vrule width 0.4pt}\hrule height 0.4pt}}}$}}
\def\subsec(#1|#2){\medskip\noindent{\it #1}\hskip8pt{\it #2}\quad}
\def\eq#1\endeq
{$$\eqalignno{#1}$$}
\def\leq#1\endeq
{$$\leqalignno{#1}$$}
%
%\qbox
%
\def\qbox#1{\quad\hbox{#1}\quad}
%
%\qqbox
%

%
%\nbox
%

%
%\akete
%

%
%\kitte
%

%
%\Proof
%
\def\Proof{\noindent {\sl Proof.\quad}}
\def\Remark{\smallskip\noindent {\sl Remark.\quad}}
%
%\Example
%
\def\Example{\smallskip\noindent {\sl Example.\quad}}
%
%\Definition
%
\def\Definition#1.#2{\smallskip\noindent {\sl Definition #1.#2\quad}}
%
%\subsec
%
\def\subsec(#1|#2){\medskip\noindent#1\hskip8pt{\sl #2}\quad}
%
%\qed
%
%\inpu box
%\def\qed{\qquad$\Fsquare(.2cm,{})$}
%
%\abstract
%
\def\abstract#1\endabstract{
\bigskip
\itemitem{{}}
{\bf Abstract.}
\quad
#1
\bigskip
}
%
%
%\section
%
\def\sec(#1){Sect.\hskip2pt#1}
\par
\par
%\input macro
%\onebox-----one box
%\twobox-----vertical two boxes
%\threebox---vertical three boxes
%\fourbox----vertical four boxes
%\four-------horizontal four boxes
%\fourtwo----1st line (four boxes) + 2nd line (two boxes)
%\threeone---1st line (three boxes) + 2nd line (one box)
%\threetwo---1st line (three boxes) + 2nd line (two boxes)
%\twoone-----1st line (two boxes) + 2nd line (one box)
%
\def\pn{\par\noindent}
\def\m@th{\mathsurround=0pt}
\par
\def\Fsquare(#1,#2){
\hbox{\vrule$\hskip-0.4pt\vcenter to #1{\normalbaselines\m@th
\hrule\vfil\hbox to #1{\hfill$#2$\hfill}\vfil\hrule}$\hskip-0.4pt
\vrule}}
\par
\def\Addsquare(#1,#2){\hbox{$
	\dimen1=#1 \advance\dimen1 by -0.8pt
	\vcenter to #1{\hrule height0.4pt depth0.0pt\vss%
	\hbox to #1{\hss{%
	\vbox to \dimen1{\vss%
	\hbox to \dimen1{\hss$~#2~$\hss}%
	\vss}\hss}%
	\vrule width0.4pt}\vss%
	\hrule height0.4pt depth0.0pt}$}}
\par
\def\bl{
        \dimen3=0.4cm \advance\dimen3 by -0.8pt
        \vrule height\dimen3 width\dimen3 depth0cm}
\par
\def\fourbox(#1,#2,#3,#4){%
%	\hbox{
	\normalbaselines\m@th\offinterlineskip
	\vcenter{\hbox{\Fsquare(0.4cm,#1)}
	      \vskip-0.4pt
	      \hbox{\Fsquare(0.4cm,#2)}
	      \vskip-0.4pt
	      \hbox{\Fsquare(0.4cm,#3)}
	      \vskip-0.4pt
	      \hbox{\Fsquare(0.4cm,#4)}}}%}
\par
\def\threebox(#1,#2,#3){%
	\normalbaselines\m@th\offinterlineskip
	\vcenter{\hbox{\Fsquare(0.4cm,#1)}
	      \vskip-0.4pt
	      \hbox{\Fsquare(0.4cm,#2)}
	      \vskip-0.4pt
	      \hbox{\Fsquare(0.4cm,#3)}}}
\par
\def\twobox(#1,#2){%
	\normalbaselines\m@th\offinterlineskip
	\vcenter{\hbox{\Fsquare(0.4cm,#1)}
	      \vskip-0.4pt
	      \hbox{\Fsquare(0.4cm,#2)}}}
\par
\def\onebox(#1){%
	\normalbaselines\m@th\offinterlineskip
	\vcenter{\hbox{\Fsquare(0.4cm,#1)}}}
\par
%---- yoko ---
\def\Htwobox(#1,#2){%
	\Fsquare(0.4cm,#1)\Addsquare(0.4cm,#2)}
\par
\def\Hthreebox(#1,#2,#3){%
	\Fsquare(0.4cm,#1)\Addsquare(0.4cm,#2)\Addsquare(0.4cm,#3)}
\par
\def\four(#1,#2,#3,#4){%
\Fsquare(0.4cm,#1)\Addsquare(0.4cm,#2)\Addsquare(0.4cm,#3)\Addsquare(0.4cm,#4)}
\par
\def\twoone(#1,#2,#3){%
%	\hbox{
	\normalbaselines\m@th\offinterlineskip
	\vcenter{\hbox{\Htwobox({#1},{#2})}
	      \vskip-0.4pt
	      \hbox{\Fsquare(0.4cm,#3)}}}
\par
\def\threeone(#1,#2,#3,#4){%
	\normalbaselines\m@th\offinterlineskip
	\vcenter{\hbox{\Hthreebox({#1},{#2},{#3})}
	      \vskip-0.4pt
	      \hbox{\Fsquare(0.4cm,#4)}}}
\par
\def\threetwo(#1,#2,#3,#4,#5){%
        \normalbaselines\m@th\offinterlineskip
	\vcenter{\hbox{\Hthreebox({#1},{#2},{#3})}
	      \vskip-0.4pt
	      \hbox{\Htwobox({#4},{#5})}}}
\par
\def\fourtwo(#1,#2,#3,#4,#5,#6){%
	\normalbaselines\m@th\offinterlineskip
	\vcenter{\hbox{\four({#1},{#2},{#3},{#4})}
	      \vskip-0.4pt
	      \hbox{\Htwobox({#5},{#6})}}}
\par
\par
\numberby{\beginsection}
\par
\def\eq#1\endeq{$$\eqalignno{#1}$$}
\def\XXX{{X\hskip-2pt X\hskip-2pt X}}
\def\XXZ{{X\hskip-2pt X\hskip-2pt Z}}

\def\H{H_{\XXZ}}
\def\slt{\goth{sl}(2)}
\def\slth{\widehat{\goth{sl}}(2)\hskip 1pt}
\def\Up{U'_q\bigl(\slth\bigr)}
\def\u{U_q\bigl(\slth \bigr)}
\def\uf{U_q\bigl(\slt\bigr)}
\def\pn{\par\noindent}
\def\o{\otimes}
\def\La{\Lambda}
\def\s(#1){\sigma^#1_{k+1}\sigma^#1_k}
\def\subsec(#1|#2){\medskip\noindent{\it #1}\hskip8pt{\it #2}\quad}
\def\Aff{\mathop{\rm A\hskip-1pt ff\/}}
\par
\par
%{\bf Abstract}\quad
%We diagonalize the anti-ferroelectric XXZ-Hamiltonian directly
%in the thermodynamic limit, where the model becomes invariant
%under the action of $U_q\bigl(\widehat{\goth{sl}}(2)\bigr)$.
%Our method is based
%on the representation theory of q-affine algebras, the related
%q-vertex operators, and q-KZ equation, and bypasses starting
%from a finite lattice, taking the thermodynamic limit and filling
%the Dirac sea.
%
%From the known algebraic structure of the eigenvectors of the
%corner transfer matrix of the model, we obtain the vacuum vector of
%the row transfer matrix. The rest of the eigenvectors are obtained by
%applying q-vertex operators, that act as particle creation
%operators in the space of eigenvectors.
%
%We check the agreement of our results with those obtained using
%the Bethe Ansatz, in a number of cases, and others obtained
%in the scaling limit: the $su(2)$-invariant Thirring model. As by-products,
%we obtain the eigenvectors of the corner transfer matrix as infinite
%correlation functions of certain q-vertex operators, and give
%a one-line derivation of the known one-point functions of the model.
\par

\par

\par
{\narrower\bigskip{\noindent\bf Abstract.\quad}
We diagonalize the anti-ferroelectric XXZ-Hamiltonian directly
in the thermodynamic limit, where the model becomes invariant
under the action of $U_q\bigl(\widehat{\goth{sl}}(2)\bigr)$.
Our method is based
on the representation theory of quantum affine algebras, the related
vertex operators and KZ equation,
and thereby
bypasses the usual process of
starting from a finite lattice, taking the thermodynamic limit and filling
the Dirac sea. From recent results on the algebraic structure of the
corner transfer matrix of the model, we obtain the vacuum vector of
the Hamiltonian. The rest of the eigenvectors are obtained by
applying the vertex operators,
which act as particle creation
operators in the space of eigenvectors.
\par
We check the agreement of our results with those obtained using
the Bethe Ansatz in a number of cases, and with others obtained
in the scaling limit --- the $su(2)$-invariant Thirring model.

\bigskip}
\par
\par
\beginsection \S0. Introduction
\par
\catcode`@=11
\s@cno=0
\catcode`@=12
\par
\subsec(0.1|A diagonalization scheme)
In this paper we give a new scheme for diagonalizing
the 1-dimensional $\XXZ$ spin chain
\eq
\H=-{1\over2}
&\sum_{k=-\infty}^\infty\bigl(\s(x)+\s(y)+\Delta\s(z)\bigr),&(XXZ)\cr
\endeq
for $\Delta<-1$,
directly in the thermodynamic limit, using
the representation theory of the quantum affine
algebra $\u$: we consider the infinite tensor product
\eq
W=&\cdots\o\C^2\o\C^2\o\C^2\o\C^2\o\cdots,&(InfTen)\cr
\endeq
on which $\H$ formally acts, rather than starting with a finite product
and subsequently taking the thermodynamic limit. The reason is that
the model is $\u$-symmetric only in the thermodynamic limit,
and the presence of that symmetry is central to our approach.
On the other hand, working directly in that limit
makes it difficult to give
rigorous analytic proofs, and a number of our statements
will be given as conjectures, supported by explicit calculations.
\par

Our scheme is a direct descendant of the recent works
\refto{FR,(KMN)^2,FM,DJO},
and
Smirnov's picture of the $su(2)$-invariant Thirring model \refto{Sm},
which is the continuum limit of the $\XXZ$ model.
(See also \refto{BerLeC,FelLeC}.)
More broadly, it originates
in many of the developments that took place
in the past two decades in the field of integrable models in
quantum field theory and statistical mechanics.
It is remarkable that in the context of this simple off-critical model,
one can recognize many of these developments, and for the rest of this
introduction, we wish to recall certain aspects that are relevant to our work.
\par
The peculiarities of this branch of science are twofold:
mathematically, its significance is that the systems we are dealing with
have infinite degrees of freedom (unlike those in conventional mathematics,
in which we manipulate finite degrees of freedom),
while physically, it is the fact that they are integrable
(unlike most of the physical problems) because of their infinite,
and sometimes hidden symmetries.
The symmetries serve as
the magical word in Arabian Nights that opens the door
between the two worlds of infinite and finite degrees of freedom.
Thus, through the study of the integrable models
we have been discovering many unexpected links between
different branches of mathematics such as representation theory,
differential equations, combinatorics, topology, algebraic geometry, and so on.
\par
Let us discuss specifically the integrability and the symmetry
of lattice systems.
For two-dimensional lattice systems, the integrability is commonly understood
as the existence of a family of commuting transfer matrices.
In the language of one-dimensional quantum chains, it is understood as
the existence of infinitely many mutually commuting conservation laws.
Let us call them the infinite abelian symmetries.
Because they are commuting we can simultaneously diagonalize them, and
in most cases the spaces of common eigenvectors are finite-dimensional.
In other words, the infinite abelian symmetries reduce
the degeneracies of the spectrum from infinite to finite.
This is the reason for the solvability. But it is not the whole story.
In this introduction, we will discuss other types of symmetries that
we will refer to as the non-abelian symmetries and the dynamical symmetries.
\par
\subsec(0.2| The Bethe Ansatz)
The Bethe Ansatz was invented in order to solve
the $\XXX$ model, then it was
applied to the $\XXZ$ and many other models.
Let us discuss the $\XXX$ model specifically.
The Bethe Ansatz reduces the diagonalization of the Hamiltonian,
a $2^N\times2^N$ (huge!) matrix where $N$ is the size of the system, to a
system of $m$ coupled algebraic equations,
where $m$ is less than or equal to $N/2$.
For a finite but large $N$, this system is
very complicated, and far from ``solvable''. (It is not
completely settled even for $m=2$ \refto{EKS}.)
On the other hand, and surprisingly, in the thermodynamic
limit ($N\rightarrow\infty$) the system of algebraic equations
changes into a system of linear integral equations
that can be solved. In this way, the ground state and
the low-lying excitations have been extensively studied for the $\XXX$ model,
and by a similar manner for many other models.
\par
In \refto{DesL, Bab} the anti-ferroelectric $\XXX$ and $\XXZ$
chains are studied in the thermodynamic limit $N=\infty$
and at the physical temperature $T=0$. We also deal with the same problem
in this paper, but by a completely different method.
Our diagonalization scheme is totally
independent of the Bethe Ansatz, and goes further:
it makes it possible to describe all the eigenvectors
by using the powerful tools of representation theory.
\par
\subsec(0.3| Non-abelian symmeries and particle picture)
In a remarkable paper \refto{FaddeevT},  Faddeev and Takhtajan
conjectured the structure of the eigenvectors of the $\XXX$ Hamiltonian.
Using the Bethe Ansatz,
they conclude that all the excitations are composed of elementary ones,
called particles, and that the physical space is
\eq
&{\cal{ F}}=\left[\sum_{n=0}^\infty\int_0^\pi
\cdots\int_0^\pi d\/k_1\ldots d\/k_{n}\otimes
^{n}{\bf C}^2\right]_{{\rm symm}}, &(3)\cr
\endeq
where $[\quad]_{{\rm symm}}$ stands for an appropriate symmetrization.
In \refto{FaddeevT} the summation is only for even $n$.
In our work, we consider both even and odd n.
\par
The issue in \refto{FaddeevT} is as follows:
The $\XXX$ model possesses an
$\slt$ symmetry. The Bethe Ansatz produces only
the highest weight vectors of this $\slt$.
Therefore, spin-1/2 two-particle states (i.e., the case
$n=2$, $\C^2\otimes \C^2$), which were correctly observed in \refto{FaddeevT}
in the thermodynamic limit, appear in the finite chain to
decompose into spin-1 plus spin-0 particles (i.e., $\C^3\oplus\C$).
In fact, the energy levels of the triplet $\C^3$ and the singlet $\C$
are different for finite
lattice and become equal only in the thermodynamic limit.
This gives an example of how a system can achieve
extra and big symmetries in the infinite volume limit, and
shows one of the advantages of a direct diagonalization
of the Hamiltonian in the infinite lattice.
\par
Let us briefly consider the particle picture in the $\XXZ$ case.
By particle picture we mean the structure of common eigenspaces
as irreducible modules of the algebra of non-abelian symmetries.
Here, we mean by non-abelian symmetries operators that are
commuting with the Hamiltonian. As opposed to the abelian
symmetries, they are non-abelian, so they change eigenvectors
of the Hamiltonian without changing their eigenvalues.
The finite $\XXZ$ chain has no $\slt$ symmetry, but as
Pasquier and Saleur \refto{PS} pointed out, if we add certain boundary
terms, then the $\XXZ$ Hamiltonian acquires the $\uf$
symmetry. But the representations are still highly reducible.
\par
In our approach, instead of adding boundary terms to the finite
chain, we consider the infinite chain from the outset, so that the
$\u$ symmetry (which is much larger than $\uf$) is manifest.
As $\u$-module the common eigenspaces of the $\XXZ$ and the higher
Hamiltonians become the irreducible tensor products $\otimes^{n}\C^2$
parametrized by $n$ quasi-momentum variables.
This is exactly the point of Faddeev and
Takhtajan (in the $\XXX$-case), though they did not explain it in these words.
\par
\subsec(0.4| Dynamical symmetries and creation and annihilation operators)
Since Onsager's solution of the Ising model,
a number of alternative methods of solving that model were developed.
A method using the infinite abelian symmetries is given, e.g., in Baxter's book
\refto{Baxbk}.
A major difference, when compared with the anti-ferroelectric
$\XXZ$ model, is that
the Ising model has non-degenerate common eigenvectors
of the commuting transfer matrices.
In other words, the Ising model has no non-abelian symmetries which commute
with the Hamiltonian.
However we may also consider symmetries in a broader sense:
those which do not commute with the Hamiltonian but map eigenvectors
to other eigenvectors with different eigenvalues
(let us call them the dynamical symmetries).
The Ising model has such symmetries: they are generated by
the creation and annihilation operators of free fermions.
\par
A natural question arises: what are the dynamical
symmetries of the $\XXZ$ model? Our answer is that the
vertex operators for $\u$ give an appropriate mathematical language
for them. We will explain this statement later.
\par
Of course, the lesson of the Ising model is not
restricted to the fact that the space of particles
is the Fock space $\cal{F}$ of free fermions. The most
striking fact observed in the Ising model is that the spin operator $s_n$ is
completely characterized by its adjoint action on $\cal{F}$,
i.e., ${\rm Ad}\,s_n\in{\rm End}\,(\cal{F}$), or on even
smaller linear subspace spanned by the free fermions.
($2^N\times2^N$ reduces to $2N\times2N$!) This was the key (or the magical
word)
to the world of the monodromy preserving deformations, the
Painlev\'e transcendents, etc. \refto{WMTB,SMJ}.
\par
Around the same time as the above developments in the Ising model,
the theory of the $S$-matrix was developed by
Zamolodchikov and Zamolodchikov,
on the basis of the fact that if a field theory has infinite abelian
symmetries, then it has a factorized $S$-matrix.
They defined the algebra of the
creation and the annihilation operators as the key to the
bootstrap program for the determination of the $S$-matrices.
The creation and annihilation operators, which we mentioned above as
generators of the dynamical symmetries of the $\XXZ$ Hamiltonian,
give a lattice realization of the
Zamolodchikov algebra \refto{FR}.
\par
\subsec(0.5| QISM)
The quantum inverse scattering method (QISM),
initiated in Leningrad when the city was called by that name,
is a large-scale project to understand integrable quantum systems as a whole.
Among its many achievements, we recall two that are
related to this work. One is, of course, the
discovery of quantum groups, or the $q$-analogue of the
universal enveloping algebras, by Drinfeld and
Jimbo.
The other is Smirnov's work, to which we come later.
The reason for the $\u$ symmetry of the $\XXZ$
Hamiltonian is simply that the Boltzmann weight of the
six-vertex model is the $R$-matrix for the two-dimensional
representation of $\u$ (i.e., $\C^2$ in \refeq{XXZ}).
Also, the existence of the infinite dimensional abelian symmetries
is a corollary of this fact.
The discovery of quantum groups came as a harvest of QISM,
not conversely. All the developments in the earlier days of
QISM were made in the absence of quantum groups.
Therefore, they are lacking in understanding of the true
nature of the symmetries. The QISM project should be further pushed forward
on the basis of the developments in the theory of quantum groups.
\par
In the early days of QISM, Smirnov invented a bootstrap program
for computing the matrix elements of local fields starting
from a factorized $S$-matrix \refto{Sm2,Sm}.
He found a system of
difference equations which ensures the locality of the
field operators, and obtained the form factors (i.e., the matrix elements
of the off-shell currents). This is a deep result.
We are only beginning to understand its true meaning.
Frenkel and Reshetikhin obtained the $q$-deformed
Knizhnik-Zamolodchikov equation, and established that the
$n$-point correlation functions of the vertex operators
satisfy this system of $q$-difference equations \refto{FR}.
Smirnov's equation was the double Yangian version of that.
\par
As we already mentioned, the creation and the annihilation operators
of the $\XXZ$ model are given in terms of the vertex operators of $\u$.
(They are not equal. See section 7 for a detailed discussion on this point.)
The latter are exactly the same
as special cases of those considered in \refto{FR}. So, their $n$-point
functions satisfy the $q$-KZ equation. This is the key to the
calculation of the energy and the momentum, and the commutation relations
of the creation and annihilation operators.
\par
\subsec(0.6| Conformal field theory)
We have considered the symmetries of the Ising model and the
$\XXZ$ model. The importance of such a viewpoint was established
in the monumental work of Belavin, Polyakov and Zamolodchikov
\refto{BPZ}. To formulate the symmetry picture of conformal field theory
in a way that is suitable to our purposes, we
will consider the $\slt$ Wess-Zumino model,
following Tsuchiya and Kanie \refto{TK}:
\smallskip
\item{(i)} The space of states of this model is a direct
sum $\cal{H}$ of the irreducible highest weight representations of
$\widehat{\goth{sl}}(2)$.
\item{(ii)}
The Hamiltonian of the model is the Virasoro generator $L_0$
in the Sugawara form. It allows the $\slt$ symmetry.
The total ${\widehat {\goth{sl}}}(2)$
acts on $\cal{H}$ as dynamical symmetries.
\item{(iii)}
The local fields of this model are given by the vertex operators:
\eq
&\Phi(z):V_z\o{\cal{H}}\rightarrow{\cal{H}}.&(4)\cr
\endeq
Here $V_z$ is a level 0 representation of $\widehat{\goth{sl}}(2)$
depending on a parameter $z$, and $\Phi$ intertwines the representations
on both sides.
\item{(iv)}
The vacuum-expectation values of the vertex operators are
characterized by the Knizhnik-Zamolodchikov equations, which
is a system of linear holonomic differential equations.
\par
Now we can give the definition of the vertex
operators for $\u$. They are exactly \refeq{4} in the context of
the quantum affine Lie algebra $\u$.
Note, however, that there is a difference between the
interpretations of the spectral parameter $z$ in  \refeq{4},
in the WZ-model and in our $\XXZ$ model. In the former,
it is the coordinate variable, and in the latter,
it is the momentum variable. The mathematical
extension from $q=1$ to $q\not=1$ of the vertex operators has
a different physical content. (This is already pointed out
by Smirnov in a different context \refto{Sm}.)
\par
Conformal field theory has had many achievements in
the subject of integrable models.
In fact, we should say that it has
changed the definition of the game we are playing.
However, it has not made progress in all of
the subjects studied in earlier days. The reason is obvious:
its basic principle, the conformal symmetry, is valid
only in massless theories.
Zamolodchikov made the first attempt to attack the massive
theories from the conformal viewpoint. He proposed to
define integrable deformations of conformal field theories by the
existence of infinite abelian symmetries embedded in
the conformal symmetry \refto{Zamol,FeiFre}. This motivated a reconsideration
of the factorized $S$-matrix theory \refto{CorDor}.
\par
\subsec(0.7| CTM and beyond)
Let us return to the $\XXZ$ model.
The point we wanted to make is the following:
The deformation parameter $q$ appears in the $\XXZ$ model as
the anisotropy parameter $\Delta=(q+q^{-1})/2$. This is the departure
from criticality, or simply the mass. So, there is
more than a good reason to expect that the quantum affine
algebras are relevant in massive integrable models.
In fact, many of the recent developments are related
to this point. For the continuum theory, we refer the reader
to Smirnov's paper \refto{Sm}.
\par
Now let us come to one of the masterpieces of Baxter,
the CTM (corner transfer matrix). In \refto{BaxCTM},
after saying that there is no Ising-like reduction from
$2^N\times2^N$ to $2N\times2N$ in the six or eight vertex
models, he wrote
\smallskip
{\narrower\noindent\it A rather ambitious hope is that by examining the CTM's
we may stumble on such a group, that the solution of the
models may thereby be simplified ...\smallskip}
\pn
A few years later, he succeeded in computing
the 1-point functions of the hard-hexagon
model and its generalizations, without giving an answer to
the original question of his earlier paper. Nevertheless
he has given us the magical word --- ``{\it Open Sesame}'',
by inventing the CTM method. When the cave-door opened,
there appeared the characters of the affine Lie
algebras \refto{DJMO,DJKMO}, the theory of crystals \refto{Ka,Ka2},
the irreducible highest weight representations \refto{FM}, the
$q$-vertex operators \refto{FR, DJO} --- the passage to the
representation theory of the affine quantum groups. And in every
case, the use of an infinite system, already assumed in
Baxter's original work, is an essential ingredient.
\par
In this paper, we proceed further along this route.
We propose that the mathematical picture that
bridges \refeq{InfTen} and \refeq{3} is
\eq
&{\rm End}\,{}_{\C}\bigl(V(\La_0)\bigr)=V(\La_0)\o V(\La_0)^*&(MathP),\cr
\endeq
(in the even particles sector; the odd particles sector is
${\rm Hom}\,{}_\C\bigl(V(\La_0),V(\La_1)\bigr)=V(\La_1)\o V(\La_0)^*$),
where the  $V(\La_i)$ are the level 1 highest weight representations of $\u$.
By this, the world of the infinite degrees of freedom (i.e., the infinite
tensor product) opens to the world of the finite degrees of freedom
(i.e., the representation theory of $\u$).
The vacuum, the lowest eigenvector of the Hamiltonian,
is the identity operator in
${\rm End}\,{}_{\C}\bigl(V(\La_0)\bigr)$, or the canonical element in
$V(\La_0)\o V(\La_0)^*$. The particle, as given in \refeq{3}, are created
by the vertex operators acting on the left half of
$V(\La_i)\o V(\La_0)^*\ (i=0,1)$.
Thus, the space ${\cal{F}}$ \refeq{3} of the eigenvectors of
the Hamiltonian \refeq{XXZ}, that are created by the creation operators
upon the vacuum, lies in the level zero $\u$-module \refeq{MathP}.
\par
Let us summarize our discussion on the symmetries of
integrable models in this introduction.
\bigskip
\halign{\hfil\bf#&\quad\hfil#\hfil&\quad\hfil#\hfil&\quad\hfil#\hfil\cr
\sl Symmetries&\sl Ising&\sl \XXZ\quad\quad Thirring&\sl WZ\cr
&&&\cr
Abelian&Commuting&Infinite&None\cr
Symmetries&Transfer Matrices&Conservation Laws&\cr
&&&\cr
Non-abelian&None&$\Up$\quad The $\slt$-Yangian&$su(2)$\cr
Symmetries&&&\cr
&&&\cr
Dynamical&Free&Vertex&Virasoro algebra\cr
Symmetries&Fermions&Operators&$\widehat {su}(2)$\cr
&&&\cr
Space of States&${\cal{F}}$&${\cal{F}}\subset
{\rm End}({\cal H})={\cal H}\o{\cal H}^*$
&${\cal H}$\cr
&&&\cr
Local Fields&Clifford group&?&Vertex Operators\cr}
\par
\subsec(0.8|Plan of the paper)
The text is organized as follows. In \sec(1) we introduce the
quantum affine algebra $\uq$ as non-abelian symmetries of the
$\XXZ$ spin chain. In \sec(2) we describe
the embedding of the
highest weight modules $V(\La_i)$ into the half-infinite tensor product
space $\cdots V\o V\o V$ or $V\o V\o V\o \cdots$
given by iterating the vertex operators.
Examining the perturbation expansion at $q=0$ we observe that,
choosing the scalar multiple of vertex operators correctly
the series defining the embedding would become finite, term by term in
powers of $q$.
To ascertain the existence of such a normalization factor we need to
know the aysmptotics of the $n$ point functions of vertex operators
(cf. \sec(6.8) below) as $n\rightarrow \infty$.
Though the problem is of its own interest, it
is beyond the scope of the present paper.
In \sec(3) we study the decomposition of the level $0$ module
$V(\La_0)\o V(\La_0)^*$ at $q=0$.
We show that its crystal
naturally decomposes to `$n$-particle sectors',
each of which can be identified as  a certain
anti-symmetrization of the affine crystal $\Aff(B)^{\o n}$.
We study the picture for nonzero $q$ in the next sections.
In \sec(4) we introduce the vacuum vectors as canonical elements of
$V(\La_i)\o V(\La_i)^*$.
In \sec(5) we formulate the `particle picture'
by utilizing a similar but different kind of vertex operators
from those used in the embeddings.
We argue that in this setting the computation of the one-point function
(in the sense of \refto{DJMO}, not the staggered polarization \refto{Bax})
can be done trivially.
\par
Sections 1--5 are devoted to a presentation of our
ideas on the problem.
In the next two sections we reformulate the problem in the language
of representation theory.
After reviewing the $q$-deformed vertex operators following \refto{FR}
and \refto{DJO} we give formulas for their two point functions and
their commutation relations.
The case of general $n$ point functions is discussed in \sec(6.8).
In \sec(7) we study the vacuum, creation and
annihilation operators
which are defined in purely representation-theoretic terms.
We prove that the vacuum vector has
the correct invariance with respect to the translation and energy operators.
We then derive the energy-momentum of the creation-annihilation operators
using the formulas for two point functions, and show that they coincide
with the known results obtained by the Bethe Ansatz method.
The \sec(8) is devoted to a summary and open problems.
\par
In the appendices we collect some data concerning the global crystal base and
vertex operators.
We give tables for the first few terms of the actions
with respect to the global base of
$\uq$ on $V(\La_0)$ (Appendix 1), of the vertex operators (Appendix 2),
the embedding into the infinite tensor product (Appendix 3) and
the images of the vacuum, one- and two-particles (Appendix 4).
In Appendix 5 we study the $q\rightarrow 0$ limit of the Bethe vectors
and compare the results with the vertex operator computations.
\par
%\endinput
\par
\par
\def\s{\sigma}
\def\o{\otimes}
\def\D{\Delta}
\def\U{U_q\bigl(\widehat{\goth{sl}}(2)\bigr)}
\def\Up{U'_q\bigl(\widehat{\goth{sl}}(2)\bigr)}
\def\Uf{U_q\bigl({\goth{sl}}(2)\bigr)}
\def\Di{\Delta^{(\infty)}}
\par
\beginsection \S1. Quantum Affine Symmetries of the $\XXZ$ model
\par
Let
$V\simeq \C^2$
be a 2-dimensional vector space, and let
$\s^x$, $\s^y$, $\s^z$
be the Pauli matrices acting on
$V$.
We consider the infinite tensor product $W$ \refeq{InfTen}
and the
$\XXZ$
Hamiltonian $\H$
that formally acts on
$W$.
The action of $\H$ is formal in the sense that it is
{\it a priori} divergent, since we are working directly
in the thermodynamic limit, and requires renormalization.
We number the tensor components by
$k\in\Z$.
The limit
$k\rightarrow\infty$
is to the left, and
$k\rightarrow-\infty$
to the right.
Our aim is to diagonalize the Hamiltonian
$\H$, using the representation theory of quantum affine algebras.
The Bethe Ansatz provides us with a method for diagonalizing
such Hamiltonians. In this method we start from a finite
(periodic or non-periodic) chain, find the eigenvalues and the eigenvectors
of the Hamiltonian in a certain form,
and take the thermodynamic limit at the end.
In this paper we propose another method that diagonalizes
$\H$
directly on
$W$.
The applicability of our method, at the moment, is limited
to the anti-ferroelectric regime
$\D<-1$.
The Bethe Ansatz method in this regime meets with the difficulty
of starting from the wrong vacuum, and filling the
Dirac sea. Our method is free from this complication.
\par
The main ingredient of our method is the quantum affine symmetries of
$\U$.
Consider the action of
$\Up$
on
$V$;
\eq
&\pi:\Up\longrightarrow\End(V)\cr
\endeq
given by
\eq
&\pi(e_0)=\pi(f_1)=\pmatrix{0&0\cr1&0\cr},\cr
&\pi(e_1)=\pi(f_0)=\pmatrix{0&1\cr0&0\cr},\cr
&\pi(t_1)=\pi(t_0^{-1})=\pmatrix{q&0\cr0&q^{-1}}.\cr
\endeq
We set
\eq
&v_+=\pmatrix{1\cr0\cr},\ v_-=\pmatrix{0\cr1\cr}.\cr
\endeq
\par
We relate
$q$
to
$\D$
by
\eq
&\D={q+q^{-1}\over2},\qquad -1<q<0.
\endeq
By using the infinite comultiplication
$\D^{(\infty)}$,
we define the action of
$\Up$
formally on
$W$:
\eq
&\Di(e_i)=\sum_{k\in\Z}\cdots\o t_i\o e_i\o1\cdots,&(Infcoma)\cr
&\Di(f_i)=\sum_{k\in\Z}\cdots\o 1\o f_i\o t_i^{-1}\cdots,&(Infcomb)\cr
&\Di(t_i)=\cdots\o t_i\o t_i\o t_i\cdots.&(Infcomc)\cr
\endeq
One can check the commutativity
\eq
&[\H,\Up]=0\cr
\endeq
as discussed in \refto{FM}.
The major advantage of working in the infinite lattice limit directly
is having this infinite symmetry,
which inevitably breaks down on a finite lattice.
On a finite lattice, one can modify
$\H$
in such a way that it possesses the
$\Uf$-symmetries \refto{PS},
but not the
$\U$-symmetries. (Otherwise, the degeneracy of the spectrum on the finite
chain would be different.)
The difference of the size of the symmetry algebra is crucial.
For instance, we can identify the vacuum vector in
$W$
as the unique
$\Up$-singlet
(a vector which generates a 1-dimensional
$\Up$-module.)
On the other hand, singlets for
$\Uf$ are many.
\par
The full algebra
$\U$ is obtained by adding
$q^d$
to
$\Up$.
Following \refto{FM} let us identify $d$ with
$(H_{CTM}-S)/2$ where
\eq
H_{CTM}&=-{q\over 1-q^2}
\sum_{k\in\Z}k
\bigl(\s^x_{k+1}\s^x_k
+\s^y_{k+1}\s^y_k
+\D\s^z_{k+1}\s^z_k\bigr)&(CTM)\cr
\endeq
and $2S=\sum \sigma^z_k$ is the total spin operator.
This identification is justified by explicit computations
and by checking that
\eq
&[d,e_i]=\delta_{i0}e_i,\ [d,f_i]=-\delta_{i0}f_i,\ [d,t_i]=0.\cr
\endeq
\par
Consider the shift
$T$ of
$W$
which induces the outer automorphism of the operator algebra
generated by
$\s^x_k$,
$\s^y_k$,
$\s^z_k$;
\eq
&T\cdot\s_k^*\cdot T^{-1}=\s_{k-1}^*.\cr
\endeq
$T$ is a shift to the right by one lattice unit.
Because of \refeq{CTM}, the Hamiltonian \refeq{XXZ} can be written as
\eq
&{q\over1-q^2}\H=T\cdot d\cdot T^{-1} - d.&(XXZCTM)\cr
\endeq
\par
This is the key to our diagonalization procedure:
the point is that the derivation $d$ \refto{FM}, as well as $T$
can be expressed in the language of representation theory,
thus the $\XXZ$ Hamiltonian can be expressed
in terms of mathematically well-defined objects.
\par
In fact, \refeq{XXZCTM} allows us to bypass explicit references
to the higher Hamiltonians, obtained by taking higher derivatives
of the row-to-row transfer matrix of the six-vertex model.
The point is that all higher
Hamiltonians can be generated from
i.e.,
$\H$,
by taking commutators with
$d$
recursively \refto{Tetel}.
This means that the essential feature of the integrability
(i.e., the existence of the infinite conservation laws),
in this model, is incorporated in
$d$.
\par
%In the following sections we will treat the parameter $q$ as a
%formal variable (indeterminate) rather than a complex number $-1<q<0$,
%regarding matrix elements, vectors, etc. as formal power series
%in $q$.
%In other words we base our analysis on the $q$-series expansion.
\par
%\endinput
\par
\par
\par
\def\N{{\bf N}}
\def\pn{\par\noindent}
\def\qbox#1{\quad\hbox{#1}}
\par
\def\U{U_q\bigl(\widehat{\goth{sl}}(2)\bigr)}
\def\Up{U'_q\bigl(\widehat{\goth{sl}}(2)\bigr)}
\def\Uu{U_q\bigl(\goth{sl}(2)\bigr)}
\par
\def\L{\Lambda}
\def\U{U_q\bigl(\widehat{\goth{sl}}(2)\bigr)}
\def\Up{U'_q\bigl(\widehat{\goth{sl}}(2)\bigr)}
\def\V(#1){V(\L_{#1})}
\def\o{\otimes}
\def\D{\Delta}
\def\P{{\cal{P}}}
\def\bp{{\bar p}}
\par
\beginsection \S2. Embedding of the highest weight modules
\par
Let
$\V(i)\ (i=0,1)$
be the level 1 irreducible highest weight
$\U$-module
with highest weight
$\L_i$.
We give a conjecture on an embedding of
$\V(i)$
into the half infinite tensor product
\eq
&W_l=\cdots\o V\o V\cr
\endeq
by using the vertex operators
\eq
&\Phi:\V(i)\longrightarrow \V(1-i)\o V,\quad
\Phi(v)=\Phi_+(v)\o v_++\Phi_-(v)\o v_-,
&(inter)\cr
\endeq
that satisfies
\eq
&\Phi(xv)=\D(x)\Phi(v)\qbox{for all $x\in\U$ and $v\in\V(i)$}.\cr
\endeq
Precisely speaking, we need a completion of
$\V(1-i)\o V$, which we will elaborate in \sec(6).
Also, we do not define $W_l$ (but see the definition $W_l^{(i)}$ below).
The subscript $l$ in $W_l$ stands for {\it left}.
It means the left-half-infinite tensor product.
\par
The idea is to consider the composition of the vertex operators;
\eq
&\V(0){\buildrel\Phi\over\longrightarrow}
\V(1)\o V{\buildrel\Phi\o \rm id\over\longrightarrow}\V(0)\o V\o V\cr
&{\buildrel\Phi\o \rm id\o id\over\longrightarrow}\V(1)
\o V\o V\o V\longrightarrow
\cdots\longrightarrow W_l&(compo)\cr
\endeq
Our conjecture is that with a proper normalization of \refeq{inter},
the composition \refeq{compo} converges to a map
\par
\eq
&\iota:\V(i)\longrightarrow W_l \cr
\endeq
satisfying
\eq
&\iota(xv)=\D^{(\infty)}(x)\iota(v),\cr
\endeq
and that if
$v$ is a weight vector of
$\V(i)$,
then
$\iota(v)$
is an eigenvector of
$H_{CTM}$
defined by \refeq{CTM} with a certain renormalization (see \refto{FM}).
\par
Let
$\P^{(i)}_l$
be the set of paths which parametrise the affine crystal
$B(\L_i)$.
(See \refto{MM,JMMO,(KMN)^2})
We use the convention that the colors $i=0,1$ are modulo $2$
(e.g., ${\cal P}^{(i)}_l={\cal P}^{(i+2)}_l$).
A path
$p=\bigl(p(k)\bigr)_{k\ge1}
=\bigl(\cdots p(3)\,p(2)\,p(1)\bigr)\in \P^{(i)}_l$
satisfies
$p(k)=(\pm)$
(often identified with $\pm1$),
and
$p(2k+i)=(+)$,
$p(2k+i+1)=(-)$
for
$k \gg 0$.
We denote by
$|p\rangle$
the vector in
$W_l$
corresponding to
$p$;
\eq
&|p\rangle=\cdots\o v_{p(3)}\o v_{p(2)}\o v_{p(1)}.\cr
\endeq
The following are called the ground-state-paths;
\eq
&\bp_0=(\cdots+\ -\ +\ -),\
\bp_1=(\cdots-\ +\ -\ +).\cr
\endeq
\par
We consider the set of formal infinite linear
combinations of paths with coefficients in ${\bf Q}(q)$,
\eq
&W_l^{(i)}=
\{\sum_{p\in\P^{(i)}_l}c(p)|p\rangle;c(p)\in {\bf Q}(q)\}
\endeq
\par
Let
$\{G(p);p\in\P^{(i)}_l\}$
be the upper global base of Kashiwara \refto{Ka} for
$\V(i)$.
Define
$c_\pm(r,p)\ (r,p\in\P^{(0)}_l\sqcup\P^{(1)}_l)$
by
\eq
&\Phi_\pm\bigl(G(p)\bigr)=
\sum_rc_\pm(r,p)G(r)\o v_{\pm}.\cr
\endeq
If
$p\in\P^{(i)}_l$,
then
$c_\pm(r,p)=0$
unless
$r\in\P^{(1-i)}_l$
and
$s(p)=s(r)\pm{1\over2}$.
Here
$s:\P^{(0)}_l\sqcup\P^{(1)}_l\longrightarrow{1\over2}\Z$
is the spin of paths defined by
$s(p)={1\over2}\lim_{k\rightarrow\infty}
\sum^{2k+i}_{j=1}p(j).$
\par
The coefficients
$c_\pm(r,p)$
of the vertex operator satisfy the following
(see \refto{DJO}):
\eq
c_-(\bp_1,\bp_0)&=c_+(\bp_0,\bp_1)=1,&(Nor)\cr
c_\pm(r,p)&\equiv1\bmod qA\qbox{if $p(1)=\pm,\, p(k+1)=r(k) ~\forall k$}
&(Rega)\cr
&\equiv0\bmod qA\qbox{otherwise},&(Regb)\cr
\sharp\{r\in\P^{(0)}_l\sqcup\P^{(1)}_l &\mid c_\pm(r,p)\not\equiv0\bmod
q^NA\}<\infty \cr
&\qquad \qbox{for all $p\in\P^{(0)}_l\sqcup\P{(1)}_l$ and $N\in\N$}.
&(Regc)\cr
\endeq
Here
$A=\{f\in\Q(q);\ q\hbox{ has no pole at }q=0\}.$
\refeq{Nor} is the normalization and \refeq{Rega}, \refeq{Regb}, \refeq{Regc}
are the key properties which mean the compatibility
between the vertex operator and the crystal base.
\par
Define
%$\iota^{(n)}:\V(i)\rightarrow \underbrace{V\otimes\cdots\otimes}_n$
%by
\eq
%&\iota^{(n)}\bigl(G(p)\bigr)=
%\sum_{a\in\P^{(i+n)}_l}\omega^{(n)}(a,p)|a\rangle,&(Iota)\cr
&\omega^{(n)}(a,p)=\langle G^*(a_{n+1})|
\Phi_{a(n)}\circ\cdots\circ\Phi_{a(1)}|G(p)\rangle.
%\sum\prod_{k=1}^n c_{a(k)}(p_{k+1},p_k).
&(Omega)\cr
\endeq
Here $a=\bigl(a_{n+1},a(n),\ldots,a(1)\bigr)$
and $p$ are paths in $\P^{(i)}_l$,
and $\{G^*(p)\}$ is the dual base to $\{G(p)\}$.
Because of \refeq{Rega}, \refeq{Regb}, \refeq{Regc},
\refeq{Omega} is convergent in ${\bf Q}[[q]]$.
%The summation \refeq{Omega} is over all the sequences of paths
%$\{p_k\}_{k=1}^{n+1}$ such that
%\eq
%&p_1=p,\ p_{n+1}=a,\ p_k\in\P^{(i+k+1)}_l,
%\ s(p_{k+1})=s(p_k)-{1\over2}a(k)~\forall k.\cr
%\endeq
\par
\proclaim Conjecture.
\item{(i)} There exists a limit
\eq
&\omega(a,p)=\lim_{n\rightarrow\infty}
{\omega^{(n)}(a,p)\over
\omega^{(n)}({\bar p}_{i+n},{\bar p}_i)}\in{\bf Z}[[q]].
\endeq
\item{(ii)} $\omega(a,p)\in{\bf Q}(q)$.
\pn
\item{(iii)}
Setting
\eq
&\iota\bigl(G(p)\bigr)=\sum_{a\in\P^{(i)}_l}\omega(a,p)
|a\rangle ,&(Conj)\cr
\endeq
we have
\eq
&\iota\bigl(G(p)\bigr)\big|_{q=0}=|p\rangle.&(CryEmb)\cr
\endeq
\item{(iv)}
\eq
&\iota\bigl(xG(p)\bigr)=\sum_{a\in\P^{(i)}_l}\omega(a,p)
\Delta^{(\infty)}(x)|a\rangle&(ConAct)\cr
\endeq
for
$x\in\Up$
and
$p\in\P^{(i)}_l$.
\par
\par
The statements of the conjecture are independent of the normalization
of the vertex operator \refeq{inter}.
With the normalization \refeq{Nor}
we have (see 6.8 and Appendix 3)
\eq
&\lim_{n\rightarrow\infty}\omega^{(n)}
({\bar p}_{i+n},{\bar p}_i)^{1/n}=1+q^4-q^6+q^8\bmod q^{10}.
\endeq
Therefore, we also conjecture that this is actually
convergent in ${\bf Z}[[q]]$. We have no reasonable
conjecture what the limit is.
\par
In \refto{FM} an embedding $\V(0)\hookrightarrow V\o V\o\cdots$
is constructed by using a different coproduct;
\eq
&\Delta_{-}(e_i)=e_i\o t_i^{-1}+1\o e_i,\cr
&\Delta_{-}(f_i)=f_i\o 1+t_i\o f_i,\cr
&\Delta_{-}(t_i)=t_i\o t_i.\cr
\endeq
To compare these two different embeddings,
let us flip right and left in the notation of \refto{FM}.
(Namely we change the embedding of \refto{FM} into the form
$\V(0)\longrightarrow \cdots\o V\o V.$)
If we denote the infinite coproduct in \refto{FM} after the flip by
$\Delta^{(\infty)}_{FM}$, we have
\eq
&\Delta^{(\infty)}_{FM}(f_i)=q\Delta^{(\infty)}(t_if_i),\cr
&\Delta^{(\infty)}_{FM}(e_i)=q\Delta^{(\infty)}(t_i^{-1}e_i).\cr
\endeq
Therefore these two embeddings differ only by certain power
of $q$.
In Appendices 3,4 we compute $c_\pm(r,p)$ and
$\omega(a,p)$ for several cases.
\par
\par
Let us discuss on the similarity and the difference
of \refto{FM} and the present paper.
\refto{FM} studied the CTM Hamiltonian \refeq{CTM} on $W_l$
and identified its eigenvectors with the weight vectors
of $\V(0)$ embedded in $W_l$.
In this paper we study the $\XXZ$ Hamiltonian \refeq{XXZ} on $W$
and identify its eigenvectors with the weight
vectors in certain $\Up$-modules embedded in $W$.
In both cases the main aim is to give an
equivalent mathematical picture to the physical content of the problem,
which is given by the CTM Hamiltonian or the $\XXZ$ Hamiltonian.
The mathematical picture presented in \refto{FM} is
$\V(0)$, the level 1 irreducible highest weight $\U$-module.
The method employed is the $q$-perturbation.
An obvious implication of \refto{FM} to the present problem
is to consider
\eq
&\V(0)\o\V(0)^{\ast a}&(Maths)\cr
\endeq
as a mathematical model for the infinite product $W$.
A similar analysis applied to
\eq
&W_r=V\o V\o\cdots\cr
\endeq
gives the level $-1$ irreducible
lowest weight $\U$-module.
The right half of \refeq{Maths} is the most appropriate realization of that.
Here
$\V(0)^{\ast }=\Hom_{{\bf Q}(q)}(\V(0),{\bf Q}(q))$
endowed with the natural right $\U$-action $\rho_R$;
 if $f\in \V(0)^\ast$ then $\bigl(\rho_R(x)f\bigr)(v)=f(xv)$
for $x\in\U$.
Then $\V(0)^{\ast a}$ is the left $\U$-module with the left
$\U$ action $\rho_{L,a}$ in terms of the antipode $a$
(see \refto{(KMN)^2}, \sec(5));
\eq
&\bigl(\rho_{L,a}(x)f\bigr)(v)=f\bigl(a(x)v\bigr).\cr
\endeq
We will discuss why we choose the antipode to construct
the left action a bit later.
Now we want to discuss a big difference between \refto{FM}
and the present paper.
The basic tool in \refto{FM} was the $q$-series expansion.
Underlying this were two things;
\item{(i)} The success of the theory of the affine crystals.
(\refto{MM,JMMO,(KMN)^2})
\item{(ii)} The CTM magic, i.e., the discreteness of the spectrum of
the CTM Hamiltonian. (\refto{Baxbk})
\par
\par
In this paper, we further exploit (i) to analyse the $q=0$ structure.
Note that the $\XXZ$ Hamiltonian is already diagonal
at $q=0$, i.e., $\Delta=-\infty$.
The eigenvectors are nothing more than
paths $p=(p(k))_{k\in{\bf Z}}$ with
appropriate boundary conditions.
The notion of crystal gives a nice combinatorial
structure to the set of paths.
This is described in detail in section 3.
The discreteness of the spectrum is no longer
true for the $\XXZ$ Hamiltonian.
It means the breakdown of the $q$-expansion as a tool for finding the
eigenvectors.
The continuum spectrum at $q\not=0$ degenerates to
the discrete spectrum at $q=0$.
Therefore we do not know a priori the correct form
of the eigenvectors at $q=0$ with which we should start
our expansions.
In \refto{FT} the expansion of the vacuum vector
(i.e., the lowest eigenvector of $H_{\XXZ}$)
was discussed.
This was possible because the lowest eigenvalue
is not degenerate even at $q=0$
( or more precisely, it is doubly degenerate,
but the two eigenvectors generate two orthogonal sectors).
So we could start with
\eq
&|vac\rangle\big|_{q=0}=(\cdots+-+-\cdots).\cr
\endeq
Similarly it is not difficult to tell the correct
limiting form of the 1 particle state because
the 1 particle states must be eigenvectors of $T^2$
(the shift of two lattice units).
Therefore we can start with
\eq
&|u,+\rangle\big|_{q=0}=
\sum_{k}(\cdots-++-\cdots)\hbox{e}^{iku}\cr
&|u,-\rangle\big|_{q=0}=
\sum_{k}(\cdots+--+\cdots)\hbox{e}^{iku}\cr
\endeq
When we consider an even particle state,
e.g., the vacuum, we take
\eq
&p(2k)=(+),\qquad p(2k+1)=(-)\qbox{for $k\rightarrow\pm\infty$}
&(Even)\cr
\endeq
as the boundary condition.
On the other hand, we take
\eq
&p(2k)=(\mp),\qquad p(2k+1)=(\pm)\qbox{for $k\rightarrow\pm\infty$}
&(Odd)\cr
\endeq
for an odd particle state.
The ordinary approach of the Bethe Ansatz dismisses all the
odd particle states,
because the cyclic boundary condition on a lattice
with even number of sites is discussed.
Let us go back to the discussion of the $q$ expansion.
If the particle-number is greater than one,
there is no a priori choice of the limit $q=0$.
The shift $T^2$ can fix only the total
momentum (at $q=0$),
which is certainly not enough.
So we need some new idea other than (i) and (ii) above.
The idea is, of course, to use the quantum affine symmetries discussed in
\S1.
What can one do if one has the symmetries of the Hamiltonian?
The answer is to decompose the space of states into the irreducible pieces.
This idea works very well if the irreducible decomposition
is multiplicity free.
As for the $\XXZ$ model in the anti-ferroelectric regime,
this is not the case if we consider only the local symmetries and the
non-local $\Uu$-symmetries.
But if we consider the $\U$-symmetries this is what we actually get.
In the case of the CTM Hamiltonian,
the decomposability of the physical space is rather
trivial;
the space $\V(0)$ is already irreducible.
This is one big reason why the CTM magic works so nicely.
For the $\XXZ$ Hamiltonian, we should ``decompose''
$\V(0)\o \V(0)^{\ast a}$,
(or equivalently $\End_{\Q(q)}(\V(0))$)
 which is a level $0$ representation of
$\Up$, an object highly reducible and much more interesting
than the irreducible modules.
\par
%\endinput
\par
\par
\def\F{{\cal F}}

\def\eps{\varepsilon}
\def\o{\otimes}
\def\wt{\mathop{\rm wt\/}}

\def\Aff{\mathop{\rm A\hskip-1pt ff\/}}
\def\XXZ{{X\hskip-2pt X\hskip-2pt Z}}
\def\H{H_{\XXZ}}
\ifx\Uq\undefined
  \def\Uq{U_q\bigl(\widehat{\goth{sl}}(2)\bigr)}
   \else\message{* Uq already defined *}\fi
\ifx\CTM\undefined
  \def\CTM{H_{C\hskip-2pt T\hskip-2pt M}}
   \else\message{* CTM already defined *}\fi
\ifx\Hilb\undefined
  \def\Hilb{H_{phy,\Delta}}
   \else\message{* Hilb already defined *}\fi
\ifx\MP\undefined
  \def\MP{V(\Lambda_0)\otimes V(\Lambda_0)^{*a}}
   \else\message{* MP already defined *}\fi
\def\MPP{V(\Lambda_{0,1})\otimes V(\Lambda_0)^{*a}}
\par
\par
\beginsection \S3. Decomposition of Crystals
\par
At $q=0$ the Hamiltonian \refeq{*} is diagonal.
Apart from the divergence of its eigenvalue,
a vector of the form
\eq
&\cdots\o v_{p(2)}\o v_{p(1)}\o v_{p(0)}\o v_{p(-1)}\o\cdots\cr
\endeq
for an arbitrary map $p:{\bf Z}\rightarrow \{+,-\}$ is
an eigenvector of the Hamiltonian at $q=0$.
Among these, we consider only those which satisfy appropriate
boundary conditions. We call them paths. The set of paths is equal to
the product of two affine crystals, those corresponding to
$V(\La_{0,1})$ and $V(\La_0)^{*a}$. (See \refto{(KMN)^2} for affine crystals).
We also consider the $q=0$ limit of
the creation operators $\psi^*_j$. (We borrow the commutation relation
\refeq{CA:b} from section 7.) We introduce the algebra generated
by $\psi^*_j$, and show that it is isomorphic to the crystal of the paths.
By using this isomorphism we give the $q=0$ limit of the $n$-particle
eigenvectors.
\par
\subsec(3.1|Crystal $\P^i$)
Recall that the highest weight module $V(\La_i)\ (i=0,1)$
has the crystal which are realized as paths \refto{MM} (see below).
Let us denote it by $\P^i_{left}$.
Similarly, we have the crystal $\P^i_{right}$ for $V(\La_i)^{*a}$.
The crystal $\P^{0,1}$ of $\MPP$ is the tensor product
$\P^{0,1}_{left}\o\P^0_{right}$ in the sense of crystals.
As a set it is just a direct product
$\P^i=\P^i_{left}\times \P^0_{right}$, equipped with the action of
the modified Chevalley generators $\eti,\fti$
according to the tensor product rule \refto{Ka}.

Practically the crystal $\P^i$ is described as follows.
An element of $\P^{i}$ is represented as a path extending to both directions:
\eq
&p=\bigl(p(k)\bigr)_{k\in\Z},\quad p(k)\in\{+,-\}, &(path)\cr
&p(k)=(-)^k ~\hbox{for}~ k\ll 0,
\qquad p(k)=(-)^{k+i} ~\hbox{for}~ k\gg 0. \cr
\endeq
The labeling $k$ is from right to left, $k\ge 1$ (resp. $k\le0$)
for $\P^{0,1}_{left}$ (resp. $\P^0_{right}$).
Alternatively $\P^i$ can be described as the set of arbitrary decreasing
sequence of integers $l_1>\cdots>l_n$ with $n\equiv i$ mod $2$.
The two pictures are related via
\eq
&\{k\in\Z \mid p(k)=p(k+1)\}=\{l_1,\cdots,l_n\}, \quad l_1>\cdots>l_n.
\endeq
We call $l_1,\cdots,l_n$ `domain walls' of the path $p$,
and write $[[l_1,\ldots,l_n]]$ to represent $p$.
We have the obvious decomposition
\eq
&\P^0=\bigcup_{n:even} \P(n), \qquad
\P^1=\bigcup_{n:odd} \P(n), &(3grad)
\endeq
where $\P(n)$ is the set of paths with $n$ domain walls.
\par
There are two types of domain walls: adjacent $(++)$ and $(--)$ pairs.
By definition the spin $s(p)$ of a path $p$ is simply
$(\sharp(++)-\sharp(--))/2$.
The total weight of \refeq{path} has the form
\eq
&\wt p = s(p)\alpha_1 - h(p)\delta. \cr
\endeq
The $h(p)\in\Z$ is given in terms of the energy function  $H(\eps,\eps')$
\eq
&H(+,-)=-1, \quad H(\eps,\eps')=0 \quad \hbox{otherwise} \cr
\endeq
as follows:
\eq
&h(p)= \sum_{k\in\Z}k\Bigl(H\bigl(p(k),p(k+1)\bigr)-
H\bigl(\bar{p}_i(k),\bar{p}_i(k+1)\bigr)\Bigr). \cr
\endeq
Here $p\in \P^i$ and
$\bar{p}_i(k)=(-)^k$ for $k\le 0$, $\bar{p}_i(k)=(-)^{k+i}$ for $k>0$.
\par
The rules for the action of $\eti,\fti$ may be expressed as
\eq
&\tilde{f}_0:(-) \mapsto (+), \qquad
 \tilde{f}_0:(-+) \mapsto 0, \cr
&\tilde{f}_1:(+) \mapsto (-), \qquad
 \tilde{f}_1:(+-) \mapsto 0. \cr
\endeq
Their application is as follows (we shall describe $\tilde{f}_0$).
$\tilde{f}_0$ has no action on any $(-+)$ pair (a singlet).
Given a path $p$ we first cut its left tail $\cdots-+-+-+$ and right tail
$-+-+-+\cdots$ to make it a finite sequence.
We reduce it further by the rule that each time there is an
adjacent singlet pair $(-+)$ we drop that pair.
For example, $(\cdots-++--++-+\cdots)$ reduces to
$(+)$.
The above procedure is not unique but the
result does not depend on the particular choice of it,
reflecting the coassociativity of the coproduct.
If no spins are left then the path as an entity is a singlet
with respect to $\tilde{f}_0$ and it is annihilated.
Otherwise we necessarily have a sequence
$\underbrace{(+\cdots+}_{k_1}\underbrace{-\cdots-)}_{k_2}$,
and it is mapped to
$\underbrace{(+\cdots+}_{k_1+1}\underbrace{-\cdots-)}_{k_2-1}$.
Namely, if $k_2\not=0$, one and only one $-$ is changed to $+$
by the action of $\tilde{f}_0$.
Note that the configuration of the path (before the reduction of singlets),
at the next left and the next right to this $(-)$ is $(+--)$.
This is changed into $(++-)$.
Thus  $\tilde{f}_0$ shifts a domain wall to the left and
cannot cause two domain walls to coalesce.
In the above situation $\tilde{e}_0$ produces
$\underbrace{(+\cdots+}_{k_1-1}\underbrace{-\cdots-)}_{k_2+1}$.
The rules for $\tilde{e}_1,\tilde{f}_1$ are given
by exchanging the roles of $+$ and $-$.
\par
Consider first the case $n=1$. The paths in $\P(1)$ have spin
$\pm 1/2$ according as $l$ is even or odd. They are a crystal of
$\Aff(B)$ according to Definition 2.2.3 of \refto{(KMN)^2} with
non-zero action ($l\in2\Z$)
\eq
&\tilde{f}_1:[[l]] \mapsto [[l+1]],\qquad
 \tilde{f}_0:[[l-1]] \mapsto [[l]],\qquad
 l\in2\Z, &(3act2) \cr
\endeq
and weights
\eq
&\wt[[l]]   =  -(l/2)\delta + (1/2)\alpha_1, \qquad
\wt[[l+1]] =  -(l/2)\delta - (1/2)\alpha_1. \cr
\endeq
Thus $\P(1)$ is a connected crystal.
\par
Now consider $n=2$. Corresponding to \refeq{3act2} we have (with
$l_1,l_2\in2\Z$, $l_1 > l_2$)
\eq
\tilde{f}_1: [[l_1-1, l_2   ]]\mapsto
             [[l_1,  &l_2   ]]\mapsto
             [[l_1,   l_2+1 ]]\mapsto  0 &(3act4:a)\cr
\tilde{f}_0: [[l_1,   l_2+1 ]]\mapsto
             [[l_1+1,&l_2+1 ]]\mapsto
             [[l_1+1, l_2+2 ]]\mapsto  0 &(3act4:b) \cr
\endeq
and the weights of these triplets are respectively
\eq
-{l_1+l_2\over 2}\delta\ +\ & \alpha_1,\,0,\,-\alpha_1
\qquad \hbox{ for \refeq{3act4:a}}, \cr
-{l_1+l_2+2\over 2}\delta\ +\ &\alpha_0,\,0,\,-\alpha_0
\qquad \hbox{ for \refeq{3act4:b}}. \cr
\endeq
So $\P(2)$ has an infinite number of disjoint connected components:
each such component may be labeled by the fact that it is generated
from the spin-1 path $[[l,0]]\  (l\in2\Z_{\ge0}+1)$ by \refeq{3act4}.
$\P(2)$ is only ``half'' of $\Aff(B)\otimes\Aff(B)$, the latter being
labeled by a pair of integers without restriction.
\par
Generally, in $\P(n)$ the maximum spin of a path is $n/2$,
and such a path has all $(++)$ walls. Typical is
\eq
&[[l_1,l_2,\cdots,l_n]], \qquad
l_j-l_{j+1}\ \in2\Z_{\ge0}+1 &(3path1)
\endeq
with wall labels alternating between even and odd integers and weight
\eq
\wt[[l_1,l_2,\cdots,l_n]]
&= -{(l_1+\cdots+l_n+\left[{n\over2}\right])\over 2}\delta
+ {n\over2}\alpha_1. &(wtpls)\cr
\endeq
Here $[n/2]$ denotes integer part.
Because of the ordering of the wall labels, there are in general only
``$1/n!$ times as many'' connected components in $\P(n)$ as in
$\Aff(B)^{\otimes n}$.
The precise formulation will be given in the next subsection.
\par
\def\sbtitle{Paths as a quotient of $\sqcup_{n=0}^\infty\Aff(B)^{\o n}$}
\def\tf{{\tilde f}}
\def\te{{\tilde e}}
\par
\subsec(3.2|\sbtitle)
Let ${\cal{Z}}$ be the $\Z$-algebra generated by $\psi^*_j\ (j\in\Z)$
satisfying the relations
\eq
&\psi^*_j\psi^*_k+\psi^*_k\psi^*_j=0,\quad
\psi^*_{j+1}\psi^*_k+\psi^*_{k+1}\psi^*_j=0,&(Zrel)\cr
\endeq
for all $j$, $k\in\Z$ such that $j\equiv k\bmod2$.
This algebra arises from the commutation relation \refeq{CA:b} of the creation
operators expanding then formally in $q$ and $z$.
\eq
&\varphi^*_+(z)=\sum_{j\in\Z}\psi^*_{2j}z^j+O(q),\cr
&\varphi^*_-(z)=\sum_{j\in\Z}\psi^*_{2j+1}z^j+O(q).\cr
\endeq
Here we assume that the creation operators preserve the crystal lattice,
i.e., there is no negative powers in $q$ (see Appendix 4). We have also removed
the fractional powers in $z$ from $\varphi^*_\pm$.
\par
Consider the $\Z$-module
$\Z{\cal{P}}$
spanned by the set of paths
${\cal{P}}^0\sqcup{\cal{P}}^1$.
We define an action of ${\cal{Z}}$ on
$\Z{\cal{P}}$ as follows.
Take $[[l_1,\ldots,l_n]]\in{\cal{P}}(n)$.
If $n=0$ we set
\eq
&\psi^*_j[[\ ]]=[[j]].\cr
\endeq
If $j>l_1+n$, then we define
\eq
&\psi^*_j[[l_1,\ldots,l_n]]=
[[j-n,l_1,\ldots,l_n]].\cr
\endeq
If $j=l_1+n$ or $l_1+n-1$, we define
\eq
&\psi^*_j[[l_1,\ldots,l_n]]=0.\cr
\endeq
Finally, if $j<l_1+n-1$, then we define
\eq
\psi^*_j[[l_1,\ldots,l_n]]&
=-\psi^*_{l_1+n-1}\psi^*_j[[l_2\ldots\l_n]]
\qbox{if $j\equiv l_1+n-1\bmod2$},\cr
&=-\psi^*_{l_1+n}\psi^*_{j-1}[[l_2\ldots\l_n]]
\qbox{if $j\not\equiv l_1+n-1\bmod2$}.\cr
\endeq

{}From this action we see that the following is
a linear base of ${\cal{Z}}$.
\eq
&{\cal{B}}=\sqcup_{n=0}^\infty{\cal{B}}(n),
\quad
{\cal{B}}(n)=\{\psi^*_{j_1}\cdots\psi^*_{j_n};
j_1-n+1>j_2-n+2>\cdots>j_n\}.\cr
\endeq
There is a bijection $\omega$ from ${\cal{Z}}$
to $\Z{\cal{P}}$ which maps ${\cal{B}}(n)$ to ${\cal{P}}(n)$.
\eq
&\omega:{\cal{B}}(n)\rightarrow{\cal{P}}(n),&(CryCre)\cr
&\psi^*_{j_1}\cdots\psi^*_{j_n}\mapsto[[j_1-n+1,j_2-n+2,\ldots,j_n]].\cr
\endeq
This is obtained from the action of ${\cal{Z}}$ on the vector
$[[\ ]]\in{\cal{P}}(0)$. (Inparticular, $\omega(1)=[[\ ]]$.)
\par
We make ${\cal{B}}$ an affine crystal as follows.
We define the weight by
\eq
\wt\bigl(\psi^*_{j_1}\cdots\psi^*_{j_n}\bigr)
&=\sum_{k=1}^n\wt(\psi^*_{j_k}),\cr
\wt\bigl(\psi^*_{2j}\bigr)=-j\delta+{1\over2}\alpha_1,\
&\wt\bigl(\psi^*_{2j+1}\bigr)=-j\delta-{1\over2}\alpha_1.\cr
\endeq
We define the crystal structure on ${\cal{B}}(1)$ by identifying it
with $\Aff(B)$ by $\psi^*_{j}=[[j]]$.
\par
\def\tens#1{\psi^*_{#1_1}\otimes\cdots\otimes\psi^*_{#1_n}}
\par
To define the crystal structure on ${\cal{B}}(n)\ (n>1)$, let us consider
the map
\eq
&\sqcup_{n=0}^\infty\Aff(B)^{\otimes n}\rightarrow
\sqcup_{n=0}^\infty\bigl({\cal{B}}(n)\sqcup-{\cal{B}}(n)\bigr)\sqcup0,\cr
&\tens{j}\mapsto
\psi^*_{j_1}\cdots\psi^*_{j_n}.\cr
\endeq
Starting from the product $\psi^*_{j_1}\cdots\psi^*_{j_n}$
with an arbitrary set of indices $j=(j_1,\ldots,j_n)\in\Z^n$,
we can modify it by using the rule
\refeq{Zrel} to $\pm\psi^*_{j'_1}\cdots\psi^*_{j'_n}$
where $j'_k>j'_{k+1}+1$ for all $k$, or to 0.
This process is compatible with the arrows in the crystal
$\sqcup_{n=0}^\infty\Aff(B)^{\otimes n}$ in the following sense.
Suppose that an index set $j$ changes to $j'$ by applying the rule
\refeq{Zrel} once.
Suppose also that an arrow goes from
$\tens{j}$ to $\tens{{\bar j}}$, and
$\tens{j'}$ to $\tens{{\bar j'}}$. Then, by case checking one can show that
${\bar j}$ changes to ${\bar j'}$ by applying \refeq{Zrel} once.
Therefore we can induce a crystal structure on ${\cal{B}}$ consistently
from $\sqcup_{n=0}^\infty\Aff(B)^{\otimes n}$.
\par
The map $\omega$ from ${\cal{B}}(n)$ to ${\cal{P}}(n)$
is an isomorphism of crystals, i.e., it is a bijection and it
commutes with $\tf_i$ and $\te_i$.
This is not an isomorphism of affine crystals, because
the affine weights in ${\cal{B}}(n)$ and ${\cal{P}}(n)$ differ
by a mutiple of $\delta$:
\eq
\wt\bigl(\omega(\tens{j})\bigr)-\wt\bigl(\tens{j}\bigr)
&=n^2/4\qbox{if $n$ is even}\cr
&=(n^2-1)/4\qbox{if $n$ is odd}.\cr
\endeq
\par
\subsec(3.3|The $q=0$ limit of the eigenvectors)
\par
We discussed in Section 2 that there is no {\it a priori} method
to forsee what the $q=0$ limit of the eigenvectors of the Hamiltonian is.
Here we give a prediction using the map $\omega$ \refeq{CryCre}.
Note that the definition \refeq{CryCre} is based on the commutation relation
of the creation operators which will be discussed in Section 7.
\par
Define
\eq
&\psi^*_+(z)=\sum_{j\in\Z}z^j\psi^*_{2j},
\quad
\psi^*_-(z)=\sum_{j\in\Z}z^j\psi^*_{2j+1}\cr
\endeq
We conjecture that the $q=0$ limit of the $n$-particle eigenstates
(see Section 5 and 7) are given by
\eq
&\omega\bigl(\psi^*_{\varepsilon_1}(z_1)
\cdots\psi^*_{\varepsilon_n}(z_n)\bigr).\cr
\endeq
The validity of this conjecture is checked for $n=1,2$ in Appendix 2.
It is also consistent with the Bethe Ansatz calculation
for $n=2h$ and $s=h,h-1$ in Appendix 5.
\par
%\endinput
\par
\par
\def\End{\hbox{End}}
\def\ad{\hbox{ad}}
\def\id{\hbox{id}}
\par
\beginsection \S4. The vacuum state
\par
The vacuum state is the unique $\Up$-singlet in $W$
under the boundary condition \refeq{Even}.
To find the vacuum in $W$ means to find a $\Up$-linear map
\eq
&{\bf Q}(q)\longrightarrow \V(0)\o\V(0)^{\ast a}.&(*)\cr
\endeq
It is given by the canonical element;
\eq
&|vac\rangle=\sum_k v_k\o v_k^\ast.&(Vac)\cr
\endeq
Here $v_k$ and $v_k^\ast$ are dual bases.
Since $\V(0)$ is infinite dimensional,
 the above sum is an infinite sum.
Therefore we need a completion of
$\V(0)\o \V(0)^{\ast a}$
in order that the vacuum vector actually
belong to it.
For our purpose the way of completion is not very important.
We take the largest one, i.e.,
the direct product of all the spaces
$\V(0)_\lambda\o \V(0)_\mu^{\ast a}$
where $\lambda$ and $\mu$ runs the weights
of $\V(0)$ and $\V(0)^{\ast a}$ respectively.
We do not use any particular notation for the completion.
We just use $\V(0)\o \V(0)^{\ast a}$ as being completed.
It is convenient to use another realization
of $\V(0)\o \V(0)^{\ast a}$.
Namely we use the canonical isomorphism
\eq
&\V(0)\o \V(0)^{\ast a}\simeq
\End_{{\bf Q}(q)}\bigl(\V(0)\bigr).\cr
\endeq
$\bigl($Strictly speaking, the right hand side is the set of
${\bf Q}(q)$-linear homomorphism
from $\V(0)$ to the completion of $\V(0)$.$\bigr)$
The $\U$-action on the left hand side
is given by the coproduct.
Then the action on the right hand side
is given by the adjoint action which we denote by $\ad \ x$;
suppose that
\eq
&f\in \End_{{\bf Q}(q)}\bigl(\V(0)\bigr),\ v\in\V(0),\  x\in\U\cr
\endeq
and
\eq
&\Delta(x)=\sum_k x_k^{(1)}\o x_k^{(2)}.\cr
\endeq
Then we have
\eq
&\bigl((\ad\ x)f\bigr)(v)=
\sum_k x_k^{(1)}f\bigl(a(x_k^{(2)})v\bigr).\cr
\endeq
If $f=\hbox{id}$, then $(\ad\ x)f=\varepsilon(x)$
$(\varepsilon: \hbox{the counit})$.
Here we used the special choice of the left module structure
$V^{*a}$ in \refeq{*}.
Therefore $\hbox{id}\in \End_{{\bf Q}(q)}\bigl(\V(0)\bigr)$
generates a singlet.
Or equivalently, the canonical element \refeq{Vac}
actually realizes the vacuum.
In Appendix 4  we give some explicit
computation of the embedding of the canonical
element in $W$.
\par
One may raise a question about the translational invariance
of the vacuum.
Because of the splitting of the whole line
into the right and the left pieces,
it is not obvious that the definition
of the vacuum is independent of the choice
of the position of the splitting.
Let us distinguish the vacuum in
$\V(0)\o \V(0)^{\ast a}$ and the vacuum in
$\V(1)\o \V(1)^{\ast a}$
by denoting the former by $|vac\rangle_0$ and
the latter by $|vac\rangle_1$.
We want to prove
\eq
&T|vac\rangle_i=|vac\rangle_{1-i}\quad i=0,1.&(Invar)\cr
\endeq
In order to prove this we need the interpretation
of the translation $T$ in our mathematical picture.
For this purpose we use the vertex
operators
\eq
&\Phi:\V(i)\longrightarrow \V(1-i)\o V\cr
\endeq
and
\eq
&\Psi^\ast:V\o\V(i)^{\ast a}\longrightarrow \V(1-i)^{\ast a}.\cr
\endeq
We introduced $\Phi$ in \S2 with the normalization
\eq
&\Phi(u_{\Lambda_0})=u_{\Lambda_1}\o(-)+\cdots,\
\Phi(u_{\Lambda_1})=u_{\Lambda_0}\o(+)+\cdots.\cr
\endeq
We define $\Psi^\ast$ as the intertwiner with the normalization
$\Psi^\ast((-)\o u_{\Lambda_0}^\ast)=u_{\Lambda_1}^\ast+\cdots$
and
$\Psi^\ast((+)\o u_{\Lambda_1}^\ast)=u_{\Lambda_0}^\ast+\cdots$,
where $u_{\Lambda_i}^\ast$ is the lowest weight vector of
$\V(i)^{\ast a}$ dual to $u_{\Lambda_i}$.
We define
\eq
&T:\V(i)\o \V(i)^{\ast a}
\longrightarrow
\V(1-i)\o \V(1-i)^{\ast a}\cr
\endeq
by the composition
\eq
&\V(i)\o \V(i)^{\ast a}
{\buildrel\Phi\o{\scriptstyle \id}
\over\longrightarrow}
\V(1-i)\o V \o \V(i)^{\ast a}
{\buildrel {\scriptstyle \id}\o\Psi^\ast\over\longrightarrow}
\V(1-i)\o \V(1-i)^{\ast a}
\endeq
up to a constant multiple.
This constant is determined in \S7 and the proof
of the assertion \refeq{Invar}
is also given (see \refeq{Trs}, Proposition \refprop{Vac}).
Once we establish the statement \refeq{Invar}
then
\eq
&H_{\XXZ}|vac\rangle_i=0&(Zero)\cr
\endeq
follows immediately,
because by the definition it is obvious that
\eq
&d|vac\rangle_i=0.&(RowCtm)\cr
\endeq
\par
Before ending this section, we wish to
comment on \refto{FT}, where it was first conjectured,
on the basis of direct perturbative calculations
up to a low order in $q$, that the vacuum vector of
the $\XXZ$ Hamiltonian is also an eigenvector of \refeq{CTM}.
In our algebraic scheme, this follows
directly from \refeq{RowCtm}. Our scheme also
makes it clear that the excited states of the $\XXZ$
Hamiltonian are not eigenstates of $d$.
\par
%\endinput
\par
Before ending this section, we will make a comment on \refto{FT}.
In this paper a conjecture is given that the vacuum vector for the
$\XXZ$ Hamiltoian is also an eigenvector of \refeq{CTM}.
This is trivially correct in our scheme by \refeq{RowCtm}.
On the other hand, it was rather surprising when it was
first pointed out in \refto{FT}. We note that the excited states of
the $\XXZ$ Hamitonians are not eigenstates of $d$.
\par
\par
\def\L{\Lambda}
\def\V#1{V(\La_{#1})}
\def\o{\otimes}
\def\Q{{\bf Q}}
\def\d{d\/}
\def\ten{V_{z_n}\o\cdots\o V_{z_1}}
\def\cre{\varphi^*_\pm(z)}
\def\hwv#1{u_{\La_{#1}}}
\par
\beginsection \S5. Particle Picture
\par
By a particle we mean a finite-dimensional
$\Up$-module consisting of eigenvectors of the
$\XXZ$-Hamiltonian.
\par
Starting from the infinite tensor product
$W$,
we have argued how the
$\Up$-module
\eq
&\V0\o\V0^{*a}\simeq\hbox{ {\rm End}}{}_{\Q(q)}\bigl(\V0\bigr)\cr
\endeq
is embedded therein, and how the vacuum (i.e., the zero particle state)
is understood as a vector of the latter.
Now, guided by the crystal decomposition of the even and the odd paths,
we reach the following; Modulo statistics to be discussed later,
the particle picture for the $\XXZ$-Hamiltonian in the anti-ferroelectric
regime is
\eq
&{\cal{F}}=\left[\oplus_{n=0}^\infty\int\cdots\int\ten\d u_n
\cdots\d u_1\right]_{{\rm symm}},&(FockS)\cr
\endeq
where
$u_i$
is a quasi-momentum of the 2-dimensional
$\Up$-module
$V_{z_k}\ \bigl(z_k=e^{iu_k}\bigr)$.
We call ${\cal{F}}$ the Fock space.
(For the meaning of the symmetrization, see \refeq{par}.)
The even particle ($n$: even) are contained in the even sector
$\V0\o\V0^{*a}$,
and the odd particles ($n$: odd) in the odd sector
$\V1\o\V0^{*a}$.
To see this we want to find a
$\Up$-linear map
\eq
&\ten\rightarrow\V{0\hbox{ or }1}\o\V0^{*a},\cr
\endeq
or equivalently, a
$\Up$-linear map
\eq
&\ten\o\V0\rightarrow\V{0\hbox{ or }1}.\cr
\endeq
Therefore, the problem reduces to finding the vertex operators
\eq
&\Phi(z):V_z\o\V i\rightarrow\V{1-i}\ (i=0,1).\cr
\endeq
The existence and the uniqueness of such vertex operators are given in
[DJO] in a general setting.
In Appendix 2, we give some explicit calculation of the above vertex operator
in terms of the global base of Kashiwara.
\par
Let us argue the physical content of our particle picture.
The
$\XXZ$-Hamiltonian possesses the infinite hierarchy of the abelian
higher order Hamiltonians, and the infinite dimensional non-abelian
symmetries of
$\Up$.
These are the symmetries of the
$\XXZ$-Hamiltonian in the strict sense, i.e., they commute with
$\H$.
The abelian symmetries do not change the eigenvectors, and the non-abelian
symmetries do not change the eigenvalues. The Lorentz boost
$e^{\varepsilon d}\ (\varepsilon:\hbox{ a scalar parameter})$
actually changes
the energy, but never creates new particles nor annihilates them.
Now we introduce the third symmetry of the
$\XXZ$-Hamiltonian, the dynamical symmetries which create and annihilate
particles.
\par
Define the creation operator
\eq
&\cre:\V i\o\V0^{*a}\rightarrow\V{1-i}\o\V0^{*a}\cr
\endeq
by
\eq
&\cre(v\o v^*)=\Phi(z)(v_\pm\o v)\o v^*.\cr
\endeq
Then, it is easy to see that
$\cre$
acting on an
$n$-particle state create an
$(n+1)$-particle state.
In this way, the Fock space ${\cal{F}}$ is embedded in
$\bigl(V(\L_0)\oplus V(\L_1)\bigr)\otimes V(\L_0)^{*a}$.
\par
In \S 6 and \S 7, we give a mathematical treatment of the creation and the
annihilation operators. Here, we give rather heuristic discussions on
several points.
\par
We argued the embedding of $\V0\o\V0^{*a}$ to the infinite tensor product.
The vectors $G(p)$ in $\V0$ are expanded in terms of
the paths $|p\rangle$, where we consider
the latter as vectors in the infinite tensor product.
A question arises: Is it possible to expand the paths in terms of the vectors
in $\V0$. The answer is no. The infinite matrix which expresses the trasition
from $|p\rangle$ to $G(p)$ is of the form
$1+qA_1+q^2A_2+\cdots$. If we invert this, we get formally
$1-qA_1+q^2(A_1^2-A_2)+\cdots$. But $A_1^2$ has divergence on the diagonal.
So, the transition matrix is not invertible. This means that our definition
of the creation operators does not apply to the paths. Suppose we were to apply
the creation operator to the bare vacuum $(\cdots+\ -\ +\ -\ \cdots)$.
Since the action of the creation operators changes only the left half,
the right half is unchanged. This contradicts with the fact that the creation
operator satisfies the proper commutation relations with the shift operator
(see section 7).
\par
%Let us argue the breakdown of the crystal picture in a naive sense.
%The action of the creation operator is not well-defined on arbitrary paths.
%Suppose that one can define the action on the bare vacuum, i.e.,
%${\bar p}_0\o\pi({\bar p}_0)$
%in
%$W$.
%Note that this is translationally invariant, i.e.,
%$T^2\bigl({\bar p}_0\o\pi({\bar p}_0)\bigr)=
%{\bar p}_0\o\pi({\bar p}_0)$.
%Since
%$\cre$ changes only
%the left half, it creates a state which is not an eigenvector
%of
%$T^2$.
%But, we expect, and in fact will prove, that
%$\cre$
%commute with
%$T^2$
%modulo some phase shift.
%This is a contradiction.
%
%The reason of this is that we incorrectly supposed that the bare vacuum
%is contained in
%(a certain completion of)
%$\V0\o{\V0}^{*a}$.
%We expanded weight vectors in
%$\V0$
%in terms of paths in
%$W_l$.
%The fact is this is not reversible;
%paths in
%$W_l$
%cannot be expanded in terms of vectors in
%$\V0$.
%
\par
We conjecture that the creation operators preserve the crystal
structure in the following sense. The vacuum vector embedded in
the infinite tensor product is expanded in power series of $q$.
The conjecture is that the particle states created by the creation operators
upon the vacuum also have the same property.
This is remarkable because the vertex operators used in the definition
of the creation oeprators do not preserve the crystal lattices.
In fact, if they do we have the following contradiction.
At $q=0$ the vacuum considered in $\V0\o\V0^{*a}$
reduces to $u_{\L_0}\o u^*_{\L_0}$,
where $u_{\L_0}$ is the highest weight vector in $\V0$ and $u^*_{\L0}$
the dual lowest weight vector in $\V0^{*a}$ (see Appendix 4).
So, if the crystal lattice were preserved by the vertex operators,
then to get the $q=0$ limit of a one-particle state, we only have to
apply the vertex operator to $u_{\L0}$. But, this breaks the translational
covariance which is expected for the one-particle state. In fact,
we will see in Appendix 4, that we have contributions to
the $q^0$-term in the one-particle state from the higher order terms in
the vacuum state.
\par
\par
Now we come to a more subtle point.
In Appendix 4, we computed 1 and 2 particle states by applying the vertex
operators successively.
In the computation of
\eq
&\Phi(z_2)\bigl(v_-\o\Phi(z_1)(v_+\o\hwv0)\bigr)\cr
\endeq
(which is a part of the computation of
$\varphi^*_-(z_2)\varphi^*_+(z_1)|vac\rangle$),
we find terms having poles at
$q=0$.
But these terms seem to be summed up to a meromorphic
function in
$z_1$, $z_2$ and $q$
that has no pole (actually has a zero) at
$q=0$.
So, even though the vertex operator
$\Phi(z):V_z\o\V i\rightarrow\V{1-i}$
does not preserve the crystal lattice, the particles created by
$\cre$
may not (and, in fact, do not) have poles at
$q=0$.
Namely, the crystal picture survives.
The necessity of this summation also tells that
the Fourier components of the creation and the annihilation operators are not
equal to those of the vertex operators. We will further discuss this
in Section 7.
\par
\par
\def\o{\otimes}
\def\La{\Lambda}
\def\bra{\langle\L_0|}
\def\ket{|\L_0\rangle}

\def\lket{|-\L_0\rangle}
\def\Uh(#1){U^{#1}_q\bigl(\widehat{\goth{sl}}(2)\bigr)}
\par
\def\V#1{V(\L_{#1})}
\par
Let us mention a few words about the statistics of our particles.
The tensor products
$V_{z_2}\o V_{z_1}$
and
$V_{z_1}\o V_{z_2}$
are different but isomorphic.
%(if $z_2/z_1\neq q^{\pm2}$).
They are intertwined by the $R$-matrix.
In fact, the embedded images of these tensor products in
$\V0\o\V0^{*a}$ are equal.
This follows from the uniqueness (up to scalar multiple)
of the vertex operator
$V_{z_1}\o V_{z_2}\o\V0\rightarrow\V0$.
So the key point is to determine this scalar multiple.
We will give the answer to this question in \S 6 by using the quantum
K-Z equation.
\par
Do particles with spin higher than $1\over2$ exist?
Let
$V^{(j)}\ (j\in{1\over2}\Z)$
be the $(2j+1)$-dimensional
$U_q\bigl(sl(2)\bigr)$-module,
and
$V_z^{(j)}$
its
$\Up$-extension. If we understand this question as the
existence of the vertex operator
\eq
&V_z^{(j)}\o\V0\rightarrow\V0\hbox{ or }\V1,\cr
\endeq
one can show the non-existence by an argument of crystals.
If we treat the problem honestly, we should start from the
finite-lattice Bethe Ansatz and examine the thermo-dynamic limit.
%We could not find any comprehensive treatment for the
%$\XXZ$-Hamiltonian in the anti-ferroelectric regime along this line
%in the literature.
There are papers \refto{FaddeevT}
for the $\XXX$-case ($\Delta=-1$)
and \refto{DesL}
for the $\XXZ$-case ($\Delta<-1$),
which are in support of our picture.
%These two papers even suggest the
%completeness of our particle picture.
\par
%Note that
%$\V0$
%and
%$\V0^*$
%are highest weight modules with the highest weight
%$\La_0$
%($\V0^{*a}$
%and
%$\V0^{**a}$
%are lowest weight modules with the lowest weight
%$-\La_0$).
%Let us denote the highest weight vectors by
%$\bra$
%and
%$\ket$
%(the lowest weight vectors by
%$\lbra$
%and
%$\lket$).
%We use Dirac's bracket notation;
%\eq
%&\bra\La_0\rangle=
%\lbra-\La_0\rangle=1.\cr
%\endeq
\par
%Let us choose weight vectors
%$P_k\in\Uh(-)$
%and
%$Q_k\in\Uh(+)$
%in such a way that
%$P_k\ket$
%and
%$\bra Q_k$
%constitute dual bases;
%\eq
%&\bra Q_jP_k\ket=\delta_{jk}.\cr
%\endeq
%Note that by the canonical identification
%$\bra Q_k\in\V0^*$
%goes to
%$a^{-1}(Q_k)\lket\in\V0^{*a}$
%($P_k\ket\in\V0^{**}=\V0$
%to
%$\lbra a(P_k)\in\V0^{**a^{-1}}$).
\par
%What is the value of
%$\lbra a(P_j)a^{-1}(Q_k)\lket$ ?
%The answer is
%\eq
%&\lbra a(P_j)a^{-1}(Q_k)\lket=q^{-4(\rho,\La_0-\mu)}\delta_{jk}\cr
%\endeq
%where
%$\mu=-{\rm wt}P_j$.
%In fact, the above argument is valid for the pair
%$V(\La')\o V(\La)^{*a}$
%and
%$V(\La')^*\o V(\La)^{**a}$,
%where
%$\La$
%and
%$\La'$
%are arbitrary highest weight and lowest weight, respectively.
Let us consider the dual space
$V^*$
of
$V$.
The
$\XXZ$-Hamiltonian
formally acts on the infinite tensor product of
$V^*$,
too.
So, we can formulate the theory in a completely analogous way by using
right modules.
In that case, the (even) particle picture will be developed on the tensor
product of two right modules
$\V0^*\o\V0^{**a^{-1}}$.
Let us, in general, define a natural pairing
$\langle\quad|\quad\rangle$
\eq
&\bigl(V(\La')^*\o V(\La)^{**a^{-1}}\bigr)\o
\bigl(V(\La')\o V(\La)^{*a}\bigr)\rightarrow{\bf Q}(q),
\endeq
where
$\La$
and
$\La'$
are arbitrary dominant integral weights.
The left
$\U$-module
$V(\La')\o V(\La)^{*a}$
and the
right
$\U$-module
$V(\La')^*\o V(\la)^{**a}$
have the canonical pairing $\langle\quad,\quad\rangle$
induced from the pairing between
$V(\La')$ and $V(\La')^*$ and the pairing between
$V(\La)^*$ and $V(\La)^{**}$.
This pairing $\langle\quad,\quad\rangle$
satisfies
\eq
&\langle fx,g\rangle=\langle f,xg\rangle,
\endeq
for $f\in V(\La')^*\o V(\la)^{**a}$,
$g\in V(\La')\o V(\La)^{*a}$ and $x\in\U$.
Note that there is an isomorphism of $\U$-modules
$V(\La)^{**a^{-1}}\simeq V(\La)^{**a}$ given by
\eq
&\bigl(V(\La)^{**a^{-1}}\bigr)_{\lambda}\ni v
\mapsto q^{-4(\rho,\lambda)}v\in \bigl(V(\La)^{**a}\bigr)_{\lambda}.
\endeq
We define $\langle~|~\rangle$ by setting
\eq
&\langle u'\o w'| u\o w\rangle=
\langle q^{-4(\rho,wt(w'))}u'\o w', u\o w\rangle,
\endeq
for weight vectors $u',w',u,v$.
%If we make the identification
%\eq
%&V(\La')\o V(\La)^{*a}\subset {\rm Hom}\bigl(V(\La),V(\La')\bigr),\cr
%&V(\La')^*\o V(\La)^{**a^{-1}}\subset{\rm Hom}\bigl(V(\La'),V(\La)\bigr),\cr
%\endeq
%and induce the pairing between
%${\rm Hom}\bigl(V(\La),V(\La')\bigr)$
%and
%${\rm Hom}\bigl(V(\La'),V(\La)\bigr)$
%from left-hand-sides,
If we regard $f\in V(\La')^*\o V(\La)^{**a^{-1}}$
(resp.  $g\in V(\La')\o V(\La)^{*a}$)
as an element of ${\rm Hom}\bigl(V(\La),V(\La')\bigr)$
(resp.  ${\rm Hom}\bigl(V(\La'),V(\La)\bigr)$)
then we have
\eq
&\langle f|g\rangle=
{\rm tr}{}_{V(\La)}\bigl(q^{-4\rho}f\circ g\bigr).\cr
\endeq
Obviously this pairing satisfies
\eq
&\langle fx|g\rangle=\langle f|xg\rangle
\qbox{for $x\in\U$}.
\endeq
In particular, we have
\eq
&\langle vac|vac\rangle={\rm tr}{}_{\V0}\bigl(q^{-4\rho}\bigr),\cr
\endeq
which is the specialized character for
$\V0$.
\par

\def\La{\Lambda}
\def\bra{\langle\La_0|}
\def\ket{|\La_0\rangle}

\def\lket{|-\La_0\rangle}
\par
\def\V#1{V(\La_{#1})}
\def\lv{\langle vac|}
\def\rv{|vac\rangle}
\def\e{\varepsilon}
\par
Getting the character expression for the one-point function of the
six-vertex model \refto{DJMO,(KMN)^2} is a simple corollary of this formula.
Define a non-local operator
\eq
&\tau_k=\sum_{j>k}\sigma_j^z\cr
\endeq
on
$\V0\o\V0^{*a}$.
We understand the meaning of the infinite sum by the normalization
$\tau_0\bigl(\ket\o\lket\bigr)=0$.
The one-point function
\eq
&P(m)={\lv {\rm Proj}(\tau_0=m)\rv\over\lv vac\rangle}\quad(m\in\Z)\cr
\endeq
of the six-vertex model is by definition the expectation value of the
projection operator to the
$\tau_0=m$
eigenspace. Therefore, it is given by
\eq
&P(m)={{\rm tr}{}_{\V0_m}q^{-4\rho}\over
{\rm tr}{}_{\V0}q^{-4\rho}},\cr
\endeq
where
$\V0_m$
is the spin $m$ subspace of
$\V0$.
\par
By a similar construction, the dual Fock space
$\cal{F}^*$ is embedded in
$\bigl(V(\L_0)^*\oplus V(\L_1)^*\bigr)\otimes V(\L_0)^{**a^{-1}}$.
The bilinear coupling between
$\V i\o\V0^{*a}$ and
$\V i^*\o\V0^{**a^{-1}}$
induces a non-degenerate coupling between
${\cal{F}}$ and ${\cal{F}}^*$.
The $n$-particle states $V_{z_n}\o\cdots\o V_{z_1}$ in $\cal{F}$
and the $m$-particle states
$V^*_{z'_m}\o\cdots\o V^*_{z'_1}$
in ${\cal{F}}^*$ are orthogonal unless $n=m$ and $\{z_i\}=\{z'_i\}$.
\par
Finally, we will consider the annihilation operator.
As we defined the creation operator in the frame of left modules,
we define the operator which create finite-dimensional right
$\Up$-modules, i.e., the dual Fock space, in
$\V0^*\o\V0^{**a^{-1}}$.
It is given by means of the vertex operator of the form
\eq
&\Phi^*(z):V_z^*\o\V i^*\rightarrow\V{1-i}^*.\cr
\endeq
Let
$v^*_\pm\in V_z^*$
be the dual elements to
$v_\pm;\langle v^*_\e,v_{\e'}\rangle=\delta_{\e\e'}$.
Define
\eq
&\varphi_\pm(z):\V i^*\o\V0^{**a^{-1}}
\rightarrow\V{1-i}^*\o\V0^{**a^{-1}}\cr
\endeq
by
\eq
&\varphi_\pm(z)(w^*\o w)=\Phi^*(z)\bigl(v^*_\pm\o w^*\bigr)\o w,\cr
\endeq
where
$w^*\in\V i^*$ and $w\in\V0$.
We define the annihilation operator $\varphi_\pm(z)$
acting on ${\cal{F}}$ by the dual action of
$\varphi_\pm(z):{\cal{F}}^*\rightarrow{\cal{F}}^*$.
Since $\varphi_\pm(z)$ creates $(n+1)$-particle
states from $n$-particle states
in ${\cal{F}}^*$, $\varphi_\pm(z)$ annihilates $(n+1)$-particle states
to $n$-particle states in ${\cal{F}}$.
\par
\par
\par
\par
\par
%\endinput
\par
\par
\par
\par
%\input sec6
%
%
%		Section 6
%
%
%
%
%
%\hsize=5.0truein
%\vsize=7.8truein
%\parindent=15pt
%\input vanilla.sty
%\input macro1
\def\C{{\bf C}}
\def\R{{\bf R}}
\def\bR{\overline{R}}
\def\Q{{\bf Q}}
\def\Z{{\bf Z}}

\def\H{{\cal H}}

\def\F{{\cal F}}

\def\P{{\cal P}}
\def\eL{{\cal  L}}
\def\la{\lambda}
\def\La{\Lambda}
\def\Ga{\Gamma}

\def\Th{\Theta}

\def\Vh{\widehat{V}}
\def\Lh{\widehat{L}}
\def\ep{\varepsilon}

\def\Hom{\hbox{Hom}}
\def\End{\hbox{End}}
\def\id{\hbox{id}}
\def\tr{\hbox{tr}}

\def\slt{\goth{sl}_2}
\def\slth{\widehat{\goth{sl}}(2)\hskip 1pt}
\def\uq{U_q(\goth{g})}
\def\goto#1{{\buildrel #1 \over \longrightarrow}}
\def\br#1{\langle #1 \rangle}
\def\bra#1{\langle #1 |}
\def\ket#1{|#1\rangle}
\def\brak#1#2{\langle #1|#2\rangle}

\def\vac{|\hbox{vac}\rangle}
\def\dvac{\langle \hbox{vac}|}

\def\wt{\hbox{wt}\,}

\def\Phih{\widetilde{\Phi}}

\def\ft{\tilde{f}}
\def\et{\tilde{e}}
\def\fti{\tilde{f}_i}
\def\eti{\tilde{e}_i}
%
%\goth
\font\germ=eufm10
\def\goth#1{\hbox{\germ #1}}
%
%Figure
%
\def\Figure(#1|#2|#3)
{\midinsert
\vskip #2
\hsize 9cm
\raggedright
\noindent
{\bf Figure #1\quad} #3
\endinsert}
%
%Table
%
\def\Table #1. \size #2 \caption #3
{\midinsert
\vskip #2
\hsize 7cm
\raggedright
\noindent
{\bf Table #1.} #3
\endinsert}
%\eq
%
\def\sectiontitle#1\par{\vskip0pt plus.1\vsize\penalty-250
 \vskip0pt plus-.1\vsize\bigskip\vskip\parskip
 \message{#1}\leftline{\bf#1}\nobreak\vglue 5pt}
\def\qed{\hbox{${\vcenter{\vbox{
    \hrule height 0.4pt\hbox{\vrule width 0.4pt height 6pt
    \kern5pt\vrule width 0.4pt}\hrule height 0.4pt}}}$}}
\def\subsec(#1|#2){\medskip\noindent{\it #1}\hskip8pt{\it #2}\quad}
\def\eq#1\endeq
{$$\eqalignno{#1}$$}
\def\leq#1\endeq
{$$\leqalignno{#1}$$}
%
%\qbox
%
\def\qbox#1{\quad\hbox{#1}\quad}
%
%\qqbox
%

%
%\nbox
%

%
%\akete
%

%
%\kitte
%

%
%\Proof
%
\def\Proof{\noindent {\sl Proof.\quad}}
\def\Remark{\smallskip\noindent {\sl Remark.\quad}}
%
%\Example
%
\def\Example{\smallskip\noindent {\sl Example.\quad}}
%
%\Definition
%
\def\Definition#1.#2{\smallskip\noindent {\sl Definition #1.#2\quad}}
%
%\subsec
%
\def\subsec(#1|#2){\medskip\noindent#1\hskip8pt{\sl #2}\quad}
%
%\qed
%
%\inpu box
%\def\qed{\qquad$\Fsquare(.2cm,{})$}
%
%\abstract
%
\def\abstract#1\endabstract{
\bigskip
\itemitem{{}}
{\bf Abstract.}
\quad
#1
\bigskip
}
%
%
%\section
%
\def\sec(#1){Sect.\hskip2pt#1}
\par
\par
\beginsection \S6. Vertex operators
\par
So far we have argued how one can treat the elementary excitations
in the anti-\break
ferroelectric $\XXZ$ model in the framework of representation
theory.
The motivations being given, we now turn to the mathematics of
vertex operators.
\par
\subsec(6.1|Notations) Let us restart by fixing the notations.
Thus let $P=\Z\La_0\oplus\Z\La_1\oplus\Z\delta$ and
$P^*=\Z h_0\oplus\Z h_1\oplus\Z d$
be the weight lattice of $\slth$ and its dual lattice respectively.
We have $\br{\La_i,h_j}=\delta_{ij}$, $\br{\La_i,d}=0$, $\br{\delta,h_i}=0$
and $\br{\delta,d}=1$.
We set $\alpha_1=2\La_1-2\La_0$, $\alpha_0=\delta-\alpha_1$ and
$\rho=\La_0+\La_1$.
We normalize the invariant symmetric bilinear form $(~,~)$ on $P$ by
$(\alpha_i,\alpha_i)=1$. Explicitly we have
$(\La_i,\La_j)=\delta_{i1}\delta_{j1}/4$, $(\La_i,\delta)=1/2$
and $(\delta,\delta)=0$.
We shall regard $P^*$ as a subset of $P$ via $(~,~)$, so that
$2\alpha_i=h_i$ and $4\rho=h_1+2d$.
\par
The quantized affine algebra $U=U_q(\slth)$
is defined  on generators
$e_i$, $f_i$ ($i=0,1$), $q^h$ ($h\in P^*$) over the base field $\Q(q)$.
The defining relations are as given in \refto{Ka,(KMN)^2}, e.g.
$[e_i,f_j]=\delta_{ij}(t_i-t_i^{-1})/(q-q^{-1})$ where $t_i=q^{h_i}$.
We let $U'=U'_q(\slth)$ denote the subalgebra generated by $e_i$, $f_i$ and
$t_i$ ($i=0,1$).
We shall take the coproduct $\Delta$ to be
\eq
&\Delta(e_i)=e_i\otimes 1+t_i\otimes e_i, \quad
\Delta(f_i)=f_i\otimes t_i^{-1}+1\otimes f_i, \cr
&\Delta(q^h)=q^h\otimes q^h\quad (h\in P^*). \cr
&&(COPROD)
\endeq
Accordingly the formula for the antipode $a$ reads
\eq
&a(e_i)=-t_i^{-1}e_i,\quad a(f_i)=-f_it_i, \quad a(q^h)=q^{-h}.
\endeq
\par
\subsec(6.2|Modules)
Given a left $U$-module $M$ we write its weight space as
$M_\nu=\{v\in M\mid q^hv=q^{\br{\nu,h}}v\quad\forall h\in P^*\}$.
For $u\in M_\nu$ we write $\wt(u)=\nu$.
Suppose $M=\oplus_\nu M_\nu$, and let $\phi$ be an anti-automorphism of
$U$. Then the restricted dual $M^*=\oplus_\nu M_\nu^*$ is endowed
with a left module structure $M^{*\phi}$ via
\eq
&\br{u,xv}=\br{\phi(x)u,v}\quad \hbox{ for }x\in U, u\in M, v\in M^*.
\endeq
We have $M\simeq (M^{*\phi})^{*\phi^{-1}}$ (canonically).
Similar convention is used for right modules and $U'$-modules.
Taking $\phi=a$ we have the canonical identification
\eq
\Hom_{U'}(L,M\otimes N)&= \Hom_{U'}(M^{*a}\otimes L, N)\cr
\Hom_{U'}(L\otimes N, M)&=\Hom_{U'}(L, M\otimes N^{*a})\cr
&&(hom)
\endeq
for left modules (for right modules we replace $a$ by $a^{-1}$).
\par
In what follows we set
\eq
&\la=\La_i,\quad \mu=\La_{1-i} \qquad \hbox{for}~i=0~\hbox{or}~1.
&(\la\mu)
\endeq
As before let $V(\la)$ (resp. $V^r(\la)$) denote
the integrable left (resp. right) highest weight module with highest weight
$\la$.
%One may identify $V^r(\la)$ with $V(\la)$ equipped with the right $U$-action
%
%$ux=\varphi(x)u$ ($x\in U$, $u\in V(\la)$),
%where $\varphi$ signifies the anti-automorphism of $U$ given by
%\eq
%&\varphi(e_i)=f_i,\quad \varphi(f_i)=e_i,\quad \varphi(q^h)=q^h.&(\varphi)
%\endeq
To distinguish the left and right structures we shall
use the bra-ket notation $\bra{u}$, $\ket{u}$ for vectors in
$V^r(\la)$ and $V(\la)$, respectively.
We fix nonzero highest weight vectors $\bra{u_\la}\in V^r(\la)$,
$\ket{u_\la}\in V(\la)$ once for all.
Then there is a unique non-degenerate symmetric bilinear pairing
$V^r(\la)\times V(\la) \rightarrow \Q(q)$ such that
\eq
&\brak{u_\la}{u_\la}=1,\quad \brak{ux}{u'}=\brak{u}{xu'}
\hbox{ for any }\bra{u}\in V^r(\la), \ket{u'}\in V(\la).
\endeq
\par
We shall also consider the simplest two-dimensional $U'$-module
$V=\Q(q)v_+ + \Q(q)v_-$ given by
\eq
&e_1v_+=0,~e_1v_-=v_+,~f_1v_+=v_-,~f_1v_-=0,~t_1v_{\pm}=q^{\pm 1}v_{\pm},\cr
&e_0=f_1,~f_0=e_1,~t_0=t_1^{-1}\quad \hbox{on} ~V.\cr
\endeq
For fixed $m\in\Z$ we equip $V\otimes \Q(q)[z,z^{-1}]$ with a
$U$-module structure by letting
\eq
&e_i ~\hbox{ act as }~e_i\otimes (zq^m)^{\delta_{i0}},\quad
f_i\hbox{ act as }~f_i\otimes (zq^m)^{-\delta_{i0}}, \cr
&\hbox{ the weight of }~v_{\pm}\otimes z^n =n\delta\pm (\La_1-\La_0).\cr
\endeq
The resulting $U$-module will be denoted by $V_{zq^m}$.
(This conflicts with the notation for weight spaces,
but the meaning will be clear from the context.)
\par
Let $v_\pm^*$ be the basis of $V^*$ dual to $v_\pm$:
$\br{v_i,v_j^*}=\delta_{ij}$.
One checks readily that the following give isomorphisms
(`charge conjugation') of $U$-modules
\eq
&C_\pm~:~V_{zq^{\mp 2}}\quad \buildrel \sim \over \longrightarrow
V_z^{*a^{\pm 1}} &(CCg)\cr
&C_\pm v_+=v_-^*,\quad C_\pm v_-=-q^{\pm 1}v_+^*.
\endeq
\par
\subsec(6.3|Basic vertex operators)
%Definition, existence, normalization, fractional powers
By vertex operators (VOs) we will mean the intertwiners of $U'$-modules
of the type
\eq
\Phih_\la^{\mu V}~&:~V(\la)\goto{} \Vh(\mu)\otimes V, &(BV1)\cr
\Phih_\la^{V \mu}~&:~V(\la)\goto{} V\otimes \Vh(\mu). &(BV2)\cr
\endeq
Here $\Vh(\mu)=\prod_\nu V(\mu)_\nu$ is a completion of $V(\mu)$.
Since the weight is preserved modulo $\delta$,
for given $v\in V(\la)_\nu$ one can write
$\Phih_\la^{\mu V}~v=\sum_{n\in\Z}
\bigl(u_{+,n}\otimes v_+ + u_{-,n}\otimes v_-\bigr)$
where $u_{\pm,n}\in V(\mu)_{\nu\mp (\La_1-\La_0)+n\delta}$.
(Because the weights of $V(\mu)$ are bounded from above,
$u_{\pm,n}=0$ for $n$ large enough.)
Thus one can define the weight components
$\bigl(\Phih_\la^{\mu V}\bigr)_{\pm,n}$,
$\bigl(\Phih_\la^{V \mu}\bigr)_{\pm,n}$  by
\eq
&\Phih_\la^{\mu V}=\sum_{n\in\Z,\pm}
\bigl(\Phih_\la^{\mu V}\bigr)_{\pm,n}\otimes v_\pm, \quad
\Phih_\la^{V \mu}=\sum_{n\in\Z,\pm}
v_\pm\otimes \bigl(\Phih_\la^{V \mu}\bigr)_{\pm,n}, \cr
&\bigl(\Phih_\la^{\mu V}\bigr)_{\pm,n},
\bigl(\Phih_\la^{V \mu}\bigr)_{\pm,n}~
:~V(\la)_\nu \longrightarrow V(\mu)_{\nu\mp (\La_1-\La_0)+n\delta}.
&(WCP) \cr
\endeq
We shall fix the normalization as follows:
\eq
\Phih_{\la}^{\mu V}\ket{u_\la}& = \ket{u_\mu}\otimes v_\mp+\cdots,&(VN:a)\cr
\Phih_{\la}^{V \mu}\ket{u_\la}& = v_\mp\otimes \ket{u_\mu}+\cdots.&(VN:b)\cr
\endeq
Here $v_-$ (resp. $v_+$) is chosen for $\la=\La_0$ (resp. $\la=\La_1$).
In \refeq{VN:a} $\cdots$ means terms of the form $\ket{u}\otimes v$ with
$\ket{u}\not\in V(\mu)_\mu$, and similarly for \refeq{VN:b}.
The existence and uniqueness of such VOs are shown in \refto{FR,DJO}.
\par
The VOs can be equivalently formulated as intertwiners of $U$-modules of
the form \refto{FR}
\eq
&\Phi_\la^{\mu V}(z)=\Phih_\la^{\mu V}(z) z^{\Delta_\mu-\Delta_\la},
\quad \Phi_\la^{V \mu}(z)
=\Phih_\la^{V \mu}(z) z^{\Delta_\mu-\Delta_\la},\cr
&\Phih_\la^{\mu V}(z)=
\sum\bigl(\Phih_\la^{\mu V}\bigr)_{\ep,n}\otimes v_\ep z^{-n}
&(FR:a)\cr
&\Phih_\la^{V \mu}(z)
=\sum v_\ep z^{-n}\otimes \bigl(\Phih_\la^{V \mu}\bigr)_{\ep,n}
&(FR:b)\cr
\endeq
Here we set $\Delta_\la=(\la,\la+2\rho)/(k+h^{\vee})$
where $k=1$ is the level and $h^{\vee}=2$ is the dual Coxeter number;
explicitly $\Delta_{\La_0}=0$, $\Delta_{\La_1}=1/4$.
The right hand sides of \refeq{FR:a}, \refeq{FR:b} mean e.g.
$V(\mu)\widehat{\otimes} V_z=\oplus_\nu \prod_\xi V(\mu)_\xi
\otimes (V_z)_{\nu-\xi}$.
The fractional powers are so designed as
to put the $q$-KZ equation in neater form, see below.
\par
By abuse of notation we let $d\in \End(V(\la))$ denote the operator
\eq
&d \ket{u}=\br{d, \nu} \ket{u} \quad \ket{u}\in V(\la)_\nu. &(D)
\endeq
It is easy to see that
$[d,\bigl(\Phih_\la^{\mu V}\bigr)_{\pm,n}]
=n \bigl(\Phih_\la^{\mu V}\bigr)_{\pm,n}$
and hence that
\eq
&(d\otimes\id) \Phi_\la^{\mu V}(z)- \Phi_\la^{\mu V}(z)d
=-\bigl(z{d\over dz}-\Delta_\mu+\Delta_\la\bigr)\Phi_\la^{\mu V}(z).
\endeq
Similar relation holds for $\Phi_\la^{V \mu}(z)$.
\par
\Remark
In \refto{(KMN)^2,DJO} the coproduct of $U$ is chosen to be
\eq
&\Delta_-(e_i)=e_i\otimes t_i^{-1}+ 1\otimes e_i, \quad
\Delta_-(f_i)=f_i\otimes 1+ t_i \otimes f_i, \cr
&\Delta_-(q^h)=q^h\otimes q^h\quad (h\in P^*). \cr
\endeq
The present formulation using
the coproduct $\Delta=\Delta_+$ \refeq{COPROD} is
related to the references above as follows.
\par
Let $M$, $N$ be $U$-modules such that
$\wt(M)\subset \la_0+\sum\Z\alpha_i$,
$\wt(N)\subset \mu_0+\sum\Z\alpha_i$
for some $\la_0, \mu_0\in P$.
We define operators $\beta_M$, $\gamma_{MN}$ by
\eq
&\beta_M u=q^{-(\la,\la)+(\la_0,\la_0)}u\quad u\in M_\la, \cr
&\gamma_{MN} u\otimes v=q^{2(\la,\mu)-2(\la_0,\mu_0)}
u\otimes v, \quad u\in M_\la,v\in N_\mu.
\endeq
Then $\Delta_+(x)=\gamma_{MN}\circ\Delta_-(x)\circ\gamma_{MN}^{-1}$
($x\in U$) and $\beta_M\otimes\beta_N=\beta_{M\otimes N}\circ\gamma_{MN}
=\gamma_{MN}\circ\beta_{M\otimes N}$.
We extend $\gamma_{MN}$ also to $M\widehat{\otimes} N$.
It is known \refto{Ka} that
\medskip\noindent
\item{(i)} $(L,B)$ is a lower crystal base of $M$ if and only if
$\bigl(\beta_M(L),\beta_M(B)\bigr)$ is an upper crystal base of $M$.
\medskip
\def\Ll{L^{\rm low}}
\def\Bl{B^{\rm low}}
\def\Lu{L^{\rm up}}
\def\Bu{B^{\rm up}}
\def\Phil{\Phi^{\rm low}}
\def\Phiu{\Phi^{\rm up}}
\par
\noindent Suppose $M_i$  have lower crystal bases $(\Ll_i,\Bl_i)$ ($i=1,2,3$),
and set $\Lu_i=\beta_{M_i}(\Ll_i)$, $\Bu_i=\beta_{M_i}(\Bl_i)$.
For $\Phil:M_1\rightarrow M_2\widehat{\otimes} M_3$ we put
$\Phiu=\gamma_{M_2M_3}\circ\Phil$. Then we have
\medskip\noindent
\item{(ii)}$\Phil x=\Delta_-(x)\Phil$ if and only if
$\Phiu x=\Delta_+(x)\Phiu$ ($x\in U$),
\item{(iii)} $\Phiu\beta_{M_1}=\beta_{M_2}\otimes\beta_{M_3}\Phil$.
Hence $\Phil\bigl(\Ll_1)\subset \Ll_2\widehat{\otimes}\Ll_3$
if and only if
$\Phiu\bigl(\Lu_1)\subset \Lu_2\widehat{\otimes}\Lu_3$
\medskip\noindent
Similar statements are valid for the intertwiners of the type
$\Psi:M_1\otimes M_2 \rightarrow M_3$.
\par
\subsec(6.4|Variants of vertex operators)
The identification \refeq{hom} along with the isomorphisms \refeq{CCg}
gives rise to the following variants of intertwiners.
\medskip\noindent Type I:
\eq
&\Phih_{\la}^{\mu V}: V(\la)\longrightarrow \Vh(\mu)\otimes V,&(TI:a)\cr
&\Phih_{\mu V}^{\la}: V(\mu)\otimes V\longrightarrow \Vh(\la),&(TI:b)\cr
&\Phih_{V \mu}^{*\la}: V\otimes V(\mu)^{*a}
\longrightarrow \Vh(\la)^{*a},&(TI:c)\cr
&\Phih_{\la}^{*V \mu}: V(\la)^{*a}\longrightarrow V\otimes\Vh(\mu)^{*a}.
&(TI:d)\cr
\endeq
\noindent Type II:
\eq
&\Phih_{\la}^{V \mu}: V(\la)\longrightarrow V\otimes\Vh(\mu),&(TII:a)\cr
&\Phih_{V \mu}^{\la}: V\otimes V(\mu) \longrightarrow \Vh(\la),&(TII:b)\cr
&\Phih_{\mu V}^{*\la}: V(\mu)^{*a}\otimes V
\longrightarrow \Vh(\la)^{*a},&(TII:c)\cr
&\Phih_{\la}^{*\mu V}: V(\la)^{*a}\longrightarrow \Vh(\mu)^{*a}\otimes V.
&(TII:d)\cr
\endeq
That is, \refeq{TI:b} is obtained from
\eq
&V(\mu)\longrightarrow \Vh(\la)\otimes V^{*a^{-1}},
\endeq
and \refeq{TI:c}, \refeq{TI:d} are transpose of
\eq
&V(\la)\longrightarrow \Vh(\mu)\otimes V^{*a^{-1}} \quad
V(\mu)\otimes V^{*a^{-1}} \longrightarrow \Vh(\la) \cr
\endeq
respectively. The case of Type II is similar.
We define
\eq
&\Phi_{\la V}^\mu(z)(u\otimes v)=(\id\otimes \br{v,~})
\Phi_\la^{\mu V^{*a}}(z)u,\quad
\Phi_\la^{\mu V^{*a}}(z)=(\id\otimes C_+)\Phi_\la^{\mu V}(zq^{-2}),\cr
&&(Red:a)\cr
&\Phi_{V \la}^\mu(z)(v\otimes u)=
(\br{v,~}\otimes \id)\Phi_\la^{V^{*a^{-1}} \mu}(z)u,\quad
\Phi_\la^{V^{*a^{-1}} \mu}(z)=
(C_-\otimes \id)\Phi_\la^{V \mu}(zq^2). \cr
&&(Red:b)\cr
\endeq
This implies the normalization
\eq
&\Phi_{\la V}^{\mu}(z)\bigl(\ket{u_\la}\otimes v_{\pm}\bigr)
=\mp z^{\pm 1/4}q^{1/2} \ket{u_\mu} +\cdots, \cr
&\Phi_{V \la}^{\mu}(z)\bigl(v_{\pm}\otimes \ket{u_\la}\bigr)
=\mp z^{\pm 1/4}q^{-1/2} \ket{u_\mu} +\cdots, \cr
\endeq
where the upper (resp. lower) sign is chosen for $\la=\La_0$
(resp. $\la=\La_1$).
For example $z^{-\Delta_\mu+\Delta_\la}\Phi_{\la V}^{\mu}(z)$ is
a $U\otimes \Q(q)[z,z^{-1}]$-linear map
\eq
&V(\la)\widehat{\otimes}V_z \goto{} \Vh(\mu)\otimes \Q(q)[z,z^{-1}]
\endeq
where on the right hand side $x\in U'$ acts as $x\otimes \id$
and $q^d$ acts as $q^d\otimes q^{zd/dz}$, $(q^{zd/dz}f)(z)=f(qz)$.
\par
We have distinguished the two types of intertwiners. The main difference is
that the type I operators preserve the crystal lattice (see \sec(6.7)), while
the type II operators do not. We shall see explicit examples of the latter
phenomenom in Appendix 4.
\par
\subsec(6.5|Two point functions)
%KZ equation, list of various types of 2 point functions,
%meromorphy of matrix elements
%
Frenkel and Reshetikhin \refto{FR} showed that the correlation functions of the
vertex operators satisfy a $q$ analog of the Knizhnik-Zamolodchikov ($q$-KZ)
equation. For our subsequent discussions we need mostly the case of two
point functions for various combinations of VOs. The general case
will be discussed in \sec(6.8).
\par
Let $\Psi(z_1,z_2)$ be one of the following correlations:
\eq
&\bra{u_\la}\Phi_\mu^{\la V_2}(z_2)\Phi_\la^{\mu V_1}(z_1)\ket{u_\la},
&(Ps:a) \cr
&\bra{u_\la}\Phi_\mu^{V_2 \la}(z_2)\Phi_\la^{\mu V_1}(z_1)\ket{u_\la}.
&(Ps:b) \cr
&\bra{u_\la}\Phi_\mu^{\la V_2}(z_2)\Phi_\la^{V_1 \mu}(z_1)\ket{u_\la},
&(Ps:c) \cr
&\bra{u_\la}\Phi_\mu^{V_2 \la}(z_2)\Phi_\la^{V_1 \mu}(z_1)\ket{u_\la}.
&(Ps:d) \cr
\endeq
Clearly
\eq
&\Psi(z_1,z_2)\in V\otimes V
\otimes (z_1/z_2)^{\Delta_\mu-\Delta_\la} \Q(q)[[z_1/z_2]]. &(BC)
\endeq
The subscripts of $V$ in \refeq{Ps:a}--\refeq{Ps:d}
indicate the tensor components.
In \refeq{BC} the left $V$ is $V_1$ and the right one is $V_2$.
Thus, for example, if we replace $\ket{u_\la}$ in
\refeq{Ps:a} by $f_i\ket{u_\la}$ then the result becomes
\eq
&\left(f_i\otimes 1 + t_i^{-1}\otimes f_i\right)
\bra{u_\la}\Phi_\mu^{\la V_2}(z_2)\Phi_\la^{\mu V_1}(z_1)\ket{u_\la}
\in V\otimes V. \cr
\endeq
(Note that in the discussion of the embedding of $V(\la)$ into
$V^{\otimes\infty}$ in \sec(3) and in \sec(6.8) we use the opposite ordering
of the components.)
In what follows we introduce the element $q^{h_1/4}$ in $U$ and
extend the base field $\Q(q)$ by adding $q^{1/4}$. (We could avoid
using fractional powers, but the formulas would become slightly more
cumbersome.)
\par
We need also prepare the $R$ matrix. Define
$\bR(z)$, $R^{+}(z)$ by
\eq
& \bR(z)v_\pm\otimes v_\pm=v_\pm\otimes v_\pm,\cr
& \bR(z)v_+\otimes v_-={1-z\over 1-q^2z}q\,v_+\otimes v_-
                     +{1-q^2\over 1-q^2z}z\,v_-\otimes v_+, \cr
& \bR(z)v_-\otimes v_+={1-q^2\over 1-q^2z}\,v_+\otimes v_-
                     +{1-z\over 1-q^2z}q\,v_-\otimes v_+. \cr
&&(RB) \cr
&R^{+}(z)=\rho(z) \bR(z),\quad
\rho(z)=
q^{-1/2}{(q^2z)_\infty^2 \over (z)_\infty(q^4z)_\infty},\cr
\endeq
where we put
\eq
&(z)_\infty=(z;q^4)_\infty,\quad
(z;p)_\infty=\prod_{j=0}^{\infty}(1-zp^j).
\endeq
Let further $P\in \End(V\otimes V)$ be $P v\otimes v'=v'\otimes v$.
Then $P\bR(z_1/z_2):$$V_{z_1}\otimes V_{z_2}
\rightarrow V_{z_2}\otimes V_{z_1}$ is an intertwiner of $U$-modules,
and $R^{+}(z_1/z_2)$ is the image of the universal $R$ matrix of $U$
in $\End\bigl(V_{z_1}\otimes V_{z_2}\bigr)$.
The scalar factor $\rho(z)$ is determined by the argument in \refto{FR}.
\par
With these notations the $q$-KZ equations read as follows:
\eq
&\Psi(q^6z_1,z_2)=A(z_1,z_2)\Psi(z_1,z_2), \quad
\Psi(q^6z_1,q^6z_2)
=\left(q^{-\phi}\otimes q^{-\phi}\right)\Psi(z_1,z_2), \cr
&&(KZ)\cr
\endeq
where $\phi=4\bar{\la}+h_1$, $\overline{\La}_0=0$,
$\overline{\La}_1=h_1/4$, and
\eq
A(z_1,z_2)
&=R^{+}(q^6z_1/z_2)\bigl(q^{-\phi}\otimes 1\bigr)\qquad\hbox{for \refeq{Ps:a}},
\cr
&=\bigl(q^{-2\bar{\la}}\otimes 1\bigr)R^{+}(q^5z_1/z_2)
(q^{-\phi+2\bar{\la}}\otimes 1\bigr)\qquad\hbox{for \refeq{Ps:b}},\cr
&=\bigl(q^{2\bar{\la}-\phi}\otimes 1\bigr)
R^{+}(qz_1/z_2)(q^{-2\bar{\la}}\otimes 1)
\qquad\hbox{for \refeq{Ps:c}},\cr
&=\bigl(q^{-\phi}\otimes 1\bigr)
R^{+}(z_1/z_2) \qquad\hbox{for \refeq{Ps:d}}.\cr
\endeq
\par
In the present case the property \refeq{BC} and the normalization
\refeq{VN:a}-\refeq{VN:b}
specify the solutions of \refeq{KZ} uniquely.
We list the answers  below.
%list of two point functions
\medskip\noindent $\la=\La_0$:
\eq
(z_1/z_2)^{-1/4}\Psi(z_1,z_2)
&={(q^6z_1/z_2)_\infty\over(q^4z_1/z_2)_\infty}
(v_-\otimes v_+ -qv_+\otimes v_-) \qquad\hbox{for \refeq{Ps:a}},\cr
&={(q^5z_1/z_2)_\infty\over(q^3z_1/z_2)_\infty}
(v_-\otimes v_+ -qv_+\otimes v_-) \qquad\hbox{for \refeq{Ps:b}},\cr
&={(qz_1/z_2)_\infty\over(q^{-1}z_1/z_2)_\infty}
(v_-\otimes v_+ -q^{-1}v_+\otimes v_-) \qquad\hbox{for \refeq{Ps:c}},\cr
&={(z_1/z_2)_\infty\over(q^{-2}z_1/z_2)_\infty}
(v_-\otimes v_+ -q^{-1}v_+\otimes v_-) \qquad\hbox{for \refeq{Ps:d}}.\cr
\endeq
\noindent $\la=\La_1$:
\eq
(z_1/z_2)^{1/4}\Psi(z_1,z_2)
&={(q^6z_1/z_2)_\infty\over(q^4z_1/z_2)_\infty}
(v_+\otimes v_- -qz_1/z_2 v_-\otimes v_+) \qquad\hbox{for \refeq{Ps:a}},\cr
&={(q^5z_1/z_2)_\infty\over(q^3z_1/z_2)_\infty}
(v_+\otimes v_- -qz_1/z_2 v_-\otimes v_+) \qquad\hbox{for \refeq{Ps:b}},\cr
&={(qz_1/z_2)_\infty\over(q^{-1}z_1/z_2)_\infty}
(v_+\otimes v_- -q^{-1}z_1/z_2 v_-\otimes v_+) \qquad\hbox{for
\refeq{Ps:c}},\cr
&={(z_1/z_2)_\infty\over(q^{-2}z_1/z_2)_\infty}
(v_+\otimes v_- -q^{-1}z_1/z_2 v_-\otimes v_+) \qquad\hbox{for \refeq{Ps:d}}.
\cr
\endeq
\par
\subsec(6.6|Commutation relations)
The general theory of $q$-difference equations tells \refto{Bir,Ao}
that the $n$ point
functions of VOs can be continued  meromorphically to the entire space
$(\C^{\times})^n$, apart from overall powers or logarithms in $z_i$.
In our case this is apparent from the explicit formulas.
It follows that the same is true of all the matrix elements
of compositions of VOs (see the proof of Proposition \refprop{COM} below.)

{}From the knowledge of the two point functions one can derive the commutation
relations of VOs \refto{FR}.
To write down the relations which will be used later, we need to
modify the scalar multiple of the $R$ matrix and define
\eq
&R_{VV}(z)=r_0(z)\bR(z),\quad
R_{V^*V^*}(z)
=\bigl(C_-\otimes C_-\bigr)R_{VV}(z)\bigl(C_-\otimes C_-\bigr)^{-1},\cr
&R_{VV^*}(z)=
\bigl(\id\otimes C_-\bigr)R_{VV}(zq^{-2})\bigl(\id\otimes C_-\bigr)^{-1}.\cr
&&(Rmat)
\endeq
Here
\eq
z^{-1/2}r_0(z)
&={(z^{-1})_\infty(q^2z)_\infty\over(z)_\infty(q^2z^{-1})_\infty}
={\Ga_{q^4}\bigl({1\over 2}+{\beta\over 2\pi i}\bigr)
 \Ga_{q^4}\bigl(-{\beta\over 2\pi i}\bigr)
\over
\Ga_{q^4}\bigl({1\over 2}-{\beta\over 2\pi i}\bigr)
\Ga_{q^4}\bigl({\beta\over 2\pi i}\bigr)
}&(qGam) \cr
\endeq
with $z=q^{-2\beta/i\pi}$ and
$\Ga_p(x)=(p;p)_\infty/(p^x;p)_\infty (1-p)^{1-x}$
denoting the $q$-gamma function.
\par
We have the unitarity and crossing symmetry:
\eq
&R_{VV}(z)P R_{VV}(z^{-1})P=\id, \cr
&\bigl(R_{VV}(z)^{-1}\bigr)^{t_2}=-R_{VV^*}(z). \cr
\endeq
Notice that $R_{VV}(z)$, $R_{V^*V^*}(z)$ and $R_{VV^*}(z)$ have no poles
in the neighborhood of $|z|=1$.
\par
In the following theorem, we list the commutation relations of
VOs of type \refeq{TII:a}-\refeq{TII:b}.
The commutation relations and the holomorphy
properties are in the sense of matrix elements.
For example \refeq{COM:c} below states that for each $\bra{u}\in V^r(\la)$,
$\ket{u'}\in V(\la)$ the following hold as meromorphic functions in $z_1$,
$z_2$ times $(z_1/z_2)^{\pm 1/4}$:
\eq
\bra{u}\Phi_\mu^{V_1\la}(z_1)\Phi_\la^{V_2^{*a^{-1}}\mu}(z_2)\ket{u'}
&=R_{VV^*}(z_1/z_2)
\bra{u}\Phi_\mu^{V_2^{*a^{-1}}\la}(z_2)\Phi_\la^{V_1\mu}(z_1)\ket{u'}.
&(COM:x)\cr
\endeq
\par
\proclaim Proposition \prop{COM}.
\item{(i)}The following commutation relation holds:
\eq
\Phi_\mu^{V_1\la}(z_1)\Phi_\la^{V_2\mu}(z_2)
&=R_{V_1V_2}(z_1/z_2)
\Phi_\mu^{V_2\la}(z_2)\Phi_\la^{V_1\mu}(z_1),&(COM:a)\cr
\Phi_\mu^{V_1^{*a^{-1}}\la}(z_1)\Phi_\la^{V_2^{*a^{-1}}\mu}(z_2)
&=R_{V_1^*V_2^*}(z_1/z_2)
\Phi_\mu^{V^{*a^{-1}}_2\la}(z_2)\Phi_\la^{V^{*a^{-1}}_1\mu}(z_1),&(COM:b)\cr
\Phi_\mu^{V_1\la}(z_1)\Phi_\la^{V_2^{*a^{-1}}\mu}(z_2)
&=R_{V_1V_2^*}(z_1/z_2)
\Phi_\mu^{V_2^{*a^{-1}}\la}(z_2)\Phi_\la^{V_1\mu}(z_1). &(COM:c)\cr
\endeq
\item{(ii)}In the neighborhood of $|z_1/z_2|=1$
both sides of \refeq{COM:a}-\refeq{COM:b}
are holomorphic, while
\refeq{COM:c} has a simple pole at $z_1=z_2$.
The residue of the latter is given by
\eq
&\hbox{Res}_{z=1}
\Phi_\mu^{V_1\la}(z_1)\Phi_\la^{V_2^{*a^{-1}}\mu}(z_2)dz
=g_{\la}\id_{V(\la)}\otimes\left(v_+\otimes v^*_+ + v_-\otimes v^*_-\right),
&(COM:d)
\endeq
where $z=z_1/z_2$ and
\eq
g_{\la}&=\mp q^{-1/2}{(q^2)_\infty\over (q^4)_\infty} &(g)
\endeq
with the sign $-$ (resp. $+$) being taken for $\la=\La_0$ (resp. $\La_1$).
%\endproclaim
\par
\Proof The argument being similar, we concentrate on \refeq{COM:c}.
\par
If $\bra{u}=\bra{u_\la}$ and $\ket{u'}=\ket{u_\la}$ the assertions (i),(ii)
follow from the explicit formulas \refeq{Ps} and \refeq{Red:a}-\refeq{Red:b}.
Suppose $\ket{u'}=\ket{xu''}$ with $x\in U'$.
Then the intertwining property of vertex operators entails
\eq
&P\bra{u}\Phi_\mu^{V_1\la}(z_1)\Phi_\la^{V_2^{*a^{-1}}\mu}(z_2)\ket{xu''}\cr
&\quad=\sum x_{(1)}\otimes x_{(2)}
P\bra{ux_{(3)}}
\Phi_\mu^{V_1\la}(z_1)\Phi_\la^{V_2^{*a^{-1}}\mu}(z_2)\ket{u''}
\in V_2^{*a^{-1}}\o V_1.
&(COMaux)\cr
\endeq
Here $\Delta^{(2)}(x)=\sum x_{(1)}\otimes x_{(2)}\otimes x_{(3)}$.
Analogous formula holds for $\bra{u}=\bra{u''x}$.
Since the action of $U'$ on $V^{*a^{-1}}_{z_2}\otimes V_{z_1}$ involves
only powers of $z_i$,
the analyticity property follows by induction on the weight of $\bra{u}$,
$\ket{u'}$.
Because $P R(z)$ is an intertwiner the relation \refeq{COM:c}
is unchanged in the process.
To see (ii) note that if we take the residue of
\refeq{COMaux} then the left hand side reduces to
\eq
&=P\bra{ux}\Phi_\mu^{V_1\la}(z_1)\Phi_\la^{V_2^{*a^{-1}}\mu}(z_2)\ket{u''}.
\endeq
This is because the vector
$w=v_+^*\otimes v_+ + v_-^*\otimes v_-\in V_{z}^{*a^{-1}}\otimes V_z$
belongs to the trivial representation.
We then obtain by the induction hypothesis
$\langle ux|u''\rangle g_\la w=\langle u|xu''\rangle g_\la w$ as desired.
This completes the proof. \qed
\par
We shall also need the following commutation relations,
which can be proved in a similar manner.
\eq
&R_{V_1V_2}(z_1/z_2)\Phi_{\mu}^{\la V_1}(z_1)\Phi_{\la}^{\mu V_2}(z_2)
=-\Phi_{\mu}^{\la V_2}(z_2)\Phi_{\la}^{\mu V_1}(z_1), &(TI_I) \cr
&\Phi_{\mu}^{V_1\la}(z_1)\Phi_{\la}^{\mu V_2}(z_2)
=\Phi_{\mu}^{\la V_2}(z_2)\Phi_{\la}^{V_1 \mu}(z_1)
\times ((z_1/z_2)^{1/2}){\Th(q z_2/z_1)\over \Th(qz_1/z_2)},&(TI_II) \cr
&\Phi_{\mu}^{V_1^{*a^{-1}}\la}(z_1)\Phi_{\la}^{\mu V_2}(z_2)
=\Phi_{\mu}^{\la V_2}(z_2)\Phi_{\la}^{V_1^{*a^{-1}} \mu}(z_1)
\times (-(z_1/z_2)^{-1/2}){\Th(q z_1/z_2)\over \Th(q z_2/z_1)}.&(TII_I)\cr
\endeq
Here we set
\eq
&\Th(z)=(z)_\infty(q^4z^{-1})_\infty(q^4)_\infty.
\endeq
\par
It is known \refto{DJO} that
the composition $\Phih_{\mu V}^{\la}\circ \Phih_{\la}^{\mu V}$ is convergent
in the $q$-adic topology and is proportional to the identity.
The proportionality scalar can be determined from the two point function
by setting $z_1=z_2$.
Using the results \refeq{Ps} we find
\eq
&\Phi_{\mu V}^{\la}(z)\circ \Phi_{\la}^{\mu V}(z)
=-{1\over g_\la}\times \id &(INV)
\endeq
where $g_\la$ is given in \refeq{g}.
\par
\subsec(6.7|Crystal lattice)
%preservation of crystal lattice,
%$\Phi^\la_{\mu V}\Phi^{\mu V}_\la=g\times \id$,
%$q$-adic completion of lattice, composition of VOP
%V(\la) \subset V^{\infty}
%
Recall that on an integrable $U$-module one has the operators
$\eti^{up}$, $\fti^{up}$, $\eti^{low}$, $\fti^{low}$
and the notion of the upper/lower crystal lattice.\refto{Ka}.
The finite dimensional module $V$ has $L=Av_+\oplus Av_-$
as a (both upper and lower) crystal lattice.
\par
The integrable highest weight module $V(\la)$
has the standard lower/upper crystal lattice
$L^{low}(\la)$, $L^{up}(\la)$ given as follows \refto{Ka}.
Let $\varphi$ denote the anti-automorphism of $U$ given by
\eq
&\varphi(e_i)=f_i,\quad \varphi(f_i)=e_i,\quad \varphi(q^h)=q^h,&(varphi)
\endeq
and let $(~,~)$ be the unique symmetric bilinear form on
$V(\la)$ such that
\eq
&(\ket{u_\la},\ket{u_\la})=1, \quad
(\ket{xu},\ket{u'})=(\ket{u},\ket{\varphi(x)u'}). &(inner)
\endeq
Then
\eq
L^{low}(\la)&=~\hbox{ the smallest}~ A\hbox{-module containing }
\ket{u_\la} ~\hbox{and stable under }~\ft^{low}_i,\cr
L^{up}(\la)&=\{u\in V(\la)\mid (u,L^{low}(\la))\subset A\}. &(Lup)\cr
L^{low}(\la)&=\{u\in V(\la)\mid (u,L^{up}(\la))\subset A\}. &(Llow)\cr
\endeq
Here $A=\{f\in\Q(q)\mid f~\hbox{ has no pole at }~q=0\}$.
We have further
\eq
L^{up}(\la)_\nu&=q^{(\la,\la)-(\nu,\nu)}L^{low}(\la)_\nu.
\endeq
\par
The crystal lattices of integrable lowest weight modules are defined
similarly by replacing $\ket{u_\la}$ by the lowest weight vector and
$\ft^{low}_i$ by $\et^{low}_i$.
Let $T_\la\in\End\bigl(V(\la)\bigr)$ denote the linear map
\eq
&T_\la u=q^{(2\rho,\la-\nu)}u\quad \hbox{for }u\in V(\la)_\nu.
\endeq
\par
\proclaim Proposition \prop{cry}.
The upper crystal lattice $L^{*a\, up}(\la)$
of $V(\la)^{*a}$ is characterized as
\eq
&L^{*a\, up}(\la)=\{u\in V(\la)^{*a}\mid \br{u,T_\la L^{up}(\la)}\subset A\},
&(La*)
\endeq
where $\br{~,~}$ denotes the canonical pairing.
\par
\par
\Proof Let $\eta'$ denote the map
$V(\la) \rightarrow V(\la)^{*a}$ given by $\br{\eta'(u),v}=(u,v)$, and set
\eq
&\eta(u)=(-1)^{ht(\la-\nu)}q^{(\nu,\nu+2\rho)-(\la,\la+2\rho)}\eta'(u)
\quad u\in V(\la)_\nu. &(V*V)
\endeq
Here for $\xi=m_0\alpha_0+m_1\alpha_1$ we set $ht(\xi)=m_0+m_1$.
One can verify that
\eq
&\eta(e_i u)=f_i\eta(u),\quad \eta(f_i u)=e_i\eta(u),
\quad\eta(q^h u)=q^{-h}\eta(u).
\endeq
This implies $\eta(\ft^{low}_i v)=\et^{low}_i\eta(v)$,
$\eta(\et^{low}_i v)=\ft^{low}_i\eta(v)$, and hence
\eq
&L^{*a\,low}(\la)=\eta\bigl(L^{low}(\la)\bigr).
\endeq
Using the relation
$\bigl(L^{*a\,up}(\la)\bigr)_{-\nu}
=q^{(\la,\la)-(\nu,\nu)}\bigl(L^{*a\,low}(\la)\bigr)_{-\nu}$
we find
\eq
&\bigl(L^{*a\,up}(\la)\bigr)_{-\nu}=
\eta'\bigl(T_\la^{-1}\bigl(L^{low}(\la)\bigr)_{\nu}\bigr).
\endeq
The assertion follows from this and \refeq{Llow}.
\qed
\par
Hereafter crystal lattice  will mean the {\it upper} crystal lattice
at $q=0$ (we drop the superscript).
\par
A basic property of the VOs \refeq{TI:a}, \refeq{TI:c} is that they
preserve the crystal lattice \refto{DJO}.
Fixing the lowest weight vector $\ket{u_\la^*}\in V^{*a}(\la)$ such
that $\langle\ket{u_\la},\ket{u_{\la}^*}\rangle=1$
we normalize \refeq{TI:c} as
\eq
&\Phih_{V \mu}^{*\la}\bigl(v_\pm\otimes \ket{u^*_\mu}\bigr)
=\ket{u^*_{\la}}+\cdots.
\endeq
\par
\def\eps{\varepsilon}
\par
Let $\Phi_{\eps,n}$ be the weight components of
$\Phi=\Phih_\la^{\mu V}$, and similarly let
$\Phih_{V \mu}^{*\la}(v_{\pm}\otimes \cdot)=\sum_n \Phi^*_{\pm\, n}$,
$\Phi^*_{\pm\, n}:V^{*a}(\mu)_\nu \rightarrow
V^{*a}(\la)_{\nu\mp \alpha_1/2 +n\delta}$.
\par
\proclaim Proposition \prop{CryL}.
Notations being as above, we have for all $\eps$ and $n$
\item{(i)}
\eq
&\Phi_{\eps\,n} L(\la)\subset L(\mu), &(PCL:a)\cr
&\Phi^*_{\eps\,n} L^{*a}(\mu)\subset L^{*a}(\la). &(PCL:b)\cr
\endeq
\item{(ii)} For some $s>0$ we have
\eq
&\Phi_{\eps\,n} \bigl(L(\la)_{\la-\xi} \bigr) \subset
q^{-3n-2(\rho,\xi)-s} L(\mu) &(pwes:a)\cr
&\Phi^*_{\eps\,n} \bigl( L^{*a}(\mu)_{-\mu+\xi}\bigr) \subset
q^{-3n-2(\rho,\xi)-s} L^{*a}(\la) &(pwes:b)\cr
\endeq
\par
\Proof Assertion \refeq{PCL:a} is proved in \refto{DJO}. The argument in
\refto{DJO} shows also that \refeq{pwes:a} is true if it holds for
$\xi=0$. In Appendix 3 we verify the latter statement explicitly.

{}From the normalization of $\Phih_{V \mu}^{*\la}$ we have the relation
\eq
&\Phih_{V \mu}^{*\la}=(\Psi^*)^t,\quad
\Psi^*=(\id\otimes C_-)\Phi(q^2)\times (-q) ~\hbox{or}~1 \cr
\endeq
according as $\la=\La_0$ or $\La_1$.
We can verify further that
\eq
&(T_\mu^{-1}\otimes \id)\circ \Psi^*\circ T_\la
=\pm (\Phi_+\otimes v_-^* -\Phi_-\otimes v_+^*).
\endeq
where we put $\Phi=\Phi_+\otimes v_+ +\Phi_-\otimes v_-$.
In view of \refeq{La*} the assertions \refeq{PCL:b}, \refeq{pwes:b}
follow from these relations.
\qed
\par
Thanks to (ii),
$\Phih_\la^{\mu V}$
is well defined on the $q$-adic completion
\eq
&\lim_{\longleftarrow} \bigl(L(\mu)\otimes L\bigr)
/\bigl(q^n L(\mu)\otimes L\bigr),\cr
\endeq
i.e., when applied to an infinite sum $u=\sum_n u_n$
such that $u_n\in q^{M_n}L(\la)$, $\lim_{n\rightarrow \infty}M_n$
$=\infty$, then for any $N$ there are only a finite number of non-zero terms
modulo $q^N$ in $\Phih_\la^{\mu V}u$.
Hence we can iterate the VOs of type \refeq{TI:a} finitely many times
(we drop the symbols for completion)
\eq
&L(\la)~\buildrel \Phi_\la^{\mu V} \over \longrightarrow L(\mu)\otimes L
\over \longrightarrow L(\la)\otimes L\otimes L
\longrightarrow. \cdots &(EMBD)
\endeq
As mentioned already (\sec(2)) we conjecture that
after suitably normalizing VO the infinite iteration also makes sense.
\par
\def\eps{\varepsilon}
\def\o{\otimes}
\def\Rh{\widehat{R}}
\def\so#1#2{\raise 0pt\hbox{\vbox{\tabskip0pt\offinterlineskip
    \halign{\tabskip0pt plus 1em
     ##\tabskip0pt\cr
     $\,\scriptscriptstyle {#1}\,$\cr
     $\scriptscriptstyle\smile$\cr      %\smallsmile
     $#2$\cr}}}}
\par
\subsec(6.8| $n$-point functions)
To examine the behavior of the embedding \refeq{EMBD} we need to study
the $n$ point correlators as $n$ tends to $\infty$.
We give below what is known to us about the general $n$ point functions.
In this subsection we consider VOs of the
type $\Phi(z)=\Phi_\mu^{\la V}(z)$.
\par
Let $n=2m$ or $2m-1$,
$\la=\La_0$ and $\mu=\La_0$ or $\La_1$ according to whether $n$ is even or odd.
We define the vector $w_n(z)\in V^{\o n}$ by
\eq
&\langle u_\la| \Phi(z_1)\cdots \Phi(z_n) |u_\mu\rangle \cr
&\quad =\prod_{j=1}^nz_j^{(-1)^j /4} \prod_{j=1}^{m-1}(z_{2j-1}z_{2j})^{-m+j}
\prod_{i<j}{(q^6z_j/z_i)_\infty \over (q^4z_j/z_i)_\infty}
\times w_n(z_1,\cdots,z_n). &(npt1) \cr
\endeq
A priori $w_n(z)\prod_{j=1}^{m-1}(z_{2j-1}z_{2j})^{-m+j}$
is then a formal power series in
$z_n, z_n/z_{n-1}$, $\cdots$, $z_2/z_1$.
In general, let $\Phi:V(\mu)\rightarrow V(\la)\otimes W$ be an intertwiner
(where $W$ is a finite dimensional module)
and set $\Phi u_\la= u_\la\otimes w +\cdots$.
Then $w$ must satisfy \refto{DJO} $e_i^{\br{\la,h_i}+1}w=0$ for all $i$.
In the present case $W=V_{z_1}\otimes\cdots \otimes V_{z_n}$, $\la=\La_0$.
Because $V$ has a perfect crystal, so does $V^{\otimes n}$ \refto{(KMN)^2}.
This implies that up to scalar the linear equations
\eq
&\Delta(e_0)^2 w(z)=0,\quad \Delta(e_1)w(z)=0.
\endeq
admit at most one solution
(recall that the first equation involves the indeterminates $z_i$),
and hence that our $w(z)$ is a rational function in $z_1,\cdots, z_n$ up to
an overall scalar factor.
\par
Set $\Rh(z)=P{\bar R}(z)\times (1-q^2z)/(z-q^2)$ with ${\bar R}(z)$ defined in
\refeq{RB}. Then we have
the following properties of $w_n(z)$:
\eq
&\Rh_{i\,i+1}(z_i/z_{i+1})w_n(z)=(s_iw_n)(z)
\qquad i=1,\cdots,n-1 &(NPT:a)
\endeq
where $s_i=(i\,i+1)$,
and we set
$(sf)(z_1,\cdots,z_n)=f(z_{s^{-1}(1)},\cdots,z_{s^{-1}(n)})$
for a permutation $s$.
\eq
&w_n(q^6z_1,z_2,\cdots,z_n)=
{\bar A}_1q^{3m-3}P_{1n}\cdots P_{13}P_{12}w_n(z_2,\cdots,z_n,z_1)
&(NPT:b)
\endeq
where ${\bar A}_1=-t_1^{-1}$ for $n=2m$, $=(qt_1^{-1})^{3/2}$ for $n=2m-1$.
\eq
w_{2m}(z_1,\cdots,z_{2m-1},0)=&w_{2m-1}(z_1,\cdots,z_{2m-1})\o v_{-}
-q (f_1.w_{2m-1}(z_1,\cdots,z_{2m-1}))\o v_{+}, \cr
w_{2m-1}(z_1,\cdots,z_{2m-2},0)=&w_{2m-2}(z_1,\cdots,z_{2m-2})\o v_{+}
\times \prod_{j=1}^{2m-2}z_j.\cr
&&(NPT:c)
\endeq
Property \refeq{NPT:a} is a consequence of the commutation relations of VO.
Property \refeq{NPT:b} is a reduced form of the $q$KZ equation under
\refeq{NPT:a}.
Finally property \refeq{NPT:c} follows from the normalization of the VOs
\refeq{VN:a}.
\par
Let us introduce the coefficients $a_n(\eps|z)$ by
\eq
&w_n(z_1,\cdots,z_n)=\sum_\eps a_n(\eps|z)v_{\eps_1}\o\cdots\o v_{\eps_n}.
\endeq
The $\eps=(\eps_1,\cdots,\eps_n)$ range over $\eps_i=\pm$ such
that the number of $+$'s equals $m$.
Then \refeq{NPT:a} is rewritten in the form
\eq
a(\cdots \so{i}{-} {+}\cdots|z)&
=\sigma_i a(\cdots \so{i}{+} {-} \cdots|z) &(brd)\cr
a(\cdots \so{i}{\eps} \eps \cdots|z)
&=q\sigma_i a(\cdots \so{i}{\eps} \eps \cdots|z).\cr
\endeq
Here the operator $\sigma_i$ is given by
\eq
&(\sigma_i^{\pm 1}f)(z)
=(q^{-1}z_i-q z_{i+1}){f(z)-(s_if)(z) \over z_i-z_{i+1}} - q^{\pm 1}f(z).
\endeq
One can verify that they obey the Hecke algebra relations
(Lusztig \refto{Lus})
\eq
&\sigma_i\sigma_{i+1}\sigma_i=\sigma_{i+1}\sigma_i\sigma_{i+1},\quad
\sigma_i\sigma_j=\sigma_j\sigma_i~(|i-j|>1), \quad
(\sigma_i+q)(\sigma_i-q^{-1})=0.
\endeq
Using \refeq{brd} the coefficients $a(\eps|z)$ are determined uniquely
once we know e.g. $a_0(z)=a(\eps|z)$ for $\eps_1=\cdots=\eps_m=+$,
$\eps_{m+1}=\cdots=\eps_n=-$.
{}From computations for small $n$ we expect
that $w(z_1,\cdots,z_n)$ is actually a polynomial in $z_i$ and $q$,
and that $a_0(z)$ is given by
\eq
(-q)^{m(m-1)/2}\prod_{1\le i< j\le m}(z_i-q^2z_j)
\prod_{m+1\le i< j\le n}(z_i-q^2z_j)
&\times 1 \quad \hbox{ for }n=2m, \cr
&\times z_{m+1}\cdots z_{2m-1} \quad \hbox{ for }n=2m-1. \cr
&&(npt2)
\endeq
The representation of the Hecke algebra generated by this polynomial is
irreducible for generic $q$; at $q=1$ it specializes to the
Specht module \refto{JK} of the symmetric group
associated with the Young diagram of shape $(m,m)$ or $(m,m-1)$.
We hope to be able to extract
exact information about the asymptotics of $n$ point functions
by analysing these representations.
\par
%\endinput
\par
\par
%\input sec7
%
%
%		Section 7
%
%
%
\def\Vs{{\cal V}}
\def\Vsr{{\cal V}^r}
\par
\def\dz#1{{dz_{#1} \over 2\pi i z_{#1}}}
\par
\beginsection \S7. Space of states
\par
\subsec(7.1|Vacuum, shift and energy)
%embedding into $V^{\otimes \infty}$,
%shift, Hamiltonian, CTM
%right modules, $\Vs$, $\Vsr$, duality, inner product, vacuum vectors
In \sec(1) through \sec(5) we have emphasized the role of the space
$V(\la)\otimes V(\la')^{*a}$
where $\la,\la'\in \{\La_0,\La_1\}$.
To treat the infinite sums of vectors safely,
we introduce its $q$-adic completion. Define
\eq
&\eL_{\la,\la'}=\lim_{\longleftarrow}
L(\la)\otimes L^{*a}(\la')/q^n L(\la)\otimes L^{*a}(\la'),\cr
&\Vs_{\la,\la'}=\Q((q))\otimes_{\Q[[q]]} \eL_{\la,\la'}. \quad
\endeq
Here $\Q((q))$ (resp. $\Q[[q]]$) denotes the field (resp. ring) of
formal Laurent (resp. power) series in $q$.
The space $\Vs_{\la,\la'}$ inherits the left $U$-module structure of
level $0$. Let $\Vsr_{\la,\la'}$ denote the right $U$-module
structure on $\Vs_{\la,\la'}$ given by $f.x=a^{-1}(x).f$
($f\in \Vs_{\la,\la'}, x \in U$).
If we regard an element $f\in \Vs_{\la',\la}$
(resp. $g\in \Vsr_{\la,\la'}$) as a
linear map $V(\la)\rightarrow \Vh(\la')$
(resp. $V^r(\la)\rightarrow \Vh^r(\la')$), the left
(resp. right) $U$-action is given by the adjoint action:
\eq
&ad(x).f=\sum x_{(1)}\circ f\circ a(x_{(2)}), \cr
&g.ad^r(x)=\sum a^{-1}(x_{(2)})\circ g\circ x_{(1)},\cr
\endeq
where $x\in U$.
We define the bilinear pairing
$\Vsr_{\la,\la'}\times \Vs_{\la',\la}\longrightarrow \Q((q))$
by
\eq
&\brak{g}{f}={\tr_{V(\la)}\bigl(q^{-4\rho}g\circ f\bigr)
\over \tr_{V(\la)}(q^{-4\rho})}.
&(CVF)\cr
\cr
\endeq
It is non-degenerate and enjoys the property
\eq
&\brak{g.ad^r(x)}{f}=\brak{g}{ad(x).f}\quad \forall x\in U. &(Cvt)
\endeq
Note that the denominator in \refeq{CVF}
\eq
&\tr_{V(\la)}(q^{-4\rho})={q^{-(4\rho,\la)} \over (q^2;q^4)}
\endeq
is the principally specialized character of $V(\la)$.
\par
For each $\nu$ we take bases $\{\ket{u^{(\nu)}_i}\}\subset L(\la)_\nu$,
$\{\ket{u^{(\nu)*}_i}\}\subset \bigl(L(\la)_\nu\bigr)^*$
which are dual with respect to the canonical pairing $\langle~,~\rangle$.
We call the canonical element
\eq
&{\vac}_\la=\sum_{i,\nu}\ket{u^{(\nu)}_i}\otimes
\ket{u^{(\nu)*}_i}~ \in \Vs_{\la,\la} &(vacm)
\endeq
the vacuum.
This corresponds to the identity map $V(\la)\rightarrow \Vh(\la)$.
Hence it is clear that the vacuum gives rise to
the trivial representation over $U$:
\eq
&ad(x).\vac_\la=\ep(x)\vac_\la \qquad \forall x \in U.
\endeq
Since
$
\br{\ket{u^{(\nu)*}_j},T_\la \ket{u^{(\nu)}_i}}\in q^{(2\rho,\la-\nu)}A
$
we see from \refeq{La*} that ${\vac}_\la$ in fact belongs to $\Vs_{\la,\la}$.
Define the right vacuum ${}_\la\!\dvac\in \Vsr_{\la,\la}$ analogously.
The pairing \refeq{CVF} is so normalized that
${}_\la\!\brak{\hbox{vac}}{\hbox{vac}}_\la=1$.
\par
Consider the composition of the maps
\eq
&L(\la)\otimes L^{*a}(\la')
\Lh(\mu)\otimes L \otimes L^{*a}(\la')
\Lh(\mu)\otimes \Lh^{*a}(\mu').
\endeq
By Proposition \refprop{CryL} it extends to the map
$\widehat{T}:\Vs_{\la,\la'}\rightarrow \Vs_{\mu,\mu'}$.
Now let $\la'=\la$ and $\mu'=\mu$.
We define the translation and energy operators $T$, $H$ by
\eq
&T=-g_\mu \widehat{T} &(Trs)\cr
&H=-(q-q^{-1})(T\circ d\circ T^{-1}-d) &(Ham)\cr
\endeq
where $d$ is given in \refeq{D}.
\par
The following is the first property we expect for the vacuum vector:
\par
\proclaim Proposition \prop{Vac}.
The vacuum vector is invariant under $T$, $H$:
\eq
&T{\vac}_\la={\vac}_\mu,\quad H{\vac}_\la=0.
\endeq
%\endproclaim
\par
\Proof The second equation is clear from the first and $d\vac_\la=0$.
Put $\Phih_{\la}^{\mu V}=\sum \Phi_\ep\otimes v_\ep$,
$\bigl(\Phih_{V \la}^{*\mu}\bigr)^t=\Phih_\mu^{\la V^{*a^{-1}}}
=\sum \Phi^*_\ep\otimes v^*_\ep$.
It suffices to show that
\eq
&-g_\mu\widehat{T}\vac_\la= -g_\mu
\sum \Phi_\ep \ket{u^{(\nu)}_i}\otimes (\Phi^*_\ep)^t \ket{u^{(\nu)*}_i}
\endeq
coincides with $\vac_\mu$ (note that $(\Phi^*_\ep)^t$ sends $V(\la)^*$
to $V(\mu)^*$).
This is  equivalent to showing
\eq
\sum \Phi_\ep \ket{u^{(\nu)}_i} \br{(\Phi^*_\ep)^t \ket{u^{(\nu)*}_i},\ket{v}}
&=-(g_\mu)^{-1}\ket{v}
\endeq
for any $\ket{v}\in V(\mu)$.
Since the left hand side is
\eq
&=\sum_\ep\Phi_\ep(\ket{u^{(\nu)}_i} \br{\ket{u^{(\nu)*}_i},\Phi^*_\ep \ket{v}}
\cr
&=\sum_\ep\Phi_\ep\Phi^*_\ep \ket{v},
\endeq
the assertion follows from \refeq{INV}.
\qed
\par
\subsec(7.2|Creation and annihilation operators)
%creation-annihilation, commutation relations including delta-function terms
%energy, momentum of cr-ann.
Let us define the components of the VOs of type \refeq{TII:a},
\refeq{TII:b} by
\eq
&\Phi_\la^{V \mu}(z)=v_+\otimes \Phi_+(z) +v_-\otimes \Phi_-(z),\cr
&\Phi^*_{\pm}(z)=\Phi_{V\la}^{\mu}(z)\bigl(v_\pm\o (\cdot)\bigr)
\endeq
Hence from \refeq{Red:b} we have the relations
\eq
&\Phi^*_+(z)=-q^{-1}\Phi_-(zq^2),\quad
\Phi^*_-(z)=\Phi_+(zq^2).
\endeq
Using these operators, we would like to define the $n$-particle states
to be
\eq
&\bigl(\Phi^*_{\ep_1}(z_1)\cdots \Phi^*_{\ep_n}(z_n)\otimes \id \bigr)
\vac_\la &(par)
\endeq
with $\ep_1,\cdots,\ep_n \in \{+,-\}$ and $|z_1|=\cdots=|z_n|=1$.
The $n$-particle states in the dual space are to be defined analogously, using
${}_\la\!\dvac$ and $\Phi_{\ep}(z)$ in place of
$\vac_\la$ and $\Phi_{\ep}^*(z)$ respectively.
These are eigenstates of the shift and the energy operators
(see \refprop{EgMt}).
\par
Let us examine the meaning of \refeq{par} more closely by studying the
Fourier components
\eq
\oint\cdots&\oint_{|z_1|=\cdots=|z_n|=1} \dz{1}\cdots \dz{n}
 z_1^{m_1}\cdots z_n^{m_n} \cr
&\times \bigl(\Phi^*_{\ep_1}(z_1)\cdots \Phi^*_{\ep_n}(z_n)\otimes \id \bigr)
\vac_\la, &(npar)\cr
\endeq
where the $m_i$ range over all integers.
For definiteness we take $n=2$.
Eq. \refeq{npar} is an infinite sum over $i,\nu$ of the terms
\eq
&\oint\oint_{|z_1|=|z_2|=1} \dz{1} \dz{2}
z_1^{m_1}z_2^{m_2}
\Bigl(\Phi^*_{\ep_1}(z_1)\Phi^*_{\ep_2}(z_2)\ket{u_i^{(\nu)}}\Bigr)
\otimes \ket{u_i^{(\nu)*}}. &(2par)\cr
\endeq
Now for each $\bra{u}\in V^r(\la)$,
$\ket{u'}\in V(\la)$ the function
$\bra{u}\Phi^*_{\ep_1}(z_1)\Phi^*_{\ep_2}(z_2)\ket{u'}$
is convergent in the domain $|z_1|\gg|z_2|$, $|q|<1$,
and has a meromorphic continuation with respect to
$z_1,z_2\in (\C)^\times$ with poles only
at $z_1/z_2=q^{-2},q^2,q^6,\cdots$
(see Proposition \refprop{COM}).
Hence the integration in \refeq{2par} is meaningful.
Notice that \refeq{2par} is
different from applying the weight components $\Phi_{\ep,m}$
of $\Phi_\ep(z)$
\eq
&\Bigl(\Phi_{\ep_1,m_1}\Phi_{\ep_2,m_2}\ket{u_i^{(\nu)}}\Bigr)
\otimes \ket{u_i^{(\nu)*}}.
\endeq
By the definition the latter is obtained by taking the contour in \refeq{2par}
to be $|z_1|\gg|z_2|$,
and because of the pole at $|z_1/z_2|=q^{-2}$ the two expressions give
different answers.
\par
The type II VOs do not preserve the crystal lattice, so the individual terms
\refeq{2par} comprise negative powers of $q$.
However computations suggest (see Appendix 4)
that when we sum over $i,\nu$  the negative powers all disappear, getting
thereby a well defined element of $\Vs_{\la,\la}$.
More generally it can be shown using the
$q$-KZ equation that each matrix element
$\bra{u}\Phi^*_{\ep_1}(z_1)\cdots \Phi^*_{\ep_n}(z_n)\ket{u'}$,
$\bra{u}\Phi_{\ep_1}(z_1)\cdots \Phi_{\ep_n}(z_n)\ket{u'}$
($\bra{u}\in V^r(\la)$, $\ket{u'}\in V(\la)$)
is holomorphic on $|z_1|=\cdots=|z_n|=1$.
Fixing $n$ let $\F(n)$,  $\F^r(n)$ denote the span
of the vectors \refeq{npar} and its dual counterpart, respectively.
We expect that
\medskip
\item{(1)} $\F(n)\subset \Vs$,\quad $\F^r(n)\subset \Vsr$.
\item{(2)} With respect to the inner product \refeq{CVF} the spaces
$\F^r(m)$ and $\F(n)$ with $m\neq n$ are orthogonal.
\item{(3)} \refeq{CVF} restricted to $\F^r(n)\times \F(n)$ is non-degenerate.
\medskip
\noindent
Assuming these, we define the physical spaces of states by
\eq
&\F=\oplus_{n=0}^\infty \F(n), \quad
\F^r=\oplus_{n=0}^\infty \F^r(n).
\endeq
They are analogous to the usual Fock space of free particles.
\par
Define the creation operator $\varphi^*_\ep(z):\F(n)\rightarrow \F(n+1)$
on $\F$, $|z|=1$, by
\eq
&\varphi^*_\ep(z)
\bigl(\Phi^*_{\ep_1}(z_1)\cdots \Phi^*_{\ep_n}(z_n)\otimes\id\bigr)\vac_\la \cr
&=\bigl(\Phi_\ep(z)\Phi^*_{\ep_1}(z_1)\cdots \Phi^*_{\ep_n}(z_n)\otimes\id
\bigr)\vac_\la.
&(cre)\cr
\endeq
Both sides are to be understood in the sense of the Fourier coefficients
as in \refeq{npar}.
The annihilation operator $\varphi(z):\F(n)\rightarrow \F(n-1)$ is defined
to be the adjoint of the dual counterpart
$\F^r(n)\rightarrow \F^r(n+1)$ of \refeq{cre}.
Thus $\varphi_\ep(z)\vac=0$ by definition, and
\eq
&{}_\la\!\dvac \varphi_{\ep_1}(z_1)\cdots\varphi_{\ep_n}(z_n)
\varphi^*_{\ep'_n}(z_n')\cdots\varphi^*_{\ep'_1}(z_1')\vac_\la \cr
&=\tr_{V(\la)}\Bigl(q^{-4\rho}\Phi_{\ep_1}(z_1)\cdots\Phi_{\ep_n}(z_n)
\Phi^*_{\ep'_n}(z_n')\cdots\Phi^*_{\ep'_1}(z_1')\Bigl)
/\tr_{V(\la)}\Bigl(q^{-4\rho}\Bigr).
\endeq
\par
Because $\varphi_\pm(z)$ act as identity on the second component of
$L(\la)\otimes L^{*a}(\la')$
the commutation relations \refeq{COM:a-COM:c} as operators on $V(\la)$
can be readily translated.
We have
\eq
\varphi_{\ep_1}(z_1)\varphi_{\ep_2}(z_2)
&=\sum \varphi_{\ep'_2}(z_2)\varphi_{\ep'_1}(z_1)
\bigl(R_{VV}(z_1/z_2)\bigr)_{\ep'_1\ep'_2}^{\ep_1\ep_2},&(CA:a)\cr
\varphi^*_{\ep_1}(z_1)\varphi^*_{\ep_2}(z_2)
&=\sum \varphi^*_{\ep'_2}(z_2)\varphi^*_{\ep'_1}(z_1)
\bigl(R_{V^*V^*}(z_1/z_2)\bigr)_{\ep'_1\ep'_2}^{\ep_1\ep_2},&(CA:b)\cr
\varphi_{\ep_1}(z_1)\varphi^*_{\ep_2}(z_2)
&=\sum \varphi^*_{\ep'_2}(z_2)\varphi_{\ep'_1}(z_1)
\bigl(R_{V V^*}(z_1/z_2)\bigr)_{\ep'_1\ep'_2}^{\ep_1\ep_2}
+g_\la\delta_{\ep_1\ep_2}\delta(z_1/z_2). &(CA:c)\cr
\endeq
Here the delta function $\delta(z)=\sum_{n\in \Z} z^n$ arises because
of the pole of \refeq{COM:c} at $z_1=z_2$.
Thus the VOs provide us with a
lattice realization of the Zamolodchikov algebra.
\par
\Remark In the limit $q \rightarrow 1$, $z=q^{-2\beta/i\pi}$,
the $R$ matrix \refeq{Rmat},
\refeq{qGam} reduces to the $S$ matrix of the $su(2)$ invariant
Thirring model \refto{KirSm}
\eq
&R_{VV}(\beta)=
{\Ga \bigl({1\over 2}+{\beta\over 2\pi i}\bigr)
 \Ga \bigl(-{\beta\over 2\pi i}\bigr)
\over
\Ga \bigl({1\over 2}-{\beta\over 2\pi i}\bigr)
\Ga \bigl({\beta\over 2\pi i}\bigr)}
{\bigl(\beta\cdot 1 - \pi i P\bigr) \over \beta-\pi i }.
\endeq
The commutation relations
\refeq{CA:a}-\refeq{CA:c} become those for the Zamolodchikov operators
(up to rescaling the operators $\varphi_\ep(z),\varphi^*_\ep(z)$).
\medskip
\par
Having set up the mathematical definitions of the creation-annihilation
operators we can discuss their transformation properties under
the shift and the energy operators.
\par
\proclaim Proposition \prop{EgMt}.
\eq
&T\varphi_\pm(z) T^{-1}= \tau(z) \varphi_\pm(z) ,\quad
T\varphi^*_\pm(z) T^{-1}= -\tau(z)^{-1} \varphi^*_\pm(z) , &(Mtm)\cr
&[H,\varphi_\pm(z)]=-\epsilon(z)\varphi_\pm(z),\quad
[H,\varphi^*_\pm(z)]=\epsilon(z)\varphi^*_\pm(z),&(Egy)\cr
\endeq
where
\eq
&\tau(z)=z^{-1/2}{\Th(qz)\over\Th(q z^{-1})}\quad
\epsilon(z)=-(q-q^{-1}) z{d\over dz}\log~ \tau(z). \cr
\endeq
%\endproclaim
\par
\Proof
We show \refeq{Mtm}, \refeq{Egy} for $\varphi_\pm(z)$.
To see \refeq{Mtm} we need to prove
$
\Phi_\mu^{V\la}(z)T=\tau(z)^{-1}T \Phi_\la^{V\mu}(z)
$
, where the left and right hand sides are the compositions of
\eq
&L(\la)\otimes L^{*a}(\la) \goto{} L(\mu)\otimes L\otimes L^{*a}(\la)
\goto{} L(\mu)\otimes L^{*a}(\mu) \goto{} L\otimes L(\la)\otimes L^{*a}(\mu)
\endeq
and
\eq
&L(\la)\otimes L^{*a}(\la) \goto{} L\otimes L(\mu)\otimes L^{*a}(\la)
\goto{} L\otimes L(\la)\otimes L\otimes L^{*a}(\la)
\goto{} L\otimes L(\la)\otimes L^{*a}(\mu)
\endeq
respectively.
This follows from \refeq{TI_II} by
multiplying $\Phi_{V_2\la^*}^{\mu^*}(z_2)\otimes \id$ from the left
and setting $z_2=1$.
%For $\varphi^*_\pm(z)$ we need only to
%change $z$ to $zq^2$ and use $\tau(q^2z)=-\tau(z)^{-1}$.
\par
Using \refeq{Mtm} we calculate
\eq
T\circ d\circ T^{-1}\varphi_\pm(z)
&=\tau(z)^{-1}T\circ \bigl([d,\varphi_\pm(z)]+\varphi_\pm(z)d\bigr)\circ
T^{-1} \cr
&=-\tau(z)^{-1}T\circ(z{d\over dz}-\Delta_\mu+\Delta_\la)\varphi_\pm(z))
\circ T^{-1}+\varphi_\pm(z) T\circ d\circ T^{-1}.
\endeq
The first term can be written as
\eq
&-\tau(z)^{-1}(z{d\over dz}-\Delta_\mu+\Delta_\la)
(T\circ\varphi_\pm(z)\circ T^{-1})\cr
&\quad =-\bigl(z{d\over dz}\log\tau(z)\bigr)\varphi_\pm(z)
-(z{d\over dz}-\Delta_\mu+\Delta_\la)\varphi_\pm(z).
\endeq
The relations \refeq{Egy} follow from these. \qed
\par
Let us compare them with the formulas for the energy and momentum
of `spin waves' obtained in refs. \refto{dCG,Bab} via the Bethe Ansatz method.
The result is given in terms of elliptic
functions of nome $-q=e^{-\gamma}$ as (\refto{Bab}, eqs.(26),(27))
\eq
&\varepsilon(\theta)={2K\over \pi}
\sinh\gamma \,\hbox{dn}\left({2K\over \pi}\theta,k\right),
\quad
p(\theta)=\hbox{am}\left({2K\over \pi}\theta,k\right)-{1\over 2}\pi.\cr
\endeq
Identifying $z=-e^{2i\theta}$ and using the identity (\refto{WW}, p.509)
\eq
&e^{-i p(\theta)}=z^{-1/2}{\Th(qz)\over \Th(qz^{-1})}, \quad
z{d\over dz}\log e^{-ip(\theta)} =-{\varepsilon(\theta) \over q-q^{-1}}
\endeq
we see that our \refeq{Mtm}, \refeq{Egy} are precisely the same.
This gives a strong evidence in favor of our mathematical framework of the
particle picture.
\par

\def\tr{\hbox{tr}}
\def\o{\otimes}
\par
%\subsec(7.3|Local operators)
%local operators
%one point functions
%Let us briefly indicate here how to define local operators in our framework.
%Let $\O\in\End(V)$ be an operator acting on $V$, e.g. $\O=\sigma^z$.
%Then the counterpart of $\O_1$ acting on the site $1$
%in the infinite tensor space is defined to be the composition of the
%following:
%\eq
%&L(\la)\o L^{*a}(\la) \goto{}
%\Lh(\mu)\o L\o L^{*a}(\la)
%\goto{{\scriptstyle id}\o{\scriptstyle \O}\o{\scriptstyle id}}
%\Lh(\mu)\o L\o L^{*a}(\la) \goto{}
%\Lh(\la)\o L^{*a}(\la).
%\endeq
%Here the first (resp. last) arrow is given by the vertex operator
%$\Phi_\la^{\mu V}\o\id$
%(resp. $-g_\la \Phi_{\mu V}^\la \o\id$ with $g_\la$ as in \refeq{INV}).
%For example the one point function of $\O_1$ is
%\eq
%&\dvac \O_1 \vac =
%{
%\tr_{V(\la)}
%\bigl(q^{-4\rho}(-g_\la)\Phi_{\mu V}^\la(\id\o\O)\Phi_\la^{\mu V}\bigr)
%\over
%\tr_{V(\la)}\bigl(q^{-4\rho}\bigr)
%}.
%\endeq
%More general correlations can be defined similarly.
%We hope to investigate such quantities in a future publication.
\par
%\endinput
\par
\def\slth{\widehat{\goth{sl}}(2)\hskip 1pt}
\def\u{U_q\bigl(\slth \bigr)}
\def\sk{\smallskip\noindent}
\def\pn{\par\noindent}
\par
\beginsection \S8. Discussions
\par
Let us summarize the content of this paper.
\sk
1.\quad We conjectured that the embedding \refto{FM}
of the irreducible highest weight $\u$-module $V(\La_0)$ into the semi-infinite
tensor product, is given by the limit of $n$-point
correlation functions of the $\u$ vertex operators as
$n\rightarrow\infty$. We do not know the exact
form of the correct normalization of the vertex operator.
\sk
2.\quad We postulated that the mathematical content of the
infinite tensor product on which the XXZ Hamiltonian
(after a suitable renormalization) is acting, is
\eq
&{\rm End}{}_\C\bigl(V(\La_0)\oplus V(\La_1)\bigr).\cr
\endeq
For simplicity, we mainly used the half of this, i.e.,
${\rm Hom}{}_\C\bigl(V(\La_0),V(\La_0)\oplus V(\La_1)\bigr)$.
\sk
3.\quad We made the dictionary between the physical and
mathematical pictures, which contains:
\item{(a)}
The translation and energy operators \refeq{Trs}, \refeq{Ham}.
\item{(b)}
The inner product of states \refeq{CVF}.
\item{(c)}
The vacuum states \refeq{vacm}.
\item{(d)}
The creation and annihilation operators \refeq{cre}.
\sk
4.\quad By utilising the $q$-deformed
Knizhnik-Zamolodchikov equation, we got the following.
\item{(e)}
The energy and momentum of the creation and annihilation operators
(Proposition \refprop{EgMt}).
\item{(f)}
The commutation relations of the creation and annihilation
operators among themselves, i.e., the Zamolodchikov algebra
\refeq{CA:a}-\refeq{CA:c}.
\item{(g)}
A conjectural formula for the $n$-point correlation
functions of the $q$-deformed vertex operators for
arbitrary $n$ \refeq{npt1}, \refeq{npt2}.
(This is not the same with the $n$-point
correlation functions of the creation and annihilation operators.)
\sk
5.\quad We checked the validity of our picture on the
following points.
\item{(h)}
A one-line proof of the character expression for the
one-point function (in the sense of \refto{DJMO}) of the
six-vertex model (\sec(5)).
\item{(i)}
Comparison of the energy and momentum of particles with the
result obtained by the Bethe Ansatz (\sec(6.7)).
\item{(j)}
Comparison of the vacuum and two-particle states with the
Bethe Ansatz eigenvectors in the anisotropic limit $\Delta=-\infty$
(Appendix 5).
\item{(k)}
Comparison of the commutation relations of the creation and
annihilation operators with the $S$-matrix of the
$su(2)$-Thirring model (\sec(7.2)).
\sk
6.\quad The following is a list of unsolved problems (besides
the several conjectures stated in the main body of the paper).
\pn
\sk
\item{$\bullet$}For the XXZ model (and the six-vertex model),
we want to know about the form factors
of the local operators, the staggered polarization \refto{Bax}
and other correlation functions.
\pn
\sk
Other models, for which a similar diagonalization scheme might apply,
are, especially:
\pn
\item{$\bullet$}The vertex models with perfect crystals.
The framework of the present paper admits immediate extension to
the general case.
Reshetikhin \refto{Resh} proposes that the space of
$n$-particle states in the
generalized $\XXX$ model with higher spin
is a tensor product of two factors, $(\C^2)^{\otimes n}$ and the space
of paths of length $n$ of an $RSOS$ model.
It would be an interesting problem to examine this picture from our viewpoint.
%Conceptually,
%nothing new would be required, but the actual computation
%is a different story.
\item{$\bullet$}The XYZ model, the hard hexagon model \refto{Baxbk},
Kashiwara-Miwa's model \refto{KasMi, HY},
the chiral Potts model \refto{CPot1, CPot2}.
These are massive models, each
within a different category. No obvious extension of
our scheme to any one of them is known.
\par
\bigskip\noindent
{\it Acknowledgement.}\quad
We would like to thank F. Smirnov, from whom
we learned a great deal during his stay at RIMS.
We thank T. Nakashima for checking the result in Table 1.
\par
We wish to acknowledge also discussions with
H. Araki,
M. Barber,
M. Batchelor,
I. Cherednik,
E. Date,
B. Feigin,
M. Kashiwara,
S. Kato,
A. Kirillov,
P. Kulish,
A. LeClair,
A. Matsuo,
K. Miki,
K. Mimachi,
M. Okado,
J. Perk
E. Sklyanin,
and
A. Tsuchiya.
\par
One of us (T.M.) thanks his colleagues in TIFR, Bombay, where he
stayed in February and March, 1992, in particular, E. Arbarello,
P.P. Divakaran, P. Mitter, A. Raina, T.R. Ramadas and  D.N. Verma.
%\endinput
\par
\par
\par
\par
\numberby{\beginsection}\prefixby{A}
\def\Ut{U_q\bigl(\widehat{\goth{sl}}(2)\bigr)}
\def\Ug{U_q(\goth{g})}
\def\V#1{V(\L_#1)}
\def\L{\Lambda}
\def\l{\lambda}
\def\e{\varepsilon}
\def\p{\varphi}
\def\N{{\bf N}}
\def\Gu{G^{\scriptstyle up}}
\def\Gl{G^{\scriptstyle low}}
\def\tf{\tilde f}
\def\te{\tilde e}

\def\bl{\bullet}
\par
\beginsection Appendix 1. Action of $\Ut$ on $\V0$
\par
We will recall some basic properties of the global
base of the irreducible highest weight $\Ug$-module
$V(\l)$, which are given in \refto{Ka,La}.
Then, we will compute the actions of the Chevalley generators on some vectors
of the upper global base in the case
$\goth{g}=\widehat{\goth{sl}}(2)$
and
$\l=\L_0$.
\par
Let
$B(\l)$
be the crystal for
$V(\l)$.
(See Section 2 of \refto{(KMN)^2}, especially
Definition 2.2.3 and Theorem 2.4.1.)
We use the maps
$\e_i$, $\p_i:B(\l)\longrightarrow\N$.
(See Definition 2.2.3 and (2.2.15) in \refto{(KMN)^2}.)
Set
$l_i(b)=\e_i(b)+\p_i(b)$.
This is the length of the
$\goth{sl}(2)$-string of color $i$ through $b$.
For
$b\in B(\l)$ we denote by
$\Gu(b)\in V(\l)$ the upper global base vector corresponding to
$b$,
and by
$\Gl(b)\in V(\l)$ the lower global base vector corresponding to
$b$.
(In \refto{Ka2},
$\Gu$ is written as $G$ (see \sec(5)) and $\Gl$
is written as $G_\l$ (see \sec(4)).)
\par
We use the following properties.
\par
\pn(1) Duality.
\par
Let $(\ ,\ )$ be the symmetric bilinear form on
$V(\l)$ given in \sec(6.7), \refeq{varphi}, \refeq{inner}.
We have
\eq
&\bigl(\Gu(b),\Gl(b')\bigr)=\delta_{bb'}.&(uplow)\cr
\endeq
Therefore, if
\eq
&e_i\Gu(b)=\sum_{b'}c_i(b',b)\Gu(b'),\quad
f_i\Gu(b)=\sum_{b'}c'_i(b',b)\Gu(b'),\cr
\endeq
then we have
\eq
&e_i\Gl(b)=\sum_{b'}c'_i(b,b')\Gl(b'),\quad
f_i\Gl(b)=\sum_{b'}c_i(b,b')\Gl(b').\cr
\endeq
\par
\pn(2) Crystalline action.
\par
We have
\eq
&e_i\Gu(b)=[\e_i(b)]_i\Gu(\te_ib)+\sum_{b'}E^i_{bb'}G(b'),\cr
&f_i\Gu(b)=[\p_i(b)]_i\Gu(\tf_ib)+\sum_{b'}F^i_{bb'}G(b'),\cr
\endeq
where
$E^i_{b'b}$ and $F^i_{b'b}$ are zero if
$l_i(b')\ge l_i(b)$. We also have
\eq
&\lim_{q\rightarrow0}{E^i_{b'b}\over[\e_i(b)]_i}=
\lim_{q\rightarrow0}{F^i_{b'b}\over[\p_i(b)]_i}=0.\cr
\endeq
\par
\pn(3) Positivity.
\par
We have $E^i_{b'b}$, $F^i_{b'b}\in\oplus_{n=0}^\infty\Z_{\ge0}[n]$,
where $[n]$ means $(q^n-q^{-n})/(q-q^{-1})$.
\medskip
(1) is implicitly given in \refto{Ka}.
(2) follows from Proposition 5.3.1 and (5.3.8--10) in \refto{Ka}.
(3) is proved in \refto{La}(Theorem 11.5).
\par
Now we restrict to the case
$\goth{g}=\widehat{\goth{sl}}(2)$ and $\l=\L_0$.
The crystal $B(\L_0)$
is identified with the set of paths:
$p=\{p(k)\}_{k\ge1}$
is called a path if and only if
$p(k)=(+)$ or $(-)$, and
\eq
p(k)&=(+)\qbox{if $k$ is even and $k\gg 0$,}\cr
&=(-)\qbox{if $k$ is odd and $k\gg 0$.}\cr
\endeq
For a path $p$,
we define its signature
\eq
&l=\pmatrix{l_1\cr\vdots\cr l_m\cr}\cr
\endeq
in such a way that
\eq
&l_1>\cdots>l_m>0,\cr
&p(k+1)=p(k)\qbox{if and only if $k\in\{l_1,\ldots,l_m\}$.}\cr
\endeq
We call $m$ the depth of the corresponding path.
\par
\Example
If $p=\bigl(\cdots p(4)\ p(3)\ p(2)\ p(1)\bigr)=
(\cdots+\ -\ -\ +\ +\ +)$, then $l=\pmatrix{4\cr2\cr1\cr}$.
The depth of $p$ is 3.
\medskip
\par
In Tables 1 and 2 below, if $l$ is the signature of $p$,
we will write $l$ to represent $\Gu(p)$.
It is quite convenient to associate a parity symbol
\eq
&s=\pmatrix{s_1\cr\vdots\cr s_m\cr}\cr
\endeq
with a signature $l$ in such a way that
\eq
s_k&=\onebox()\qbox{if $l_k\equiv k\bmod 2$,}\cr
&=\onebox(\bl)\qbox{if $l_k\not\equiv k\bmod 2$.}\cr
\endeq
Namely, we represent $l=\pmatrix{4\cr2\cr1\cr}$ as
\eq
&\pmatrix{4\cr2\cr1\cr}\threebox(\bl,{},{}).\cr
\endeq
In Table 1 we list the actions of
$e_i$ and $f_i (i=0,1)$ on the upper global base labeled by $l$ such that
$m\le3$.
The symbol $\phi$ means the highest weight vector. We abbreviate
signatures by the differences to the signature of the operand.
For example,
\eq
&f_1\onebox({})=[2](1)\onebox(\bl)+\pmatrix{0\cr1\cr}\twobox({},\bl)\cr
\endeq
really means
\eq
f_1(l_1)\onebox({})=[2](l_1+1)\onebox(\bl)+\pmatrix{l_1\cr1\cr}\twobox({},\bl)
\qbox{for all positive odd $l_1$.}\cr
\endeq
We have derived these formulas inductively by using the properties (1--3).
\par
In the right-hand-sides of these formulas, we allow a parity symbol
to have length greater than the actual depth of the accompanied signature.
Namely, if a signature $l$ is accompanied by a parity symbol of length $m$,
then
$l=(l_k)_{1\le k\le m}$ can be such that
$l_1>\cdots>l_j>0=l_{j+1}=\cdots=l_m$,
and it is considered as $(l_k)_{1\le k\le j}$.
If a signature which breaks even this condition does appear,
(e.g., $l_1=l_2>0$), we understand that the corresponding
$\Gu(p)$ to be zero.
\par
There are some cases such that a parity symbol to have length smaller
than the depth of the accompanied signature.
In those cases we understand the corresponding terms are zero.
\par
In the left-hand-sides of the formulas, we always assume the length
of a parity symbol is equal to the depth of the accompanied signature.
\par
%\endinput
\par
\def\s{\sigma}
\par
\beginsection Appendix 2.
Vertex operator $V_z\otimes V(\L_0)\rightarrow V(\L_1)$

Let $V(\L_i) (i=0,1)$ be the irreducible highest weight
$\Ut$-module with the highest weight $\L_0$,
and let $V_z$ be the 2 dimensional
$U$-module given in \sec(6.2).
\par
Denote by
$\widehat V(\L_i)$
the direct product of the weight spaces of $V(\L_i)$.
There is a unique
$U$-linear
\eq
&\Phi:V_z\otimes V(\L_0)\longrightarrow {\widehat V}(\L_1)\cr
\endeq
such that
$\bigl(\Phi\bigl(v_+\otimes u_{\L_0}\bigr),u_{\L_1}\bigr)=1$,
where $(\ ,\ )$ is as in Appendix 1.
(To be precise $\Phi$ is a
$U\otimes \Q(q)[z,z^{-1}]$-linear map
$V_z\widehat{\otimes} V(\L_0)\rightarrow {\widehat V}(\L_1)\otimes
\Q(q)[z,z^{-1}]$ (see \sec(6.4)),
but we shall not bother writing $\Q(q)[z,z^{-1}]$.)
We list
\eq
&\Phi\bigl(v_+\otimes\Gu(p)\bigr)=\sum_{p'}c_\pm(p',p)\Gu(p')\cr
\endeq
for the first seven paths $p$ whose signatures (see Appendix 1)
are
$\phi$, $(1)$, $(2)$, $(3)$, $\pmatrix{2\cr1\cr}$, $(4)$,
$\pmatrix{3\cr1\cr}$.
\par
For convenience, we define
$\Phi^\s:V_z\otimes V(\L_0)\longrightarrow\widehat V(\L_0)$
by
$\Phi^\s=\s\Phi$,
where $\s$ is the $\Q(q)$-linear map
$\widehat V(\L_0)\longrightarrow\widehat V(\L_1)$
induced by the Dynkin diagram automorphism of
$\Ut$ which exchanges the colors 0 and 1.
The list in Table 2 is for $\Phi^\s$.
The vertex operator
$\Phi:V_z\otimes V(\L_0)\longrightarrow \widehat{V}(\L_1)$
is given by
\eq
&\Phi\bigl(v_\pm \otimes v\bigr)=
\s\bigl(\Phi^\s\bigl((\pm)\otimes v\bigr)\bigr),\cr
\endeq
and the vertex operator
$\Phi^c:V_z\otimes V(\L_1)\longrightarrow V(\L_0)$
is given by
\eq
&\Phi^c\bigl(v_+\otimes v\bigr)=z\Phi^\s\bigl(v_-\otimes \s(v)\bigr),\quad
\Phi^c\bigl(v_-\otimes v\bigr)=\Phi^\s\bigl(v_+\otimes \s(v)\bigr).
&(PhiPrime)\cr
\endeq
The formulas below are obtained inductively by using the result of
Appendix 1.
For example,
\eq
\Phi^\s\bigl(v_-\otimes(1)\onebox({})\bigr)&=
\Phi^\s\bigl(v_-\otimes f_0\phi\bigr)\cr
&=f_1\Phi^\s\bigl(v_-\otimes\phi\bigr)-z^{-1}q^{-1}
\Phi^\s\bigl(v_+\otimes\phi\bigr)\cr
&=\sum_{n=0}^\infty z^n(2n+1)\bigr\{[2](1)\onebox(\bl)+\pmatrix{0\cr1\cr}
\twobox({},\bl)\bigr\}-z^{-1}q^{-1}\sum_{n=0}^\infty z^n(2n)\onebox(\bl)\cr
&=-z^{-1}q^{-1}\phi+\sum_{n=0}^\infty z^nq(2n+2)\onebox(\bl)
+\sum_{n=1}^\infty z^n\pmatrix{2n+1\cr1\cr}\twobox({},\bl).\cr
\endeq
\par
%\endinput
\par
\def\V#1{V(\L_{#1})}
\def\L{\Lambda}
\def\o{\otimes}
\def\P{\Phi}
\def\Ps{\P^\sigma}
\def\u#1{u_{\L_{#1}}}
\def\Onebox{Onebox}
\def\e{\varepsilon}
\def\Gl(#1){G^{\rm low}(#1)}
\def\tf{\tilde f}
\def\te{\tilde e}
\def\bmod{\mathop{\rm mod\/}}
\par
\beginsection Appendix 3. Vertex Operator
$\V0\rightarrow\V1\o V$
\par
We follow the notation in Appendix 1 except that now we use $\P$ for
\eq
&\P:\V0\rightarrow \widehat{V}(\La_1)\o V\cr
\endeq
and set
\eq
&\P(v)=\P_+(v)\o v_++\P_-(v)\o v_-.\cr
\endeq
We list
$\Ps_\pm=\sigma\P_\pm$
in Table 3, where
$\sigma$
is the Dynkin diagram automorphism. We use the normalization
$\bigl(\P_-(\u0),\u1\bigr)=1$.
Let us prove
\eq
&\Ps_+(\phi)=-\sum_{n=0}^\infty q^{3n+1}(2n+1)\onebox({}),\cr
&\Ps_-(\phi)=\sum_{n=0}^\infty q^{3n}(2n)\onebox(\bl),&(Seed)\cr
\endeq
from which the rest of formulas are inductively derived by using
\eq
&\Ps_\pm(f_0v)=q^{\pm1}f_1\Ps_\pm(v)+\delta_{\pm,+}\Ps_-(v),\cr
&\Ps_\pm(f_1v)=q^{\mp1}f_0\Ps_\pm(v)+\delta_{\pm,-}\Ps_+(v).\cr
\endeq
The proof goes as follows. We have
\eq
&\bigl(\Ps_\pm(e_0v),v'\bigr)_\pm=
n\bigl(\Ps_\pm(v),f_1v'\bigr)+\delta_{\pm,-}\bigl(\Ps_+(v),t_1v'\bigr),\cr
&\bigl(\Ps_\pm(e_1v),v'\bigr)_\pm=
\bigl(\Ps_\pm(v),f_0v'\bigr)+\delta_{\pm,+}\bigl(\Ps_-(v),t_0v'\bigr).
&(Rec)\cr
\endeq
By using these formulas for $v=\phi$, we find that
\eq
&\bigl(\Ps_\pm(\phi),v'\bigr)=0\quad
\hbox{if $v'\in{\rm Im}f_0^2\bigcup{\rm Im}f_1^2$}.\cr
\endeq
In general, the lower global base $\Gl(b)\ (b\in B)$ has the property that
$\e_i(b)\ge k$ if and only if $\Gl(b)\in{\rm Im}f_i^k$.
In our situation, from this follows that
\eq
&\bigl(\Ps_\pm(\phi),\Gl(p)\bigr)=0\quad\hbox{if $\e_0(p)\ge2$ or
$\e_1(p)\ge2$}&(StpOne)\cr
\endeq
Now, for a given $p\in{\cal P}_0$, take any sequence $(i_N,\ldots,i_1)$
such that $p=\tf_{i_N}\cdots\tf_{i_1}{\bar p}_0$.
Again, in general, the actions of $e_i$ and $f_i$
on the global base are given by
\eq
&e_i\Gl(b)=[\phi_i(b)+1]_i\Gl(\te_ib)
+\sum_{b'}F^i_{b'b}\Gl(b'),\cr
&f_i\Gl(b)=[\e_i(b)+1]_i\Gl(\tf_ib)
+\sum_{b'}E^i_{b'b}\Gl(b'),\cr
\endeq
where $F^i_{b'b}$ and $E^i_{b'b}$ are the same as those in Appendix 1.
In short, the actions create the correction terms (i.e., terms
other than the combinatorial ones $\Gl(\te_ib)$ and $\Gl(\tf_ib)$)
of the string length greater than the combinatorial term.
Therefore, we conclude that
\eq
&\bigl(\Ps_\pm(\phi),\Gl(p)\bigr)=c
\bigl(\Ps_\pm(\phi),f_{i_N}\cdots f_{i_1}\phi\bigr)\cr
\endeq
with some non zero $c\in\Q(q)$, and further that
\eq
&\bigl(\Ps_\pm(\phi),\Gl(p)\bigr)\neq0\cr
\endeq
if and only if
\eq
p&=\tf_0(\tf_1\tf_0)^k\phi\hbox{ for }+\cr
&=(\tf_1\tf_0)^k\phi\hbox{ for }-.\cr
\endeq
Finally, by using \refeq{Rec} (with $v=\phi$) recursively,
we get \refeq{Seed}.
\par
\def\M{\Phi^\sigma_-}
\def\vn{\omega^{(n)}({\bar p}_{i+n},{\bar p}_i)}
\def\ma{\buildrel\M\over\rightarrow}
\par
Let us compute
$\vn=\langle G(\phi)|
\underbrace{
\M\circ\cdots\circ\M}_n|G(\phi)\rangle$
up to $q^8$ for an arbitrary $n$ by using Table 3.
Let $v_i\ (i=1,\ldots,6)$ be the upper global base corresponding to
the following symbols.
\eq
&\matrix{v_1&v_2&v_3&v_4&v_5&v_6\cr
\phi&(2)\onebox(\bl)&(4)\onebox(\bl)&{3\choose1}\twobox( ,\bl)&
{5\choose1}\twobox( ,\bl)&{4\choose2}\twobox(\bl, )\cr}\cr
\endeq
We can extract the following from Table 3.
\eq
&\M(v_1)=v_1+q^3v_2+q^6v_3\bmod q^8,\cr
&\M(v_2)=(q-q^3+q^5)v_1+(-q^2+q^4)v_2+q^3v_3+q^2v_4\bmod q^5,\cr
&\M(v_3)=q^2v_1-qv_4\bmod q^2,\cr
&\M(v_4)=-q^3v_1+v_2-qv_3+q^2v_6\bmod q^3,\cr
&\M(v_5)=v_3\bmod q^1,\cr
&\M(v_6)=v_4\bmod q^1.\cr
\endeq
Here ``$\bmod  q^k$" is to be understood for those terms except for $v_1$.
The $v_1$-terms are given by ``$\bmod q^{k+1}$". The reason we truncated the
expansions at order less than $q^8$ is as follows. For example, starting
from $v_1$ and applying $\M$ repeatedly one will get $v_2$-terms
only with power $q^3$ at the lowest.
Then, unless one get a $v_1$-term in
the next application of $\M$,
we get at least one more power of $q$
until one does finally reach a $v_1$-term. Therefore,
$\bmod q^5$ is enough ($8-3-1=4$) except for the $v_1$-term.
\par
Let us denote by $(1,j_1,\ldots,j_k)$ a process
\eq
&v_1\ma v_{j_1}\ma\cdots\ma v_{j_k}.\cr
\endeq
For example, if $k=1$ we have two processes $(12)$ and $(13)$
that contribute to the final answer. Along with coefficients, the $k=1$ process
gives rise to $q^3(12)+q^6(13)$. We then proceed for a larger $k$ step by
step. Since we consider only up to $q^8$, this terminates in finite steps.
Picking up the $k$-processes ending at $j_k=1$, we have
\eq
&k=0\quad(1),\cr
&k=2\quad(q^4-q^6+q^8)(121)+q^8(131),\cr
&k=3\quad(-q^6+2q^8)(1221)+q^8(1231)-q^8(1241),\cr
&k=4\quad q^8(12221)+(q^6-q^8)(12421)-q^8(12431)-q^8(13421),\cr
&k=5\quad-q^8\bigl((122421)+(123421)+(124221)\bigr),\cr
&k=6\quad q^8\bigl((1242421)+(1243421)+1246421)\bigr).\cr
\endeq
Taking combinatorial factors into consideration, we get (for $n\ge5$)
\eq
\omega^{(n)}({\bar p}_n,{\bar p}_0)=
1&+(n-1)(q^4-q^6+2q^8)+(n-2)(-q^6+2q^8)\cr
&+(n-3)(q^6-2q^8)+(n-4)(-3q^8)+(n-5)(3q^8)\cr
&+{(n-2)(n-3)\over2}q^8\cr
=1&+(n-1)q^4-nq^6+{n(n-1)\over2}q^8\bmod q^{10}.\cr
\endeq
For smaller values of $n$, we get
\eq
&\omega^{(2)}({\bar p}_0,{\bar p}_0)=1+q^4-q^6+q^8\bmod q^{10},\cr
&\omega^{(3)}({\bar p}_1,{\bar p}_0)=1+2q^4-3q^6+6q^8\bmod q^{10},\cr
&\omega^{(4)}({\bar p}_0,{\bar p}_0)=1+3q^4-4q^6+9q^8\bmod q^{10}.\cr
\endeq
The results agree with those in 6.8 obtained by solving the $q$-KZ
equation. Similarly, we calculate
\eq
&\omega^{(n)}((\cdots+ - + - - +),{\bar p}_0)=-q+q^3-nq^5+2nq^7,\cr
&\omega^{(n)}((\cdots+ - + + - -),{\bar p}_0)=-q+2q^3-(n+1)q^5+(3n+5)q^7.\cr
\endeq
Thus we get
\eq
&\iota((\cdots+ - + - - +),{\bar p}_0)=-q+q^3-q^5+q^7,\cr
&\iota((\cdots+ - + + - -),{\bar p}_0)=-q+2q^3-4q^5+7q^7,\cr
\endeq
which are in agreement with the result in \refto{FM}.
\par
%\endinput
\par
\par
\def\L{\Lambda}
\def\R{V(\L_0)^{*a}}
\def\o{\otimes}
\def\U{U_q\bigl(\widehat{\goth{sl}}(2)\bigr)}
\def\om{\eta}          %\def\om{\omega}
\def\pai{\omega}	%\pi
\def\V#1{V(\L_{#1})}
\def\P{{\cal{P}}}
\def\Gu{G^{{\rm up}}(p)}
\def\Gl{G^{{\rm low}}(p)}
\def\uv(#1){G^{{\rm up}}(\cdots#1)}
\def\lv(#1){G^{{\rm low}}(\cdots#1)}
\par
\beginsection Appendix 4.
Embedding of the vacuum, 1 and 2 particle states
\par
First we give a supplement to \sec(2) about the embedding
$\R\longrightarrow V\o V\o\cdots$. Let us denote by
$\pai$
the automorphism of the algebra
$\U$
given by
\eq
&\pai(e_i)=f_i,\ \pai(f_i)=e_i,\ \pai(q^h)=q^{-h}.\cr
\endeq
Let
$V_1$
and
$V_2$
be left
$\U$-modules.
A
$\Q(q)$-linear isomorphism
$\om:V_1\rightarrow V_2$
is called
$\pai$-compatible if and only if
\eq
&\pai(x)\om(v)=\om(xv)\cr
\endeq
for
$x\in\U$
and
$v\in V_1$.
The map \refeq{V*V} in \sec(6.7)
\eq
&\om:\V0\rightarrow\R\cr
\endeq
is $\pai$-compatible.
%There exists a unique $\pai$-compatible map
%\eq
%&\om:\V0\rightarrow\R\cr
%\endeq
%such that $\om$ maps the highest weight vector of $\V0$
%to the lowest weight vector of $\R$.
The map
\eq
&\om~:~V\longrightarrow V \qquad v_\pm\mapsto v_\mp
\endeq
is also $\pai$-compatible.
Suppose that
$\om_i:V_i\rightarrow V'_i\ (i=1,2,3)$
are
$\pai$-compatible, and that
$\Phi:V_1\rightarrow V_2\o V_3$
is an intertwiner. Then the map
\eq
&\Phi'~:~V'_1 \longrightarrow V'_3\o V'_2
\qquad
\om_1(v)\mapsto(\om_3\o\om_2)\circ P\circ\Phi(v)&(Trps)\cr
\endeq
is also an intertwiner. Here,
$P$
is the transposition of the tensor components.
For a given path
$p\in\P_i$
we set
\eq
&p^*=(-p(1)\ -p(2)\ -p(3)\cdots).\cr
\endeq
The set
$\P^*_i=\{p^*;p\in\P_i\}$
is naturally the crystal of
$\V{i}^{*a}$.
Let
$\Phi:\V{i}\rightarrow\V{1-i}\o V$
be the vertex operator given in \sec(4). Then
\eq
&\Phi':\V{i}^{*a}\rightarrow V\o \V{1-i}^{*a}\cr
\endeq
is also a vertex operator. From this we can define an embedding
\eq
&\iota':\R\rightarrow V\o V\o V\o\cdots\cr
\endeq
The embeddings
$\iota$
and
$\iota'$
commute with the
$\pai$-compatible maps
$\om$;
\eq
&\matrix{
\V0&{\buildrel\iota\over\longrightarrow}&\cdots\o V\o V\cr
{\om\downarrow}&&{\om\downarrow}\cr
\R&{\buildrel\iota'\over\longrightarrow}&V\o V\o\cdots\cr}
\cr
\endeq
%We want to compute the image of the canonical element
%$\sum_{p\in\P_0}\Gu\o \Gu^*$
%by the embedding
%$\iota\o\iota'$
%of
%$\V0\o\R$
%into
%$W$.
%Instead of computing
%$\iota'\bigl(\Gu^*\bigr)$
%directly, we compute
%$\om\circ\iota\circ\om^{-1}\bigl(\Gu^*\bigr)$.
%We will show that
%\eq
%&\om^{-1}\bigl(\Gu^*\bigr)
%=(-1)^{{\scriptstyle ht}(\L_0-\mu)}q^{|\rho+\L_0|^2-|\rho+\mu|^2}\Gl.&(Low)\cr
%\endeq
%Here
%$ht(m_0\alpha_0+m_1\alpha_1)=m_0+m_1$.
%Take
%$v$, $v'\in\V0$
%and
%$x\in\U$.
%We denote the coupling between
%$\V0^*$
%and
%$\V0$
%by
%$\langle\ ,\ \rangle$.
%Then, we have
%\eq
%&\langle\om(xv),v'\rangle
%=\langle\pai(x)\om(v),v'\rangle
%=\langle\om(v),a\circ\pai(x)v'\rangle&(AOme).\cr
%\endeq
%Let us denote by
%$\om'$
%the identification map
%$\V0\rightarrow\V0^*$
%by means of the bilinear form
%$(\ ,\ )$
%discussed in Appendix 1. We have
%$\om'\bigl(\Gl\bigr)=\Gu^*$.
%Then, we have
%\eq
%&\langle\om'(xv),v'\rangle=\langle\om'(v),\varphi(x)v'\rangle.&(Phi)\cr
%\endeq
%Since
%$\om(u_{\L_0})=\om'(u_{\L0})$,
%by comparing \refeq{AOme} and \refeq{Phi}
%for
%$x=f_i$,
%we obtain
%\eq
%&\om(v)=(-1)^{{\scriptstyle ht}(\L_0-\mu)}
%q^{-|\rho+\L_0|^2+|\rho+\mu|^2}\om'(v)\cr
%\endeq
%for
%$v\in(\V0)_\mu$.
%The assertion \refeq{Low} follows from this.
\par
{}From \refeq{V*V} and \refeq{uplow} we see that
the vacuum state $|vac\rangle$ embedded in $W$ is given by
\eq
&|vac\rangle=\sum_\mu \sum_{p\in\ (P_0){}_\mu}
(-1)^{{\scriptstyle ht}({\L}_0-\mu)}q^{|\rho+{\L}_0|^2-|\rho+\mu|^2}
\iota\bigl(\Gu\bigr)\o\eta\circ\iota\bigl(\Gl\bigr).\cr
\endeq
As we have noted in \sec(2), we have
\eq
&\iota\bigl(\Gu\bigr)\big|_{q=0}=|p\rangle.\cr
\endeq
For
$\Gl(p)$
we must shift the
$q$
power;
\eq
&\iota\bigl(q^{|\L_0|^2-|\mu|^2}\Gl\bigr)\big|_{q=0}=|p\rangle.\cr
\endeq
Note that
\eq
&(-1)^{{\scriptstyle ht}(\L_0-\mu)}q^{|\rho+\L_0|^2-|\rho+\mu|^2}
q^{-|\L_0|^2+|\mu|^2}=(-q)^{{\scriptstyle ht}(\L_0-\mu)}.\cr
\endeq
Therefore, the term
$(-1)^{{\scriptstyle ht}(\L_0-\mu)}q^{|\rho+\L_0|^2-|\rho+\mu|^2}
\iota\bigl(\Gu\bigr)\o\pai\circ\iota\bigl(\Gl\bigr)$
contributes to the sum only with power
$q^{{\scriptstyle ht}(\L_0-\mu)}$.
Let us compute
$|vac\rangle$
up to order
$q^3$.
We need the following five vectors;
\eq
&v_1=\uv(+\ -\ +\ -)=\lv(+\ -\ +\ -),\cr
&v_2=\uv(+\ -\ +\ +)=\lv(+\ -\ +\ +),\cr
&v_3=\uv(+\ -\ -\ +)={1\over[2]}\lv(+\ -\ -\ +),\cr
&v_4=\uv(+\ +\ -\ +)={1\over[2]}\lv(+\ +\ -\ +),\cr
&v_5=\uv(+\ -\ -\ -)=\lv(+\ -\ -\ -).\cr
\endeq
We have
\eq
&|vac\rangle\equiv\sum_{i=1}^5(-q)^{m(i)}
\iota(v_i)\o\pai\circ\iota(v_i)\bmod q^4,\cr
\endeq
where
$\{m(i)\}_{1\le i\le5}=\{0,1,2,3,3\}$.
\par
In the following table we show the coefficients of paths in the expansion of
$\iota(v_i)$ %/\omega({\bar p}_0,{\bar p}_0)$
($1\le i\le5$),  %where $\omega$ is defined in \refeq{Omega}),
up to the relevant order for each
$i$.
If we write
$\cdots$
as a part of a path, we mean that the indicated part of the path
is identical with the ground-state-path
${\bar p}_0$.
If we write
$(\cdots)$,
we mean that the indicated part may be void; otherwise it must not be void.
We write
$\pm$
or
$\mp$,
if the indicated column of the path differs from the corresponding column of
${\bar p}_0$.
Therefore, for example,
$\cdots\pm\ \mp\cdots$
actually represents the paths
$(\cdots+\ -\ +\ +\ -\ -)$,
$(\cdots+\ -\ -\ +\ +\ -)$,
$(\cdots+\ +\ -\ -\ +\ -)$,
$(\cdots-\ +\ +\ -\ +\ -)$,
etc., simultaneously.
\medskip
\settabs 2 \columns
\+{\it Path}&{\it Coefficient}\cr
\pn$v_1$
\+$\cdots$            &1\cr
\+$\cdots-\ +$        &$-q+q^3$\cr
\+$\cdots\pm\ \mp\cdots        $&$-q+2q^3       $\cr
\+$\cdots\pm\ \mp\ \pm\ \mp(\cdots)        $&$2q^2       $\cr
\+$\cdots\pm\ \mp\cdots\pm\ \mp(\cdots)        $&$q^2       $\cr
\+$\cdots\pm\ \pm\ \mp\ \mp(\cdots)        $&$-q^3       $\cr
\+$\cdots\pm\ \mp\ \pm\ \mp\ \pm\ \mp(\cdots)        $&$-5q^3       $\cr
\+$\cdots\pm\ \mp\cdots\pm\ \mp\ \pm\ \mp(\cdots)        $&$-2q^3       $\cr
\+$\cdots\pm\ \mp\ \pm\ \mp\cdots\pm\ \mp(\cdots)        $&$-2q^3       $\cr
\+$\cdots\pm\ \mp\cdots\pm\ \mp\cdots\pm\ \mp(\cdots)     $&$-q^3       $\cr
\pn$v_2$
\+$\cdots+        $&$1-q^2       $\cr
\+$\cdots+\ -\ +        $&$-2q       $\cr
\+$\cdots\pm\ \mp\cdots+        $&$-q       $\cr
\+$\cdots+\ -\ +\ -\ +        $&$5q^2       $\cr
\+$\cdots\pm\ \mp\ \pm\ \mp\cdots+        $&$2q^2       $\cr
\+$\cdots\pm\ \mp\cdots+\ -\ +        $&$2q^2       $\cr
\+$\cdots\pm\ \mp\cdots\pm\ \mp\cdots+        $&$q^2       $\cr
\+$\cdots+\ +\ -        $&$q^2       $\cr
\pn$v_3$
\+$\cdots-\ +        $&$1       $\cr
\+$\cdots        $&$q       $\cr
\+$\cdots-\ +\ -\ +        $&$-3q       $\cr
\+$\cdots\pm \mp\cdots-\ +        $&$-q       $\cr
\pn$v_4$
\+$\cdots+\ -\ +        $&$1       $\cr
\pn$v_5$
\+$\cdots- \ -        $&$1       $\cr
\medskip
\pn From this we get the expansion of
$|vac\rangle$      % /\omega({\bar p}_0,{\bar p}_0)^2$
up to order $q^3$. The result reads as follows.
\medskip
\+{\it Path}&{\it Coefficient}\cr
\+$\cdots        $&$1       $\cr
\+$\cdots\pm\ \mp\cdots        $&$-q+2q^2       $\cr
\+$\cdots\pm \ \mp\ \pm\ \mp\cdots        $&$2q^2       $\cr
\+$\cdots\pm\ \mp\cdots\pm \mp\cdots        $&$q^2       $\cr
\+$\cdots\pm\ \mp\ \pm\ \mp\ \pm\ \mp\cdots        $&$-5q^3       $\cr
\+$\cdots\pm\ \pm\ \mp\ \mp\cdots        $&$-q^3       $\cr
\+$\cdots\pm\ \mp\cdots\pm\ \mp\ \pm\ \mp\cdots        $&$-2q^3       $\cr
\+$\cdots\pm\ \mp\ \pm\ \mp\cdots\pm\ \mp\cdots        $&$-2q^3       $\cr
\+$\cdots\pm\ \mp\cdots\pm\ \mp\cdots\pm\ \mp\cdots        $&$-q^3       $\cr
\medskip
This agrees with the perturbation expansion expansion (see \refto{FT}).
\bigskip
\par
%(Continued: The embedding of the vacuum, 1 and 2 particle states.)
\def\V#1{V(\L_{#1})}
\def\L{\Lambda}
\def\o{\otimes}
\def\P{\Phi}
\def\Ps{\P^\sigma}
\def\u#1{u_{\L_{#1}}}
\def\eq#1\endeq{$$\eqalignno{#1}$$}
\def\Onebox{\onebox}
\def\Twobox(#1,#2){\twobox(#1,#2)}
\def\b{black}
\def\e{\varepsilon}
\def\Gl(#1){G^{\rm low}(#1)}
\def\Gu(#1){G^{\rm up}(#1)}
\def\tf{\tilde f}
\def\te{\tilde e}
\def\Q{{\bf Q}}
\def\Z{{\bf Z}}
\par
The 1-particle state with the quasi momentum
$u\ (z=e^{iu})$
is an embedding of
$V_z$
into
$\V1\o\V0^{*a}$,
or further into
$W$.
We denote the latter embedding by
$\iota_z^{(1)}$
(in general,
$\iota^{(n)}_{z_n,\ldots,z_1}$
for the
$n$-particle embedding).
The quantum affine symmetry discussed in \sec(1) assures that the vectors
obtained by this embedding are doubly degenerate eigenvectors of
$\H$
and its higher order relatives.
Because of the property of the dual modules discussed in \sec(5) of
\refto{(KMN)^2}, the existence of the embedding is equivalent to the existence
of the vertex operator
$\P(z):V_z\o\V0\rightarrow\V1$.
\par
Let us compute this embedding at
$q=0$ (i.e., only the crystal)
by using the data in Table 2.
We discuss in details only on the
$v_+$-component. First consider the top term in
$\iota^{(1)}_z(v_+)$;
\eq
&\P(z)(v_+\o\u0)\o\u0\cr
&=\sum_{n=0}^\infty z^n(2n)\Onebox({})\o\phi\cr
&=(\cdots-\ +\ -\ +|+\ -\ +\ -\cdots)
+z(\cdots-\ +\ +\ -|+\ -\ +\ -\cdots)\cr
&\qquad+z^2(\cdots+\ -\ +\ -|+\ -\ +\ -\cdots)
+\cdots\cr
\endeq
\par
So, we get
$\sum_{n=0}^\infty z^n[[2n]]$
from the top term.
Because of the translational covariance,
we expect the whole limit
$\iota^{(1)}_z(v_+)$
to be
$\sum_{n\in\Z}z^n[[2n]]$.
In the lowest order in
$q$,
the term
$(-)^{{\scriptstyle ht}(\L_0-\mu)}q^{|\rho+\L_0|^2-|\rho+\mu|^2}
\eta\circ\iota\bigl(\Gl(p)\bigr)$
gives rise to
$(-q)^{{\scriptstyle ht}(\L_0-\mu)}\pai(|p\rangle)$.
Therefore, among 6 more datum in Table 2,
only
$p=(\cdots+\ -\ -\ +)$
and
$(\cdots-\ +\ -\ +)$
give non-zero contribution at
$q=0$, i.e.,
$z^{-1}[[-2]]$
and
$z^{-2}[[-4]]$,
respectively.
So, it is consistent.
Similarly, the
$v_-$-component has the limit
$\sum_{n\in\Z}z^n[[2n+1]]$.
\par
The two-particle state is
$V_{z_2}\o V_{z_1}$
embedded in
$\V0\o\V0^{*a}$
or in
$W$.
This embedding
$\iota^{(2)}_{z_2,z_1}$
is obtained by a successive application of the vertex operators.
For example, the spin 1 component is
\eq
&\sum_{p\in{\cal P}_0{}\mu}
(-)^{{\scriptstyle ht}(\L_0-\mu)}q^{|\rho+\L_0|^2-|\rho+\mu|^2}
\P_{z_2}\bigl(v_+\o \P_{z_1}\bigl(v_+\o\Gu(p)\bigr)\bigr)
\o\eta\circ\iota\bigl(\Gl(p)\bigr).&(TwoPar)\cr
\endeq
\par
Let us use signatures and parity symbols to represent the paths and the
upper global base vectors (see Appendix 1).
If
$p=\phi$,
we have
\eq
&\P(z_1)(v_+\o\phi)\equiv \phi+z_1(2)\Onebox({})
+z_1^2(4)\Onebox({})+\cdots\in\V1.\cr
\endeq
Then, by using \refeq{PhiPrime}
\eq
\P'(z_2)(v_+\o\phi)&\equiv
z_2(1)\Onebox({})+z_2^2(3)\Onebox({})+\cdots\in\V0\bmod q,\cr
\P'(z_2)(v_+\o z_1(2)\Onebox({}))
&\equiv-z_1(1)\Onebox({})\cr
&+z_1z_2^2\pmatrix{3\cr2\cr}\Twobox({},{})
+z_1z_2^3\pmatrix{5\cr2\cr}\Twobox({},{})+\cdots\in\V0\bmod q,\cr
\P'(z_2)(v_+\o z_1^2(4)\Onebox({}))
&\equiv
-z_1^2(3)\Onebox({})-z_1^2z_2\pmatrix{3\cr2\cr}\cr
&+z_1^2z_2^3\pmatrix{5\cr4\cr}\Twobox({},{})+
z_1^2z_2^4\pmatrix{7\cr4\cr}\Twobox({},{})+\cdots\in\V0\bmod q.\cr
\endeq
Therefore, the contribution to \refeq{TwoPar} at
$q=0$
from
$p=\phi$
is
\eq
&z_2[[1,0]]+z_2^2[[3,0]]+\cdots\cr
&-z_1[[1,0]]+z_1z_2^2[[3,2]]+z_1z_2^3[[5,2]]+\cdots\cr
&-z_1^2[[3,0]]-z_1^2z_2[[3,2]]+z_1^2z_2^3[[5,4]]+z_1^2z_2^4[[7,4]]+\cdots.\cr
\endeq
So, by the translational covariance, we expect
\eq
&\sum_{m\ge n}(z_1^nz_2^{m+1}-z_1^{m+1}z_2^n) [[2m+1,2n]] &(pp)\cr
\endeq
as the whole answer.
The datum in Table 2 turn out to be consistent with this.
In Appendix 5, the comparison of this result with the Bethe Ansatz calculation
is given.
\par
Similarly, we obtain
\eq
&\iota^{(2)}_{z_2,z_1}(v_-\o v_-)\big|_{q=0}=
\sum_{m\ge n}(z_1^{n-1}z_2^m-z_1^mz_2^{n-1})[[2m,2n-1]].&(mm)\cr
\endeq
\par
The case spin is equal to zero, is somewhat tricky because the limit
$q=0$
can be taken only after summing up certain series.
Let us consider
$\iota^{(2)}_{z_2,z_1}(v_-\o v_+)$.
The first term is
$p=\phi$,
and we want to compute
$\P'(z_2)\bigl(v_-\o\P(v_+\o\phi)\bigr)$,
or
$\Ps(z_2)\bigr(v_+\o z_1^n(2n)\Onebox(\bl)\bigr)$
(see \refeq{PhiPrime} and Table 2).
A cumbersome (or rather interesting) feature here is that
the terms proportional to
$\phi$
contains negative powers in
$q$.
Keeping the whole terms that are proportional to
$\phi$,
and neglecting all other terms that vanishes at
$q=0$,
we have
\eq
\Ps(z_2)(v_+\o\phi)&=\phi+\sum_{n=1}^\infty z^n(2n)\Onebox(\bl),\cr
\Ps(z_2)(v_+ \o z_1(2)\Onebox(\bl))&=
{z_1\over z_2q^2}\dot{1-q^2\over1-q^4}\phi-z_1(2)\Onebox(\bl)\cr
&+z_1\sum_{n=2}^\infty z_2^n\pmatrix{2n\cr2\cr}\Twobox(\bl,{})+\cdots,\cr
\Ps(z_2)(v_+\o z_1^2(4)\Onebox(\bl)&=
\left(z_1\over z_2q^2\right)^2{1-q^2\over1-q^4}\cdot{1-q^6\over1-q^8}\phi\cr
&-z_1^2(4)\Onebox(\bl)-z_1^2z_2\pmatrix{4\cr2\cr}\Twobox(\bl,{})
+z_1^2\sum_{n=3}^\infty z_2^3\pmatrix{2n\cr4\cr}\Twobox(\bl,{})+\cdots.\cr
\endeq
Extrapolating, we have
\eq
1+&
{z_1\over z_2q^2}\dot{1-q^2\over1-q^4}
+\left(z_1\over z_2q^2\right)^2{1-q^2\over1-q^4}\cdot{1-q^6\over1-q^8}
+\cdots\cr
&={q^2(z_2-z_1)\over z_2q^2-z_1}
{z_2-z_1q^4\over z_2-z_1q^2}
{z_2-z_1q^8\over z_2-z_1q^6}\cdots .\cr
\endeq
Therefore, after summation the terms proportional to
$\phi$
vanish at
$q=0$.
(In fact this is exact as shown in \sec(6.5).)
With this understood, we get
\eq
&\iota^{(2)}_{z_2,z_1}(v_-\o v_+)\big|_{q=0}=
\sum_{m>n}(z_1^nz_2^m-z_1^mz_2^n)[[2m,2n]],&(mp)\cr
\endeq
and similarly,
\eq
&\iota^{(2)}_{z_2,z_1}(v_+\o v_-)\big|_{q=0}=
\sum_{m>n}(z_1^nz_2^{m+1}-z_1^mz_2^{n+1})[[2m+1,2n+1]].&(pm)\cr
\endeq
\par
%\endinput
\par
\par
\par
\def\x1m{(x_1,\cdots,x_M)}
\def\vec{|x_1,\cdots,x_M\rangle}
\def\vech{|x_1,\cdots,x_{n-h}\rangle}
\def\la{\lambda}
\def\si{\sigma}
\def\ga{\gamma}
\def\reomega{a}
\def\qhq#1{\quad\hbox{#1}\quad}
\def\e{\hbox{e}}
\def\ra{\rightarrow}
\def\lajo{\la_j^{(0)}}
\def\zo{z^{(0)}}

\def\lako{\la_k^{(0)}}
\def\laio{\la_1^{(0)}}

\def\ktil{\tilde{k}}
\par
\beginsection Appendix 5. Comparison with the Bethe Ansatz States
\par
In this appendix we calculate the Bethe vectors at $q=0$
and compare them with the results in Appendix 3.
We assume the periodic boundary condition and hence
that the length $N$ of row to be even. Set $N=2n$.
Following \refto{Bax} we label the standard basis vectors
$v_{\varepsilon_1}\otimes \cdots\otimes v_{\varepsilon_N}$
of $V^{\otimes N}$ by the locations of $v_-$:
$(x_1,\cdots,x_M)$, $1\le x_1<\cdots<x_M\le N$.
We denote by $|x\rangle=|x_1,\cdots,x_M\rangle \in V^{\otimes N}$
the corresponding vector with spin $n-M$.
It is known that the vector
\eq
&v=\sum f\x1m\vec&(Bethevec)\cr
\endeq
is an eigenvector of the $\XXZ$ Hamiltonian $\H$ if $f\x1m$
is of the following form:
\eq
&f\x1m=
\sum_{\sigma\in S_M}
\prod_{1\le j<k\le M}\hbox{E}(u_{\si(j)},u_{\si(k)})
\prod_{l=1}^M(-\hbox{F}(u_{\si(l)}))^{x_l-1}
&(coeff)\cr
\endeq
where
\eq
\hbox{E}(u,v)&=
{\sin(u-v+i\ga)\over \sin(u-v)},
\qquad
\hbox{F}(u)={\sin(u+{1\over2}i\ga)\over \sin(u-{1\over2}i\ga)}
\cr
\Delta&={q+q^{-1}\over 2},\quad q=-e^{-\ga}, \quad\ga\in{\bf R}_{\geq0},
\endeq
and $\{u_j\}_{j=1}^M$ is a solution of the following Bethe Ansatz
equation (BAE)
\eq
&\hbox{F}(u_k)^{2n}=
-\prod_{j=1}^MS(u_k,u_j)
\qhq{for}1\le k\le M,&(BAE)\cr
\endeq
where
\eq
&S(u,v)=
{\sin(u-v+i\ga)\over\sin(u-v-i\ga)}.
\endeq
\vskip5pt
\par
We calculated the form of Bethe vectors at $q=0$
in the cases of $2h$-particles with spin $h$ and $h-1$.
Here the particle number is the same as the number of
holes in \refto{Bab} or \refto{DesL}.
As a consequence
we find that the Bethe vectors at $q=0$
coincide with the $q\rightarrow0$ limit of
 eigenvectors calculated in App. 4. The correspondences
are given by
\item{$\bullet$} the ground states (Example 1),
\item{$\bullet$} two particle states with spin 1 (Example 2) (\refeq{pp}),
\item{$\bullet$} two particle states with spin 0 (Example 3)
(\refeq{pm} and \refeq{mp}).
\vskip5pt
\par
\noindent (i) The case of $2h$-particles and spin $h$.
\par
\vskip5pt
Here we consider the case $M=n-h$ for some $h\in {\bf Z}_{\geq1}$.
It can be shown that there is a solution $\{u_j\}_{j=1}^{n-h}=
\{\la_j\}_{j=1}^{n-h}$
of \refeq{BAE} which has the expansion of the form
\eq
&\e^{2i\la_j}=
\la_j^{(0)}(1+\sum_{i=1}^{\infty}\la_j^{(i)}q^i)
\qhq{for}1\le j\le n-h&(exp)\cr
\endeq
where $q=-\e^{-\ga}$.
Substituting \refeq{exp} into \refeq{BAE}
and comparing the coefficients of $q^0$ %and $q^1$
we have
\eq
(\lajo)^{-(n+h)}&=
(-1)^{n-h+1}\prod_{k=1}^{n-h}\lako \qhq{for}1\le j\le n-h.
&(0and1)\cr
%\laji&=0 \qhq{for}1\le j\le n-h.\cr
\endeq
%The equation \refeq{0and1} can easily be solved and we have
\par
\vskip10pt
\proclaim Proposition \prop{N1}.
Let $\{r_j\}_{j=1}^{2h}\sqcup\{\mu_j\}_{j=1}^{n-h}$ be a
partition of $\{0,1,\cdots,n+h-1\}$
such that $r_1<\cdots<r_{2h}$ and $\mu_1<\cdots<\mu_{n-h}$.
Let $\theta$ be a real number which
satisfies
\eq
\theta&\equiv
{\pi(-2h\mu_1+\sum_{j=1}^{2h}r_j)\over n(n+h)}
\bmod{\pi\over n}{\bf Z},
&(valthe)\cr
-\pi&\le\theta<-\pi+{2\pi(\mu_1+1)\over n+h}.&(ranthe)\cr
\endeq
Set
\eq
&\la_1^{(0)}=\e^{i\theta},
\quad
\lajo=\la_1^{(0)}\reomega^{\mu_j-\mu_1}
\quad 1\le j\le n-h,\cr
\endeq
 where $\reomega=\exp({2\pi i\over n+h})$.
Then $\{\lajo\}_{j=1}^{n-h}$ is a solution of \refeq{0and1} and any solution
of \refeq{0and1} is obtained in this way.
\par
\vskip10pt
\noindent{\sl Remark}\quad
For each $\{r_j\}_{j=1}^{2h}$, if $n$ is sufficiently large,
there are $2(\mu_1+1)$ $\theta$'s which satisfy \refeq{valthe} and
\refeq{ranthe}.
More exactly let $\theta_0$  be such a $\theta$ that satisfies
\refeq{valthe} and $-\pi\le\theta<-\pi+{\pi\over n}$.
Then $\theta_{0,j}=\theta_0+{\pi j\over n}$ $(0\le j\le 2\mu_1+1)$
satisfies \refeq{ranthe} if $n$ is sufficiently large.
\par
\vskip5pt
\par
By expanding $\hbox{F}(\la_j)$ and $\hbox{E}(\la_j,\la_k)$ in $q$
at $q=0$ we have
\eq
\hbox{F}(\la_j)&=
-(\lajo)^{-1}(1+O(q)),
\cr
\hbox{E}(\la_j,\la_k)&=
{\lako\over \lajo-\lako}q^{-1}(1+O(q)).\cr
\endeq
Substituting these expressions to \refeq{coeff} we have
\par
\proclaim Proposition \prop{N2}.
The eigenvector $v$ is given by
\eq
v=const.&\sum_{\{k\}}(\laio\reomega^{-\mu_1})^{-\sum_{k=1}^{2h} k_j}\cdot
\prod_{i=1}^{2h}P_{k_i}(\reomega)\cdot V(k,\reomega)
\cr
&\times\sum_{x_1}
(-1)^{x_1(n+h-1)}(\laio)^{-(n-h)x_1}
\reomega^{-x_1(2h\mu_1+\sum_{j=1}^{2h} r_j)}
\vech
\bmod q,\cr
\endeq
where $const.$ is an overall factor, and
\eq
P_l(\reomega)&=
\prod_{-(n+h-1)<i<j\le0,i,j\not=l}(\reomega^i-\reomega^j),
\cr
V(k,\reomega)&=
\det(\reomega^{-(n-r_i)k_j})_{1\le i,j\le 2h}.
\endeq
Here $\{k_j\}_{j=1}^{2h}$ and $x_1$ varies subject to the following conditions.
\eq
&0\geq k_1>\cdots>k_{2h}\geq -(n+h-1),
\cr
&1\le x_1\le b_{n-h}+n+h+1\quad(\le 2h+1),
\cr
&\{b_j\}_{j=1}^{n-h}=\{-j\}_{j=1}^{n+h-1}\backslash\{k_j\}_{j=1}^{2h},
\quad  b_1>\cdots>b_{n-h}.
\endeq
For these data $\{x_2,\cdots,x_{2h}\}$ are determined by
\eq
&x_j=x_1+2(j-1)+i\qbox{for $-k_i-i+2\le j\le -k_{i+1}-i$
and $0\le i\le 2h$},
\endeq
where we consider $k_0=1$ and $k_{2h+1}=n-h$.
\par
\proclaim Lemma.
$P_{k+1}(\reomega)=-\reomega P_k(\reomega).$
\par
\par
Set
\eq
&l_j=2(k_j+n)-n+j-x_1\qhq{for}1\le j\le 2h.
\endeq
Now we shall take a limit $n\ra\infty$ in such a way that
$\reomega^{r_j}$ $(1\le j\le 2h)$
are finite and $(-\laio)^n$ has a limit.
By the Lemma, if
$k_1-k_{2h}$ is sufficiently small compared with $n$,
the ratio $P_{k_i}(a)/P_{k_1}(a)$ goes to one in the limit
$n\rightarrow\infty$.
Using this and setting
\eq
&z_j=\reomega^{r_j}  \quad (1\le j\le 2h)
\endeq
we have in the $n\ra\infty$ limit
\par
\vskip10pt
\proclaim Proposition \prop{N3}.
Suppose $|l_1-l_{2h}|\ll\infty$ and $|l_1|\ll\infty$.
Then in the limit $q\ra0$ and $n\ra \infty$, the coefficients $v(x)$
of $v=\sum v(x)|x\rangle$ is
\eq
&v(x)=\pm\det(z_i^{k_j})_{1\le i,j\le 2h}
\bigl(\prod_{j=1}^{2h}z_j\bigr)^{{x_1-1\over2}}.
\endeq
where $x_1=1$ or $2$ and the signature is same if $x_1$ is same.
\par
\par
\vskip20pt
\par
\noindent{\sl Example 1} ($h=0$, Ground states)\quad
In this case $\reomega=\exp({2\pi i\over n})$ ,
$\mu_j=j-1$ $(1\le j\le n)$, $\laio=-1$ or $-\exp({\pi i\over n})$ and
$\lajo=\laio\reomega^{j-1}$ $(1\le j\le n)$.
\vskip5pt
\item{(1)} $\laio=-1$\par
$\qquad$\noindent$v=-|1,3,5,\cdots,2n-1\rangle+|2,4,6,\cdots,2n\rangle$
mod $q$.
\vskip5pt
\item{(2)} $\laio=-\exp({\pi i \over n})$\par
$\qquad$\noindent$v=|1,3,5,\cdots,2n-1\rangle+|2,4,6,\cdots,2n\rangle$ mod $q$.
\par
\vskip 20pt
\par
Let us write the vector of the form $|x_1,\cdots,x_M\rangle$ by
using the chain of $+$ and $-$. Namely, for example,
the vectors
$|1,3,5,\cdots,2n-1\rangle$ and $|2,4,6,\cdots,2n\rangle$ are
$$
\matrix{
&1&2&3&4&\cdots&2n-3&2n-2&2n-1&2n&
\cr
&-&+&-&+&\cdots&-&+&-&+&
\cr
&+&-&+&-&\cdots&+&-&+&-&.
\cr
}
$$
\par
\noindent{\sl Example 2} ($h=1$, $2$-particle states, spin $1$)\quad
The coefficients of the limit vectors are
\eq
&v(1,x_2,\cdots,x_{n-1})=\pm(z_1^{k_1}z_2^{k_2}-z_1^{k_2}z_2^{k_1}),
\cr
&v(2,x_2,\cdots,x_{n-1})=
\pm(z_1^{k_1}z_2^{k_2}-z_1^{k_2}z_2^{k_1})(z_1z_2)^{{1\over2}}.\cr
\endeq
Here
\eq
&0\geq k_1>k_2 \geq -n,
\endeq
and
\eq
&l_1=2k_1+n+1-x_1,
\cr
&l_2=2k_2+n+2-x_1.
\endeq
The numbers $l_1$ and $l_2$ have the following pictorial meaning.
We shall number the place between $i-1$ and $i$ by $n+1-i$.
Then $l_1$ and $l_2$ are the numbers between $+$ and $+$
which occurs twice in the above picture.
$$
\matrix{
&-&+&-&+&\cdots&-&+&|&+&-&\cdots&-&+&|&+&-&\cdots&-&+&
\cr
&{}&{}&{}&{}&{}&{}&{}&l_1&{}&{}&{}&{}&{}&l_2&{}&{}&{}&{}&{}&.
\cr
}
$$
\vskip 20pt
\par
%\noindent {\bf 2.}The other cases\par
%\noindent
\par
\vskip10pt
\noindent (ii) The case of $2h$-particles and spin $h-1$ ($h\geq1$).
\par
\vskip5pt
Suppose that the solution of BAE consists of real quasi-momenta
and one 2-string. (\refto{Bab}).
Let us denote them by $\{\lambda_j\}_{j=1}^{n-h-1}$ and
$\{z,w\}$ respectively.
Then BAE takes the form
\eq
F(\la_j)^{2n}&=-S(\la_j,z)S(\la_j,w)\prod_{k=1}^{n-h-1}S(\la_j,\la_k),
&(BA1)
\cr
F(z)^{2n}&=S(z,w)\prod_{k=1}^{n-h-1}S(z,\la_k),
&(BA2)
\cr
F(w)^{2n}&=S(w,z)\prod_{k=1}^{n-h-1}S(w,\la_k).
&(BA3)
\cr
\endeq
It can be shown that there is a set
of solutions of \refeq{BA1}-\refeq{BA3}
which has the following expansions at $q=0$.
\eq
\e^{2i\la_j}&=
\la_j^{(0)}(1+\sum_{i=1}^{\infty}\la_j^{(i)}q^i)
\qhq{for}1\le j\le n-h-1,&(exp1)
\cr
\e^{2iz}&=
z^{(0)}q(1+z^{(n+h-1)}q^{n+h-1}+\sum_{k=n+h}^{\infty}z^{(k)}q^k),
&(exp2)\cr
\e^{2iw}&=
z^{(0)}q^{-1}(1+w^{(n+h-1)}q^{n+h-1}+\sum_{k=n+h}^{\infty}w^{(k)}q^k).
&(exp3)\cr
\endeq
Substituting these expressions into \refeq{BA1}-\refeq{BA3}
we have, at $q=0$
\eq
&(\lajo)^{-n-h+1}=(-1)^{n-h}(\zo)^2\prod_{k=1}^{n-h-1}\lako,
&(la0)
\cr
&(\zo)^{-2(h+1)}=(\prod_{k=1}^{n-h-1}\lako)^2.
&(z0w0)
\cr
\endeq
We can solve \refeq{la0} and \refeq{z0w0} easily.
\par
\proclaim Proposition \prop{N4}.
Let $\{r_j\}_{j=1}^{2h}\sqcup\{\mu_j\}_{j=1}^{n-h-1}$ be a
partition of $\{0,1,\cdots,n+h-2\}$
such that $r_1<\cdots<r_{2h}$ and $\mu_1<\cdots<\mu_{n-h-1}$.
Let $\theta_r$ and $\theta_c$ be real numbers which
satisfy
\eq
\theta_r&\equiv
{\pi(h-1)(-2h\mu_1+\sum_{j=1}^{2h}r_j)\over nh(n+h-1)}
\qhq{mod}{\pi\over nh}{\bf Z},
&(ther)\cr
-\pi&\le\theta_r<-\pi+{2\pi(\mu_1+1)\over n+h-1}.&(ranther)
\cr
\theta_c&\equiv
-(n-1)\theta_r+
{\pi(-2h\mu_1+\sum_{j=1}^{2h}r_j)\over n+h-1}
\qhq{mod}\pi{\bf Z},
&(thec)\cr
-\pi&\le\theta_c<\pi.
\cr
\endeq
Set
\eq
&\la_1^{(0)}=\e^{i\theta_r},
\quad
\zo=\e^{i\theta_c},
\quad
\lajo=\la_1^{(0)}\reomega^{\mu_j-\mu_1}
\quad 1\le j\le n-h-1,\,
\cr
\endeq
 where $\reomega=\exp({2\pi i\over n+h-1})$.
Then $\{\lajo\}_{j=1}^{n-h-1}\sqcup\{z,w\}$ is a
solution of \refeq{la0}and \refeq{z0w0}. Any set of solutions
of \refeq{la0}and \refeq{z0w0} is obtained in this way.
\par
\par
\proclaim Lemma.
Let $f(x_1,\cdots,x_{n-h+1})$ be defined by \refeq{coeff}.
Then
\eq
&f(x_1,\cdots,x_{n-h+1})
=f_0(x_1,\cdots,x_{n-h+1})q^{{1\over2}(n-h-1)(n-h-4)}
+O\bigl(q^{{1\over2}(n-h-1)(n-h-4)}\bigr).
\cr
\endeq
Here $f_0(x_1,\cdots,x_{n-h+1})=0$ unless $(x_1,\cdots,x_{n-h+1})$
is one of the following form.
\item{(1)}There exists $p\geq1$ which satisfies
\eq
&x_{p+1}=x_p+1,\quad x_{i+1}-x_i\geq2\qbox{if $i\neq p,$}
\endeq
where $x_{n-h+1}\neq 2n$ if $p=1$.
\item{(2)}$x_1=1,$ $x_2=2,$ $x_{n-h+1}=2n,$ $x_{i+1}-x_i\geq2$ if $i\neq1$.
\item{(3)} $x_1=1,$ $x_{n-h}=2n-1,$ $x_{n-h+1}=2n,$
$x_{i+1}-x_i\geq2$ if $i\neq n-h$.
\par
\vskip5pt
\par
Let us calculate $f_0(x_1,\cdots,x_{n-h+1})$ in the case of (1) in Lemma.
Let us again use the $\pm$ description and denote by
$(p(i))_{i=1}^{2n}$ be the $\pm$ chain corresponding to
$(x_1,\cdots,x_{n-h+1})$.
\par
\proclaim Lemma.
Suppose that $(p(i))_{i=1}^{2n}$ is of the following form.
\eq
p(2m_i-i-2+x_1)&=p(2m_i-i-1+x_1)=+\qbox{for $1\le i\le l-1,$}
\cr
p(2m_l-l-1+x_1)&=p(2m_l-l+x_1)=-,
\cr
p(2m_i-i+x_1)&=p(2m_i-i+1+x_1)=+\qbox{for $l+1\le i\le 2h,$}
\cr
2\le m_1<&\cdots<m_{2h}\le n+h-1.
\endeq
Set
\eq
&k_j=-m_j+1\qbox{for $1\le j\le 2h$ and $x_1=1,2.$}
\cr
\endeq
Then $p=m_l-l$ and, up to constants independent of $\{x_j\}_{j=1}^{n+h-1}$
\eq
f_0(x_1,\cdots,x_{n-h+1})=&
(-1)^{x_1(n+h)}(\zo)^{2p-x_{p+1}}
(\laio a^{\mu_1})^{-\sum_{j=1}^{n-h+1}x_j+2x_{p+1}-2p}\cr
&a^{x_1\sum_{j=1}^{2h}r_j}\prod_{i=1}^{2h}P_{k_i}(a)
D(r_1,\cdots,r_{2h}|k_1,\cdots,k_{2h}).
\cr
\endeq
where
\eq
&D(r_1,\cdots,r_{2h}|k_1,\cdots,k_{2h})=
\hbox{det}(a^{(i-1)(j-1)})_
{i\neq k_1,\cdots,k_{2h},j\neq r_1,\cdots,r_{2h}}.
\cr
\endeq
\par
\par
Let us define $\ktil_j$ $(1\le j\le 2h)$ by
\eq
\ktil_j&=n+2k_j+j-x_1\qbox{for $1\le j\le l-1,$}
\cr
\ktil_l&=n+2k_l+l-1-x_1,
\cr
\ktil_j&=n+2k_j+j-2-x_1\qbox{for $l+1\le j\le 2h.$}
\cr
\endeq
\par
By taking the limits $n\rightarrow\infty,$ $q\rightarrow0$
in a similar manner as in Proposition \refprop{N3} we have
\par
\proclaim Proposition \prop{N5}.
Suppose $|\ktil_1-\ktil_{2h}|\ll\infty$ and $|\ktil_1|\ll\infty$.
Then in the limit $q\ra0$ and $n\ra \infty$, the coefficients $v(x)$
of $v=\sum v(x)|x\rangle$ is
\eq
&v(x)=\pm\det(z_i^{k_j})_{1\le i,j\le 2h}
\bigl(\prod_{j=1}^{2h}z_j\bigr)^{x_1-{l+x_1\over 2h}}.
\endeq
where $x_1=1$ or $2$ and the signature is same as far as $x_1$ is same.
\par
\par
\par
\noindent{\sl Example 3} ($h=1$, $2$-particle states, spin $0$)\quad
The coefficients of the limit vectors are
\eq
v(1,x_2,\cdots,x_{n-1})&=\pm(z_1^{k_1}z_2^{k_2}-z_1^{k_2}z_2^{k_1})
\qbox{for $l=1,$}
\cr
v(1,x_2,\cdots,x_{n-1})&=\pm(z_1^{k_1}z_2^{k_2}-z_1^{k_2}z_2^{k_1})
(z_1z_2)^{-{1\over 2}}\qbox{for $l=2,$}
\cr
v(2,x_2,\cdots,x_{n-1})&=
\pm(z_1^{k_1}z_2^{k_2}-z_1^{k_2}z_2^{k_1})
(z_1z_2)^{{1\over 2}}
\qbox{for $l=1,$}
\cr
v(2,x_2,\cdots,x_{n-1})&=
\pm(z_1^{k_1}z_2^{k_2}-z_1^{k_2}z_2^{k_1})
\qbox{for $l=2.$}
\endeq
Here
\eq
&0\geq k_1>k_2 \geq -n,
\cr
\endeq
and
\eq
{\ktil}_j&=2k_j+n-x_1,\qbox{for $l=1,$}
\cr
&=2k_j+n+1-x_1,\qbox{for $l=2.$}
\cr
\endeq
\par
The $\pm$-chain pictures are
\eq
&\matrix{
&-&+&-&+&\cdots&+&-&|&-&+&\cdots&-&+&|&+&-&\cdots&-&+&
\cr
&{}&{}&{}&{}&{}&{}&{}&\ktil_1&{}&{}&{}&{}&{}&\ktil_2&{}&{}&{}&{}&{}&,
\cr
}
\endeq
for $l=1,$ $x_1=1$ and
\eq
&\matrix{
&-&+&-&+&\cdots&-&+&|&+&-&\cdots&+&-&|&-&+&\cdots&-&+&
\cr
&{}&{}&{}&{}&{}&{}&{}&\ktil_1&{}&{}&{}&{}&{}&\ktil_2&{}&{}&{}&{}&{}&,
\cr
}
\endeq
for $l=2,$ $x_1=1$.
%\endinput
\par
\par
\par
%\input refmacros
% Standard macros for setting out
%
\catcode`@=11
\newif\ifs@p
\def\refjl#1#2#3#4%
  {#1\def\l@st{#1}\ifx\l@st\empty\s@pfalse\else\s@ptrue\fi%
   \def\l@st{#2}\ifx\l@st\empty\else%
   \ifs@p, \fi{\frenchspacing\sl#2}\s@ptrue\fi%
   \def\l@st{#3}\ifx\l@st\empty\else\ifs@p, \fi{\bf#3}\s@ptrue\fi%
   \def\l@st{#4}\ifx\l@st\empty\else\ifs@p, \fi#4\s@ptrue\fi%
   \ifs@p.\fi\hfill\penalty-9000}
\def\refbk#1#2#3%
  {#1\def\l@st{#1}\ifx\l@st\empty\s@pfalse\else\s@ptrue\fi%
   \def\l@st{#2}\ifx\l@st\empty\else%
   \ifs@p, \fi{\frenchspacing\sl#2}\s@ptrue\fi%
   \def\l@st{#3}\ifx\l@st\empty\else\ifs@p, \fi#3\s@ptrue\fi%
   \ifs@p.\fi\hfill\penalty-9000}
\catcode`@=12
%
% abbreviations for commonly quoted journals
%
\def\APNY{Ann. Phys., NY}

\def\CMP{Commun. Math. Phys.}
\def\Duke{Duke Math. J.}

\def\JMP{J. Math. Phys.}
\def\JPA{J. Phys. A: Math. Gen.}
\def\JSP{J. Stat. Phys.}

\def\NPB{Nucl. Phys. B}

\def\PL{Phys. Lett.}

\def\PrAMS{Proc. Amer. Math. Soc.}

\def\PRB{Phys. Rev. B}

\def\RIMS{RIMS preprint}

\par
%\endinput
\par
\par
\par
\bigskip\noindent{\bf References}\medskip
\par
\refis{Sm} \refjl
{Smirnov F A,
Dynamical symmetries of massive integrable models}
{\RIMS}{772, 838}{(1991)}
\par
\refis{Sm2} \refbk
{Smirnov F A,
Form factors in completely integrable models of quantum field theory}
{Advanced Series in Mathematical Physics {\bf 14}}
{World Scientific, Singapore 1992}
\par
\refis{BerLeC} \refjl
{Bernard D and LeClair A,
Quantum Group Symmetries and Non-Local Currents in 2D QFT}
{\CMP}{142}{(1991) 99-138}
\par
\refis{FelLeC} \refjl
{Felder G and LeClair A,
Restricted Quantum Affine Symmetry of Restricted Minimal Conformal Models}
{\RIMS} {799} {(1991)}
\par
\refis{EKS} \refjl
{E\ss ler F H L, Korepin V E and Schoutens K,
Fine structure of the Bethe Ansatz for the spin-$1\over 2$
Heisenberg $XXX$ model}
{preprint}{}{(1992) }
\par
\refis{Zamol} \refbk
{Zamolodchikov A B,
Integrable field theory from conformal field theory}
{Advanced Studies in Pure Mathematics {\bf 19}}{Kinokuniya, Tokyo 1989}
\par
\refis{BPZ} \refjl
{Belavin A A, Polyakov A M and Zamolodchikov A B,
Infinite conformal symmetry in two-dimensional quantum field theory}
{\NPB}{241}{(1984) 333--380}
\par
\refis{PS} \refjl
{Pasquier V and Saleur H,
Common structures between finite size systems and conformal field theories
through quantum groups}
{\NPB}{330}{(1990) 523--556}
\par
\refis{Baxegv} \refjl
{Baxter R J,
Eight-vertex model in lattice statistics and one-dimensional
anisotropic Heisenberg chain I--III}
{\APNY}{76}{(1973), 1--24, 25--47, 48--71}
\par
\refis{BaxCTM} \refjl
{Baxter R J,
Corner transfer matrices of the eight vertex model.
Low-temperature expansions and conjectured properties}
{\JSP}{15}{(1976) 485--503}
\par
\refis{Ka} \refjl
{Kashiwara M,
On crystal bases of the $q$-analogue of universal enveloping algebras}
{\Duke}{63}{(1991) 465--516}

\par
\refis{Ka2} \refjl
{Kashiwara M,
Global crystal bases of quantum groups}
{\RIMS}{756}{(1991)}
\par
\def\uq{U_q\bigl(\widehat{\goth{sl}}\hskip2pt(n)\bigr)}
\par
\refis{MM} \refjl
{Misra K C and Miwa T,
Crystal base for the basic representation of $\uq$}
{\CMP}{134}{(1990) 79--88}
\par
\refis{JMMO} \refjl
{Jimbo M, Misra K C, Miwa T and Okado M,
Combinatorics of representations of $\uq$ at $q=0$}
{\CMP}{136}{(1991) 543--566}
\par
\refis{(KMN)^2} \refjl
{Kang S-J, Kashiwara M, Misra K, Miwa T, Nakashima T and Nakayashiki A,
Affine crystals and vertex models}
{\RIMS}{836}{1991}
\par
\refis{FM} \refjl
{Foda O and Miwa T,
Corner transfer matrices and quantum affine algebras}
{\RIMS}{836}{(1991)}
\par
\refis{DJO} \refjl
{Date E, Jimbo M and Okado M,
Crystal base and $q$ vertex operators}
{Osaka Univ. Math. Sci. preprint}{1}{(1991)}
\par
\refis{FR} \refjl
{Frenkel I B and Reshetikhin N Yu,
Quantum affine algebras and holonomic difference equations}
{preprint}{}{(1991)}
\par
\refis{FT} \refjl
{Frahm H and Thacker H B,
Corner transfer matrix eigenstates for the six vertex model}
{\JPA}{24}{(1991) 5587--5603}
\par
\refis{Tetel} \refjl
{Tetel'man M G,
Lorentz group for two-dimensional integrable lattice systems}
{Zh. Eksp. Teor. Fiz.} {82} {(1982) 528--535}
\par
\refis{Bax} \refjl
{Baxter R J,
Spontaneous staggered polarization of the $F$ model}
{\JSP}{9}{(1973) 145--182}
%Spontaneous polarization paper
\par
\refis{Baxbk} \refbk
{Baxter R J}
{Exactly solved models in statistical mechanics}
{Academic Press, London 1982}
\par
\refis{DesL} \refjl
{Destri C and Lowenstein J H,
Analysis of the Bethe-Ansatz equations of the chiral-invariant
Gross-Neveu model}
{\NPB}{205}{[FS5] (1982) 369--385}
\par
\refis{FaddeevT} \refjl
{Faddeev L and Takhatajan L A,
Spectrum and scattering of excitations in the one-dimensional
Heisenberg model}
{J. Soviet Math.}{24}{(1984) 241--167}
\par
\refis{Bab} \refjl
{Babelon O, de Vega H J and Viallet C M,
Analysis of the Bethe Ansatz equations of the $XXZ$ model}
{\NPB}{220}{[FS8] (1983) 13--34}
\par
\refis{dCG} \refjl
{des Cloiseau and Gaudin,
Anisotropic linear magnetic chain}
{\JMP}
{7}{(1966)1384--1400}
\par
\refis{KirSm} \refjl
{Kirillov A N and Smirnov F A,
Form factors in the $SU(2)$-invariant Thirring model}
{J. Soviet  Math.}{47} {(1989) 2423--2450}
\par
\refis{Bir} \refjl
{Birkhoff G D,
The generalized Riemann problem for linear differential equations and the
allied problems for linear difference and $q$-difference equations}
{Proc. Am. Acad. Arts and Sci.}{49}{(1914) 521--568}
\par
\refis{Ao} \refbk
{Aomoto K,
A note on holonomic $q$-difference systems}
{Algebraic Analysis, vol.I, Eds. Kashiwara M and Kawai T, pp.25--28}
{Academic Press, San Diego 1988}
\par
\refis{WW} \refbk
{Whittaker E T and Watson G N}
{Modern Analysis, 4th ed.}{Cambridge University Press, London 1962}
\par
\refis{Lus} \refjl
{Lusztig G,
Equivariant $K$-theory and representations of Hecke algebras}
{\PrAMS}{94}{(1985)337--342}
\par
\refis{JK} \refbk
{James G D and Kerber A}
{The representation theory of the symmetric group}
{Addison-Wesley, Reading 1981}
\par
\refis{La} \refjl
{Lusztig G,
Quivers, perverse sheaves, and quantized enveloping algebras}
{J. of A.M.S.}{4}{(1991) 365--421}
\par
\refis{DJMO} \refjl
{Date, E., Jimbo, M., Kuniba, A., Miwa, T. and Okado, M,
One dimensional configuration sums in vertex models and affine Lie
algebra characters}
{Lett. Math. Phys.} {17} {(1989) 69--77}
\par
\refis{DJKMO} \refjl
{Date E, Jimbo M, Kuniba A, Miwa T and Okado M,
Exactly solvable SOS models:
Local height probabilities and theta function identities}
{\NPB}{290}{[FS20] (1987) 231--273}
%II. Proof of the star-triangle relation and combinatorial identities,
%Adv. Stud. Pure Math. {\bf 16} (1988), 17-122.
\par
\refis{WMTB} \refjl
{Wu T T, McCoy B M, Tracy C A and Barouch E,
Spin-spin correlation functions for the two-dimensional Ising
model: Exact theory in the scaling region}{\PRB}{13}{(1976) 316--374}
\par
\refis{SMJ} \refjl
{Sato M, Miwa T and Jimbo M,
Holonomic quantum fields I--V}
{Publ. RIMS}{14--17}{(1978) 223--267, (1979) 201--278, 577--629, 871--972,
(1980) 531--584}
\par
\refis{TK} \refbk
{Tsuchiya A and Kanie Y,
Vertex operators in
conformal field theory on ${\bf P}^1$ and monodromy
representations of braid group}
{Advanced Studies in Pure Mathematics {\bf 16}}{Kinokuniya, Tokyo 1988}
\par
\refis{CorDor} \refjl
{Corrigan E and Dorey P, A representation of the exchange
relation for affine Toda field theory}{\PL}{B273}{(1991) 237--245}
\par
\refis{FeiFre} \refjl
{Feigin B and Frenkel E, Free field resolutions in affine Toda field theories}
{\PL}{B276}{(1992) 79--86}
\par
\refis{Resh} \refjl
{Reshetikhin N Yu,
$S$-matrices in integrable models of isotropic magnetic chains:I}
{\JPA}{24}{(1991) 3299-3309}
\par
\refis{KasMi} \refjl
{Kashiwara M and Miwa T,
A class of elliptic solutions to the star-triangle relation}
{\NPB}{275}{[FS17] (1986) 121-134}
\par
\refis{HY} \refjl
{Hasegawa K and Yamada Y,
Algebraic derivation of the broken $Z_N$ symmetric model}
{\PL}{146A}{(1990) 387--396}
\par
\refis{CPot1} \refjl
{Au-Yang H, McCoy B M, Perk J H H,  Tang S and Yan M-L,
Commuting transfer matrices in the chiral Potts models: Solutions of
star--triangle equations with genus $> 1$}
{\PL}{123A}{ (1987) 219--223}
\par
\refis{CPot2} \refjl
{Baxter R J, Perk J H H and Au-Yang H,
New solutions of the star--triangle relations for the chiral Potts model}
{\PL}{128A}{(1988) 138--142}
\par
\eject
\par
\par
\listreferences
\par
\par
%
%
%
%\refis{D88} Davies B (1988) Corner Transfer Matrices for the
%Ising Model, Physica A 154, 1-20.
%
%\refis{DK86} Davies B and Kieu T D (1986) The Quantum Inverse Method
%and the non-linear Schr\"odinger equation, Inverse Problems 2,
%141-159.
%
%
%
%
%
%
\par
%\endinput
\par
%\input table
%\input macro
%\onebox-----one box
%\twobox-----vertical two boxes
%\threebox---vertical three boxes
%\fourbox----vertical four boxes
%\four-------horizontal four boxes
%\fourtwo----1st line (four boxes) + 2nd line (two boxes)
%\threeone---1st line (three boxes) + 2nd line (one box)
%\threetwo---1st line (three boxes) + 2nd line (two boxes)
%\twoone-----1st line (two boxes) + 2nd line (one box)
%
\def\pn{\par\noindent}
\def\m@th{\mathsurround=0pt}
\par
\def\Fsquare(#1,#2){
\hbox{\vrule$\hskip-0.4pt\vcenter to #1{\normalbaselines\m@th
\hrule\vfil\hbox to #1{\hfill$#2$\hfill}\vfil\hrule}$\hskip-0.4pt
\vrule}}
\par
\def\Addsquare(#1,#2){\hbox{$
	\dimen1=#1 \advance\dimen1 by -0.8pt
	\vcenter to #1{\hrule height0.4pt depth0.0pt\vss%
	\hbox to #1{\hss{%
	\vbox to \dimen1{\vss%
	\hbox to \dimen1{\hss$~#2~$\hss}%
	\vss}\hss}%
	\vrule width0.4pt}\vss%
	\hrule height0.4pt depth0.0pt}$}}
\par
\def\bl{
        \dimen3=0.4cm \advance\dimen3 by -0.8pt
        \vrule height\dimen3 width\dimen3 depth0cm}
\par
\def\fourbox(#1,#2,#3,#4){%
%	\hbox{
	\normalbaselines\m@th\offinterlineskip
	\vcenter{\hbox{\Fsquare(0.4cm,#1)}
	      \vskip-0.4pt
	      \hbox{\Fsquare(0.4cm,#2)}
	      \vskip-0.4pt
	      \hbox{\Fsquare(0.4cm,#3)}
	      \vskip-0.4pt
	      \hbox{\Fsquare(0.4cm,#4)}}}%}
\par
\def\threebox(#1,#2,#3){%
	\normalbaselines\m@th\offinterlineskip
	\vcenter{\hbox{\Fsquare(0.4cm,#1)}
	      \vskip-0.4pt
	      \hbox{\Fsquare(0.4cm,#2)}
	      \vskip-0.4pt
	      \hbox{\Fsquare(0.4cm,#3)}}}
\par
\def\twobox(#1,#2){%
	\normalbaselines\m@th\offinterlineskip
	\vcenter{\hbox{\Fsquare(0.4cm,#1)}
	      \vskip-0.4pt
	      \hbox{\Fsquare(0.4cm,#2)}}}
\par
\def\onebox(#1){%
	\normalbaselines\m@th\offinterlineskip
	\vcenter{\hbox{\Fsquare(0.4cm,#1)}}}
\par
%---- yoko ---
\def\Htwobox(#1,#2){%
	\Fsquare(0.4cm,#1)\Addsquare(0.4cm,#2)}
\par
\def\Hthreebox(#1,#2,#3){%
	\Fsquare(0.4cm,#1)\Addsquare(0.4cm,#2)\Addsquare(0.4cm,#3)}
\par
\def\four(#1,#2,#3,#4){%
\Fsquare(0.4cm,#1)\Addsquare(0.4cm,#2)\Addsquare(0.4cm,#3)\Addsquare(0.4cm,#4)}
\par
\def\twoone(#1,#2,#3){%
%	\hbox{
	\normalbaselines\m@th\offinterlineskip
	\vcenter{\hbox{\Htwobox({#1},{#2})}
	      \vskip-0.4pt
	      \hbox{\Fsquare(0.4cm,#3)}}}
\par
\def\threeone(#1,#2,#3,#4){%
	\normalbaselines\m@th\offinterlineskip
	\vcenter{\hbox{\Hthreebox({#1},{#2},{#3})}
	      \vskip-0.4pt
	      \hbox{\Fsquare(0.4cm,#4)}}}
\par
\def\threetwo(#1,#2,#3,#4,#5){%
        \normalbaselines\m@th\offinterlineskip
	\vcenter{\hbox{\Hthreebox({#1},{#2},{#3})}
	      \vskip-0.4pt
	      \hbox{\Htwobox({#4},{#5})}}}
\par
\def\fourtwo(#1,#2,#3,#4,#5,#6){%
	\normalbaselines\m@th\offinterlineskip
	\vcenter{\hbox{\four({#1},{#2},{#3},{#4})}
	      \vskip-0.4pt
	      \hbox{\Htwobox({#5},{#6})}}}
\par
\eject
\par
\def\bl{\bullet}
\def\eq#1\endeq{$$\eqalignno{#1}$$}
\def\aku{\hskip1.3cm}
\def\gyoake{\noalign{\bigskip}}
\def\sk{\hskip 2cm}
\pn{\bf Table 1.}
\bigskip
\par
\eq &\sk
e_0\phi=0             \aku
e_1\phi=0              \aku
f_0\phi=(1)\onebox( )  \aku
f_1\phi=0              \aku  \cr
\gyoake &\sk
e_0\onebox( )=(-1)\onebox(\bl) \aku
e_0\onebox(\bl)=0              \aku
e_1\onebox( )=0                \aku
e_1\onebox(\bl)=(-1)\onebox( ) \aku \cr
\gyoake &\sk
f_0\onebox( )=0                   \aku
f_0\onebox(\bl)=(1)\onebox( )  \aku
f_1\onebox( )=[2](1)\onebox(\bl)+\pmatrix{0\cr1\cr}\twobox( ,\bl)  \aku
f_1\onebox(\bl)=\pmatrix{0\cr1\cr}\twobox(\bl,\bl)   \aku \cr\gyoake &\sk
e_0\twobox( , )=[2]\pmatrix{0\cr-1\cr}\twobox( ,\bl)
       +\left\{\pmatrix{-1\cr0\cr}+\pmatrix{1\cr-2\cr}\right\}
       \twobox(\bl, )          \aku
e_0\twobox(\bl, )=0            \aku \cr
\gyoake &\sk\hskip 8cm
e_0\twobox( ,\bl)=\pmatrix{-1\cr0\cr}\twobox(\bl,\bl)   \aku
e_0\twobox(\bl,\bl)=0          \aku \cr
\gyoake &\sk
e_1\twobox( , )=0              \aku
e_1\twobox(\bl, )=\pmatrix{-1\cr0\cr}\twobox( , ) \aku
e_1\twobox( ,\bl)=0            \aku \cr
\gyoake &\sk\hskip 7cm
e_1\twobox(\bl,\bl)=[2]\pmatrix{0\cr-1\cr}\twobox(\bl, )
       +\left\{\pmatrix{-1\cr0\cr}+\pmatrix{1\cr-2\cr}\right\}
       \twobox( ,\bl)          \aku \cr
\gyoake &\sk
f_0\twobox( , )=\pmatrix{0\cr0\cr1\cr}\threebox( , , )        \aku
f_0\twobox(\bl, )=\pmatrix{0\cr0\cr1\cr}\threebox(\bl, , )    \aku \cr
\gyoake &\sk
f_0\twobox( ,\bl)=[2]\pmatrix{0\cr1\cr}\twobox( , )
       +\pmatrix{0\cr0\cr1}\threebox( ,\bl, )                 \aku
f_0\twobox(\bl,\bl)=[3]\pmatrix{1\cr0\cr}\twobox( ,\bl)
       +[2]\left\{\pmatrix{0\cr1\cr}+\pmatrix{2\cr-1\cr}\right\}
       \twobox(\bl, )+\pmatrix{0\cr0\cr1\cr}\threebox(\bl,\bl, )  \aku \cr
\gyoake &\sk
f_1\twobox( , )=[2]\pmatrix{1\cr0\cr}\twobox(\bl, )
       +\left\{\pmatrix{0\cr1\cr}+\pmatrix{2\cr-1\cr}\right\}
       \twobox( ,\bl)                                           \aku
f_1\twobox(\bl, )=\pmatrix{0\cr1\cr}\twobox(\bl,\bl)            \aku \cr
\gyoake &\sk\hskip 8cm
f_1\twobox( ,\bl)=0                                             \aku
f_1\twobox(\bl,\bl)=0                                           \aku \cr
\gyoake &\sk
e_0\threebox( , , )=[3]\pmatrix{0\cr0\cr-1\cr}\threebox( , ,\bl)
       +[2]\left\{\pmatrix{0\cr-1\cr0\cr}+\pmatrix{2\cr-1\cr-2\cr}
       +\pmatrix{0\cr1\cr-2\cr}\right\}\threebox( ,\bl, ) \cr
&\sk \hskip 7cm
       +\left\{\pmatrix{-1\cr0\cr0\cr}+\pmatrix{1\cr-2\cr0\cr}
       +\pmatrix{1\cr0\cr-2\cr}\right\}\threebox(\bl, , )
       +\pmatrix{2\cr-2\cr-1\cr}\onebox( )                      \aku \cr
\gyoake &\sk
%e_0\threebox( , , )=[3]\pmatrix{0\cr0\cr-1\cr}\twobox( , )
%       +\pmatrix{-1\cr0\cr0\cr}\threebox(\bl, , )
%       +\pmatrix{2\cr-1\cr-1\cr}\onebox( )                      \aku
e_0\threebox(\bl, , )=\pmatrix{0\cr0\cr-1\cr}
       \threebox(\bl, ,\bl)                                     \aku
e_0\threebox( ,\bl, )=\pmatrix{-1\cr0\cr0\cr}
       \threebox(\bl,\bl, )                                     \aku \cr
\gyoake &\sk
e_0\threebox( , ,\bl)=[2]\pmatrix{0\cr-1\cr0\cr}
       \threebox( ,\bl,\bl)+\left\{\pmatrix{-1\cr0\cr0\cr}
       +\pmatrix{1\cr-2\cr0\cr}+\pmatrix{1\cr0\cr-2\cr}
       \right\}\threebox(\bl, ,\bl)                             \aku \cr
\gyoake &\sk
e_0\threebox(\bl,\bl, )=0                                       \aku
e_0\threebox(\bl, ,\bl)=0                                       \aku
e_0\threebox( ,\bl,\bl)=\pmatrix{-1\cr0\cr0\cr}
       \threebox(\bl,\bl,\bl)                                   \aku
e_0\threebox(\bl,\bl,\bl)=0                                     \aku \cr
\gyoake &\sk
e_1\threebox( , , )=0                                           \aku
e_1\threebox(\bl, , )=\pmatrix{-1\cr0\cr0\cr}
       \threebox( , , )                                         \aku
e_1\threebox( ,\bl, )=0                                         \aku
e_1\threebox( , ,\bl)=0                                         \aku \cr
\gyoake &\sk
e_1\threebox(\bl,\bl, )=[2]\pmatrix{0\cr-1\cr0\cr}
       \threebox(\bl, , )+\left\{\pmatrix{-1\cr0\cr0\cr}
       +\pmatrix{1\cr-2\cr0\cr}+\pmatrix{1\cr0\cr-2\cr}
       \right\}\threebox( ,\bl, )                               \aku \cr
\gyoake &\sk
e_1\threebox(\bl, ,\bl)=\pmatrix{-1\cr0\cr0\cr}
       \threebox( , ,\bl)                                       \aku
e_1\threebox( ,\bl,\bl)=\pmatrix{0\cr0\cr-1\cr}
       \threebox( ,\bl, )                                       \aku \cr
\gyoake &\sk
e_1\threebox(\bl,\bl,\bl)=[3]\pmatrix{0\cr0\cr-1\cr}
       \threebox(\bl,\bl, )+[2]\left\{\pmatrix{0\cr-1\cr0\cr}
       +\pmatrix{2\cr-1\cr-2\cr}+\pmatrix{0\cr1\cr-2\cr}
       \right\}\threebox(\bl, ,\bl) 				\cr
\gyoake &\sk\hskip 7cm
       +\left\{\pmatrix{-1\cr0\cr0\cr}+\pmatrix{1\cr-2\cr0\cr}
       +\pmatrix{1\cr0\cr-2\cr}\right\}
       \threebox( ,\bl,\bl)                                     \aku \cr
\gyoake &\sk
f_0\threebox( , , )=0                                           \aku
f_0\threebox(\bl, , )=0                                         \aku
f_0\threebox( ,\bl, )=0                                         \aku
f_0\threebox( , ,\bl)=\pmatrix{0\cr0\cr1\cr}\threebox( , , )    \aku \cr
\gyoake &\sk
f_0\threebox(\bl,\bl, )=\pmatrix{1\cr0\cr0\cr}
       \threebox( ,\bl, )                                       \aku
f_0\threebox(\bl, ,\bl)=\pmatrix{0\cr0\cr1\cr}
       \threebox(\bl, , )                                       \aku \cr
\gyoake &\sk
f_0\threebox( ,\bl,\bl)=[2]\pmatrix{0\cr1\cr0\cr}
       \threebox( , ,\bl)+\left\{\pmatrix{0\cr0\cr1\cr}
       +\pmatrix{2\cr0\cr-1\cr}+\pmatrix{0\cr2\cr-1\cr}
       \right\}\threebox( ,\bl, )                               \aku \cr
\gyoake &\sk
f_0\threebox(\bl,\bl,\bl)=[3]\pmatrix{1\cr0\cr0\cr}
       \threebox( ,\bl,\bl)+[2]\left\{
       \pmatrix{0\cr1\cr0\cr}+\pmatrix{2\cr-1\cr0\cr}
       +\pmatrix{2\cr1\cr-2\cr}\right\}\threebox(\bl, ,\bl) \cr
\gyoake &\sk \hskip7cm
       +\left\{\pmatrix{0\cr0\cr1\cr}+\pmatrix{2\cr0\cr-1\cr}
       +\pmatrix{0\cr2\cr-1\cr}\right\}\threebox(\bl,\bl, )     \aku \cr
\gyoake &\sk
f_1\threebox( , , )=[4]\pmatrix{1\cr0\cr0\cr}
       \threebox(\bl, , )+[3]\left\{\pmatrix{0\cr1\cr0\cr}
       +\pmatrix{2\cr-1\cr0\cr}+\pmatrix{2\cr1\cr-2\cr}
       \right\}\threebox( ,\bl, )				\cr
\gyoake &\sk
       +[2]\left\{\pmatrix{0\cr0\cr1\cr}
       +\pmatrix{2\cr0\cr-1\cr}+\pmatrix{0\cr2\cr-1\cr}
       \right\}\threebox( , ,\bl)+\pmatrix{0\cr0\cr0\cr1\cr}
       \fourbox( , , ,\bl)                                      \aku \cr
\gyoake &\sk
f_1\threebox(\bl, , )=[3]\pmatrix{0\cr1\cr0\cr}
       \threebox(\bl,\bl, )+[2]\left\{\pmatrix{0\cr0\cr1\cr}
       +\pmatrix{2\cr0\cr-1\cr}+\pmatrix{0\cr2\cr-1\cr}
       \right\}\threebox(\bl, ,\bl)+\pmatrix{0\cr0\cr0\cr1\cr}
       \fourbox(\bl, , ,\bl)
       +\pmatrix{1\cr1\cr-1\cr}\twobox( ,\bl)                   \aku \cr
                                                                \gyoake &\sk
%f_1\threebox(\bl, , )=[3]\pmatrix{0\cr1\cr0\cr}
%       \threebox(\bl,\bl, )+[2]\left\{\pmatrix{2\cr0\cr-1\cr}
%       +\pmatrix{0\cr2\cr-1\cr}\right\}\twobox(\bl, )
%       +\pmatrix{1\cr1\cr-1\cr}\twobox( ,\bl)+[2]
%       \pmatrix{0\cr0\cr1\cr}\threebox(\bl, ,\bl)               \aku \cr
%                                                                \gyoake &\sk
f_1\threebox( ,\bl, )=[2]\pmatrix{0\cr0\cr1\cr}
       \threebox( ,\bl,\bl)+\pmatrix{0\cr0\cr0\cr1\cr}
       \fourbox( ,\bl, ,\bl)                                    \aku
f_1\threebox( , ,\bl)=[2]\pmatrix{1\cr0\cr0\cr}
       \threebox(\bl, ,\bl)+\pmatrix{0\cr0\cr0\cr1\cr}
       \fourbox( , ,\bl,\bl)                                    \aku \cr
\gyoake &\sk
f_1\threebox(\bl,\bl, )=[2]\pmatrix{0\cr0\cr1\cr}
       \threebox(\bl,\bl,\bl)+\pmatrix{0\cr0\cr0\cr1\cr}
       \fourbox(\bl,\bl, ,\bl)                                  \aku \cr
\gyoake &\sk
f_1\threebox(\bl, ,\bl)=\pmatrix{0\cr0\cr0\cr1\cr}
       \fourbox(\bl, ,\bl,\bl)                                  \aku
f_1\threebox( ,\bl,\bl)=\pmatrix{0\cr0\cr0\cr1\cr}
       \fourbox( ,\bl,\bl,\bl)                                  \aku
f_1\threebox(\bl,\bl,\bl)=\pmatrix{0\cr0\cr0\cr1\cr}
       \fourbox(\bl,\bl,\bl,\bl)                                \aku \cr
\endeq
\par
\eject
%\endinput
\par
\def\eq#1\endeq{$$\eqalignno{#1}$$}
\def\aku{\hskip1.3cm}
\def\gyoake{\noalign{\bigskip}}
\def\P{\Phi^\sigma}
\pn{\bf Table 2.}
\par
%1
\eq
&\sk
\P\bigl(v_+\otimes\phi\bigr)=\sum^\infty_{n=0}z^n(2n)\onebox(\bl)
\aku \cr\gyoake &\sk
%2
\P\bigl(v_-\otimes\phi\bigr)=\sum^\infty_{n=0}z^n(2n+1)\onebox( )
  \aku \cr\gyoake &\sk
%3
\P\bigl(v_+\otimes(1){\onebox( )}\bigr)
=\sum^\infty_{n=1}z^n{2n\choose1}
  \twobox(\bl,\bl)                                         \aku \cr\gyoake &\sk
%4
\P\bigl(v_-\otimes(1){\onebox( )}\bigr)=-z^{-1}q^{-1}\phi
   +\sum^\infty_{n=0}z^nq(2n+2)\onebox(\bl)
   +\sum^\infty_{n=1}z^n{2n+1\choose1}
    \twobox( ,\bl)                                         \aku \cr\gyoake &\sk
%5
\P\bigl(v_+\otimes(2)\onebox(\bl)\bigr)
=z^{-1}{q^{-3}\over[2]}\phi-(2)\onebox(\bl)
   +\sum^\infty_{n=0}z^n{q\over[2]}(2n+2)\onebox(\bl)
   +\sum^\infty_{n=1}z^nq{2n+1\choose1}\twobox( ,\bl)
   +\sum^\infty_{n=2}z^n{2n\choose2}
    \twobox(\bl, )                                         \aku \cr\gyoake &\sk
%6
\P\bigl(v_-\otimes(2)\onebox(\bl)\bigr)=-z^{-1}{q^{-1}\over[2]}(1)\onebox( )
   +\sum^\infty_{n=0}z^n{q\over[2]}(2n+3)\onebox( )
   +\sum^\infty_{n=1}z^n{2n+1\choose2}
    \twobox( , )                                            \aku \cr\gyoake
&\sk
%7
\P\bigr(v_+\otimes(3)\onebox({})\bigr)=-\pmatrix{2\cr1\cr}\twobox(\bl,\bl)
   +\sum^\infty_{n=0}z^n{q\over[2]}{2n+2\choose1}\twobox(\bl,\bl)
   +\sum^\infty_{n=2}z^n{2n\choose3}
    \twobox(\bl,\bl)                                         \aku \cr\gyoake
&\sk
%8
\P\bigl(v_-\otimes(3)\onebox({})\bigr)=
    -z^{-2}{q^{-4}\over[2]}\phi
    -z^{-1}{1\over[2]}(2)\onebox(\bl)
   -\pmatrix{3\cr1\cr}\twobox({},\bl)
   +\sum^\infty_{n=0}z^n{q^2\over[2]}(2n+4)\onebox(\bl)
+\sum^\infty_{n=0}z^n{q\over[2]}{2n+3\choose1}\twobox( ,\bl)  \aku \cr\gyoake
&\sk
   \qquad+\sum^\infty_{n=1}z^nq{2n+2\choose2}\twobox(\bl, )
   +\sum^\infty_{n=2}z^n{2n+1\choose3}\twobox( ,\bl)          \aku \cr\gyoake
&\sk
%9
\P\bigl(v_+\otimes\pmatrix{2\cr1\cr}\twobox(\bl,{})\bigr)
=z^{-1}q^{-2}(1)\onebox( )
    -(3)\onebox( )
    +\sum^\infty_{n=1}z^nq^2{2n+1\choose2}\twobox( , )
    +\sum^\infty_{n=2}z^n \left(\matrix{2n\cr 2\cr 1\cr} \right)
    \threebox(\bl, , )                                  \aku \cr\gyoake &\sk
%10
\P\bigl(v_-\otimes\pmatrix{2\cr1\cr}\twobox(\bl, )\bigr)
=\sum^\infty_{n=1}z^n
    \left(\matrix{2n+1\cr 2\cr 1\cr} \right)
    \threebox( , , )                                \aku \cr\gyoake &\sk
%11
\P\bigl(v_+\otimes(4)\onebox(\bl)\bigr)=
    z^{-2}{q^{-6}[3]\over[2]^2([3]-1)}\phi
   -z^{-1}{q^{-2}\over[2]^2([3]-1)}
   (2)\onebox(\bl)-(4)\onebox(\bl)-q\pmatrix{3\cr1\cr}\twobox({},\bl)
      -z\pmatrix{4\cr2\cr}\twobox(\bl, )            \aku \cr\gyoake &\sk
   \qquad+\sum^\infty_{n=0}z^n{q^2[3]\over[2]^2([3]-1)}(2n+4)\onebox(\bl)
   +\sum^\infty_{n=0}z^n{q^2\over[2]}{2n+3\choose1}\twobox( ,\bl)
   +\sum^\infty_{n=1}z^n{q\over[2]}{2n+2\choose2}\twobox(\bl, )
            \aku \cr\gyoake &\sk
  \qquad+\sum^\infty_{n=2}z^nq{2n+1\choose3}\twobox( ,\bl)
   +\sum^\infty_{n=3}z^n{2n\choose4}\twobox(\bl, )
            \aku \cr\gyoake &\sk
%12
\P\bigl(v_-\otimes(4)\onebox(\bl)\bigr)=
   -z^{-2}{q^{-2}\over[2]^2([3]-1)}(1)\onebox( )
   -z^{-1}{[3]+q^{-3}[2]\over[2]^2([3]-1)}(3)\onebox( )
   -\pmatrix{3\cr2\cr}\twobox( , )
\aku \cr\gyoake &\sk
   \qquad+\sum^\infty_{n=0}{z^nq^2[3]\over[2]^2([3]-1)}(2n+5)\onebox( )
   +\sum^\infty_{n=0}z^n{q\over[2]}{2n+3\choose2}\twobox( , )
   +\sum^\infty_{n=2}z^n{2n+1\choose4}\twobox( , )
\aku \cr\gyoake &\sk
%13
\P\bigl(v_+\otimes\pmatrix{3\cr1\cr}\twobox({},\bl)\bigr)=
    -z^{-2}{q^{-6}\over[2]([3]-1)}\phi
   +{z^{-1}q^{-2}[3]\over[2]([3]-1)}(2)\onebox(\bl)
   -\sum^\infty_{n=0}{z^nq^2\over[2]([3]-1)}(2n+4)\onebox(\bl)
\aku \cr\gyoake &\sk
   \qquad+\sum^\infty_{n=1}z^nq{2n+2\choose2}\twobox(\bl, )
   +\sum^\infty_{n=2}z^n\left(\matrix{2n \cr 3 \cr 1\cr }\right)
   \threebox(\bl,\bl, )
\aku \cr\gyoake &\sk
%14
\P\bigl(v_-\otimes\pmatrix{3\cr1\cr}\twobox({},\bl)\bigr)
   = -{z^{-2}q^{-6}\over[2]([3]-1)}(1)\onebox( )
     +{z^{-1}q^{-2}[3]\over[2]([3]-1)}(3)\onebox( )
     -\sum^\infty_{n=0}{z^nq^2\over[2]([3]-1)}(2n+5)\onebox( )\cr
\aku \cr\gyoake &\sk
\qquad+\sum^\infty_{n=1}z^nq\left(\matrix{2n+2\cr 2\cr 1\cr }\right)
      \threebox(\bl, , )
     +\sum^\infty_{n=2}z^n\left(\matrix{2n+1\cr 3\cr 1\cr }\right)
      \threebox( ,\bl, )
\aku \cr
\endeq
\par
%\endinput
\par
\eject
\par
\def\aku{\hskip1.3cm}
\def\gyoake{\noalign{\bigskip}}
\def\P{\Phi^\sigma}
\par
\noindent{\bf Table 3.}\bigskip
%1
\eq
&
\P_+\bigl(\phi\bigr)=-\sum^\infty_{n=0}q^{3n+1}(2n+1)\onebox()
\qquad
%\aku \cr\gyoake &
%2
\P_-\bigl(\phi\bigr)=\sum^\infty_{n=0}q^{3n}(2n)\onebox(\bl)
  \aku \cr\gyoake &
%3
\P_+\bigl((1){\onebox( )}\bigr)
=\phi
-\sum^\infty_{n=0}q^{3n+1}(2n+2)\onebox(\bl)
-\sum^\infty_{n=1}q^{3n+2}{2n+1\choose1}
  \twobox( ,\bl)                                         \aku \cr\gyoake &
%4
\P_-\bigl((1){\onebox( )}\bigr)=
   \sum^\infty_{n=1}q^{3n-1}{2n\choose1}
    \twobox(\bl,\bl)                                         \aku \cr\gyoake &
%5
\P_+\bigl((2)\onebox(\bl)\bigr)
={q^{-1}\over[2]}(1)\onebox()
   -\sum^\infty_{n=0}{q^{3n}\over[2]}(2n+3)\onebox( )
   -\sum^\infty_{n=1}q^{3n+1}{2n+1\choose2}\twobox( , )
                                                 \aku \cr\gyoake &
%6
\P_-\bigl((2)\onebox(\bl)\bigr)={1\over[2]}\phi
 -{q\over[2]}(2)\onebox(\bl)
   +\sum^\infty_{n=1}{q^{3n-1}\over[2]}(2n+2)\onebox(\bl)
                                              \aku \cr\gyoake &
   +\sum^\infty_{n=1}q^{3n-1}{2n+1\choose1}
    \twobox( ,\bl)
  +\sum^\infty_{n=2}q^{3n}{2n\choose2}
    \twobox(\bl, )                                            \aku \cr\gyoake &
%7
\P_+\bigl((3)\onebox({})\bigr)=
{1\over[2]}\phi
+{q^{-1}\over[2]}(2)\onebox(\bl)
-\sum^\infty_{n=0}{q^{3n}\over[2]}(2n+4)\onebox(\bl)
                                         \aku \cr\gyoake &
+{q^3\over[2]}\pmatrix{3\cr1\cr}\twobox( ,\bl)
-\sum^\infty_{n=1}{q^{3n+1}\over[2]}{2n+3\choose1}\twobox( ,\bl)
                                         \aku \cr\gyoake &
   -\sum^\infty_{n=2}q^{3n+2}{2n+1\choose3}
    \twobox( ,\bl)
   -\sum^\infty_{n=1}q^{3n+1}{2n+2\choose2}
    \twobox(\bl, )
                                             \aku \cr\gyoake &
%8
\P_-\bigl((3)\onebox({})\bigr)=
   -{1\over[2]}\pmatrix{2\cr1\cr}\twobox(\bl,\bl)
   +\sum^\infty_{n=1}{q^{3n-2}\over[2]}{2n+2\choose1}\twobox(\bl,\bl)
      \aku \cr\gyoake &
   +\sum^\infty_{n=2}q^{3n-1}{2n\choose3}\twobox(\bl,\bl)    \aku \cr\gyoake &
%9
\P_+\bigl(\pmatrix{2\cr1\cr}\twobox(\bl, )\bigr)
=-\sum^\infty_{n=1}q^{3n}
    \left(\matrix{2n+1\cr 2\cr 1\cr} \right)
    \threebox( , , )                                \aku \cr\gyoake &
%10
\P_-\bigl(\pmatrix{2\cr1\cr}\twobox(\bl, )\bigr)
=(1)\onebox( )
    -q(3)\onebox( )
    +\sum^\infty_{n=1}q^{3n-1}{2n+1\choose2}\twobox( , )
                \aku \cr\gyoake &
    +\sum^\infty_{n=2}q^{3n+1}\left(\matrix{2n\cr 2\cr 1\cr} \right)
    \threebox(\bl, , )
                           \aku \cr\gyoake &
%11
\P_+\bigl((4)\onebox()\bigr)=
    {q^{-3}\over[2]^2([3]-1)}(1)\onebox( )
+{q^{-4}+q^{-2}+2+q^2\over[2]^2([3]-1)}(3)\onebox( )
                 \aku \cr\gyoake &
+{q^2\over[2]}\pmatrix{3\cr2\cr}\twobox( , )
   -\sum^\infty_{n=0}{q^{3n-1}[3]\over[2]^2([3]-1)}(2n+5)\onebox( )
                 \aku \cr\gyoake &
   -\sum^\infty_{n=1}{q^{3n}\over[2]}{2n+3\choose2}\twobox( , )
   -\sum^\infty_{n=2}q^{3n+1}{2n+1\choose4}\twobox( , )
            \aku \cr\gyoake &
%12
\P_-\bigl((4)\onebox(\bl)\bigr)=
{[3]\over[2]^2([3]-1)}\phi
   -{q^{-1}\over[2]^2([3]-1)}(2)\onebox(\bl)
  + {1-q[3][2]\over[2]^2([3]-1)}(4)\onebox(\bl)
\aku \cr\gyoake &
   -{1\over[2]}\pmatrix{3\cr1\cr}\twobox( ,\bl)
   -{q^4\over[2]}\pmatrix{4\cr2\cr}\twobox(\bl, )
   +\sum^\infty_{n=1}{q^{3n-2}[3]\over[2]^2([3]-1)}(2n+4)\onebox(\bl)
\aku \cr\gyoake &
   +\sum^\infty_{n=1}{q^{3n-2}\over[2]}{2n+3\choose1}\twobox( ,\bl)
   +\sum^\infty_{n=2}q^{3n-1}{2n+1\choose3}\twobox( ,\bl)
\aku \cr\gyoake &
+\sum^\infty_{n=2}{q^{3n-1}\over[2]}{2n+2\choose2}\twobox(\bl, )
   +\sum^\infty_{n=3}q^{3n}{2n\choose4}\twobox(\bl, )
\aku \cr\gyoake &
%13
\P_+\bigl(\pmatrix{3\cr1\cr}\twobox({},\bl)\bigr)=
    {q\over[2]([3]-1)}(1)\onebox( )
   -{[3]\over[2]([3]-1)}(3)\onebox( )
               \aku \cr\gyoake &
+\sum^\infty_{n=0}{q^{3n-1}\over[2]([3]-1)}(2n+5)\onebox( )
-\sum^\infty_{n=1}q^{3n}
\left(\matrix{2n+2\cr 2 \cr 1\cr }\right)
   \threebox(\bl, , )
   -\sum^\infty_{n=2}q^{3n+1}\left(\matrix{2n+1\cr 3 \cr 1\cr }\right)
   \threebox( ,\bl, )
\aku \cr\gyoake &
%14
\P_-\bigl(\pmatrix{3\cr1\cr}\twobox({},\bl)\bigr)
   = -{1\over[2]([3]-1)}\phi
     +{q^{-1}[3]\over[2]([3]-1)}(2)\onebox(\bl)
     -{q^{-2}\over[2]([3]-1)}(4)\onebox(\bl)
     +q^2{4\choose2}\twobox(\bl, )
\aku \cr\gyoake &
     -\sum^\infty_{n=1}{q^{3n-2}\over[2]([3]-1)}(2n+4)\onebox(\bl)
+\sum^\infty_{n=2}q^{3n-1}\left(\matrix{2n+2\cr 2\cr }\right)
      \twobox(\bl, )
     +\sum^\infty_{n=2}q^{3n}\left(\matrix{2n\cr 3\cr 1\cr }\right)
      \threebox(\bl,\bl, )
\aku \cr
\endeq
\par
\end